\documentclass[%
 superscriptaddress,
 twocolumn,
 amsmath,amssymb,
 aps,superscriptaddress,
 prc,longbibliography
]{revtex4-2}

\usepackage[colorlinks=true,allcolors=blue]{hyperref} % blue cross-refs
\usepackage{graphicx}% Include figure files
\usepackage{dcolumn}% Align table columns on decimal point
\usepackage{bm}% bold math
\usepackage{physics} % Physics symbols
\newcolumntype{d}[1]{D{.}{.}{#1}} % Specify float format with column d{x.x}
\usepackage{xcolor}
\usepackage{soul}
\usepackage{amsthm}

\usepackage{orcidlink}
\usepackage{algorithm}
\usepackage{algorithmic}
\usepackage[normalem]{ulem}
\usepackage{bbold}
\usepackage{multirow}
\usepackage{booktabs}
\usepackage{makecell}
\usepackage{mathtools}
\begin{document}

%\preprint{APS/123-QED}
\title{Surrogate Models for Linear Response}
%\title{Efficient surrogates for linear response theory}
%I worry using 'emulating' makes it seem like some physics-based simplification, rather than attempt to machine learn the complicated method itself

\author{L. Jin\orcidlink{0000-0002-0533-9949}}
\email[]{jinl@frib.msu.edu}
\affiliation{Facility for Rare Isotope Beams, Michigan State University, East Lansing, Michigan 48824, USA}

\author{A. Ravli\'c\,\orcidlink{0000-0001-9639-5382}}
\email[]{ravlic@frib.msu.edu}
\affiliation{Facility for Rare Isotope Beams, Michigan State University, East Lansing, Michigan 48824, USA}
\affiliation{Department of Physics, Faculty of Science, University of Zagreb, Bijeni\v cka c. 32, 10000 Zagreb, Croatia}

\author{P. Giuliani\,\orcidlink{0000-0002-8145-0745}}
\email[]{giulianp@frib.msu.edu}
\affiliation{Facility for Rare Isotope Beams, Michigan State University, East Lansing, Michigan 48824, USA}

\author{K. Godbey\,\orcidlink{0000-0003-0622-3646}}
\email[]{godbey@frib.msu.edu}
\affiliation{Facility for Rare Isotope Beams, Michigan State University, East Lansing, Michigan 48824, USA}

\author{W. Nazarewicz\,\orcidlink{0000-0002-8084-7425}}
\email[]{witek@frib.msu.edu}
\affiliation{Facility for Rare Isotope Beams, Michigan State University, East Lansing, Michigan 48824, USA}
\affiliation{Department of Physics and Astronomy, Michigan State University, East Lansing, Michigan 48824, USA}

\date{\today}

\begin{abstract}
Linear response theory is a well-established method in physics and chemistry for exploring excitations of many-body systems. In particular, the quasiparticle random-phase approximation (QRPA) provides a powerful microscopic framework by building excitations on top of the mean-field vacuum; however, its high computational cost limits model calibration and uncertainty quantification studies. Here, we present two complementary QRPA surrogate models and apply them to study response functions of finite nuclei. One is a reduced-order model that exploits the underlying QRPA structure, while the other utilizes the recently developed parametric matrix model algorithm to construct a map between the system's Hamiltonian and observables. Our benchmark applications, the calculation of the electric dipole polarizability of ${}^{180}$Yb and the $\beta$-decay half-life of ${}^{80}$Ni, show that both emulators can achieve 0.1\%--1\% accuracy while offering a six to seven orders of magnitude speedup compared to state-of-the-art QRPA solvers. These results demonstrate that the developed QRPA emulators are well-positioned to enable Bayesian calibration and large-scale studies of computationally expensive physics models describing the properties of many-body systems. 
\end{abstract}

\maketitle

\section{Introduction}
Linear response theory provides a general framework for describing the response of a system to a weak external perturbing field~\cite{Kubo1966a}. It is fundamentally tied to the fluctuation-dissipation theorem by relating the ground-state thermal fluctuations to a dissipative external field. Because of this universality, linear response theory finds applications across a wide range of disciplines of natural sciences. In condensed matter physics, it underlies the study of dielectric functions and optical response of the system~\cite{Onida2002a,Aryasetiawan1998a}; in nuclear physics, it provides a microscopic description of collective excitations governing nuclear reactions and decay  processes~\cite{Osterfeld1992a,Harakeh2001a}. In quantum chemistry, linear response theory forms the foundation for time-dependent methods that describe excited-state and ground-state correlations in molecular systems~\cite{Szabo1996a,Casida2012a}, while in molecular biology, it has been used to analyze cellular responses to external perturbations~\cite{Molinelli2013a}. Beyond these domains, it also applies to classical statistical mechanical systems \cite{Lucarini2018a,Ruelle2009a}, non-equilibrium fluctuations in complex quantum systems \cite{Guarnieri2024a}, and even to modeling fluctuations in financial markets \cite{Puertas2021a}.

A specific formulation of linear response theory was developed by Bohm and Pines \cite{Pines1952a, Bohm1951a, Bohm1953a} to describe electron-electron interactions within a degenerate electron gas embedded in a uniform positive background. In this system, the total contribution from background and interactions between electrons and positive ions cancels out, leaving the electron-electron interaction as the dominant effect. The collective motion of the electrons gives rise to plasma oscillations -- quantized as plasmons. Although not explicitly named in the original work, this framework became known as the \textit{random-phase approximation} (RPA), referring to the partial cancellation of out-of-phase density-fluctuation terms in the Fourier expansion.

% particular organization of density fluctuations in the Fourier series, which yielded s between out-of-phase contributions.
Within many-body perturbation theory, the same approximation corresponds to a resummation of ring diagrams~\cite{Gell-Mann1957a}.
%ground-state energy can be expanded through a series of ring diagrams, also called the \textit{ring approximation}~\cite{Gell-Mann1957a}. 
Today, the RPA is widely used in quantum chemistry and material science to study electronic excitations, and evaluate ground-state correlation energies~\cite{Chen2017a,Hesselmann2011a,Yu2021a}.

The RPA equations can be derived in many ways. Within the Green's function approach \cite{Thouless1961a,Fetter1971a} the starting point is the two-particle Green's function, which is treated by  the many-body perturbation theory. To derive the RPA equation, one assumes that the one-particle propagators are diagonal and single-particle levels are solutions of Hartree-Fock (HF) equations, and performs a re-summation of ring diagrams. Starting from the time-dependent Hartree-Fock (TDHF)  equation, by assuming a harmonic time-dependence of the induced density, and expanding the mean-field Hamiltonian up to linear order, one can derive the linear response equation, which has the same form as the RPA \cite{Nakatsukasa2016a}. This approach can be coupled with density functional theory (DFT) in the form of time-dependent DFT (TDDFT) \cite{Ullrich2012a}. Lastly, one can obtain RPA equations for density-independent interactions by assuming a quasi-boson approximation  \cite{Ring2004a,Rowe2010a}. Formally, the RPA is related to other many-body methods, being equivalent to the coupled-cluster doubles theory~\cite{Scuseria2008a}.

The RPA was adapted to the nuclear many-body problem in Ref.~\cite{Rowe1968a}, providing microscopic description of nuclear excitations and collective modes~\cite{Harakeh2001a,Osterfeld1992a}. In particular, an important extension of RPA is due to the presence of nucleonic pairing. First, the description of the nuclear mean field has to be extended by introducing the concept of \textit{quasiparticles}, representing a linear combination of particle and hole states~\cite{Bogolyubov1959a}. Subsequently, the RPA was formulated in the basis of two-quasiparticle excitations,
yielding quasiparticle RPA (QRPA)~\cite{Baranger1960a}. Nuclear response functions and the corresponding excited states are central to a wide range of applications, from nuclear structure theory \cite{Paar2007a,Roca-Maza2018a} and astrophysics \cite{Lund2023a,Langanke2021a,Suzuki2022a}, to the investigation of fundamental symmetries \cite{Engel2017a}. In particular, determining excitation energies and wave functions is essential for computing decay rates and transition probabilities, such as electric and magnetic transitions \cite{Hinohara2013a,Kortelainen2015a}, as well as charge-exchange processes such as  $\beta$ decay \cite{Marketin2016a,Ney2020a}. In nuclear astrophysics, QRPA is used to estimate decay rates important  for the synthesis of elements heavier than iron within the $r$-process nucleosynthesis~\cite{Horowitz2019a}. 

%The most widely used approach for calculating the nuclear response function is the random-phase approximation (RPA). It is derived by constructing a basis of all possible particle-hole ($ph$) excitations, satisfying specific selection rules, and solving the corresponding equations of motion \cite{Paar2007a}. Inclusion of nuclear pairing effects within the space of two-quasiparticle (2qp) excitations results in the quasiparticle RPA (QRPA) equations.

General methods for QRPA matrix diagonalization are discussed in Refs.~\cite{Ring2004a,Suhonen2007a}. A more computationally efficient approach, suited for deformed nuclei, is the finite-amplitude method (FAM) \cite{Nakatsukasa2007a}. Although much development has been achieved in both approaches in recent years, QRPA calculations become increasingly more expensive as the number of possible two-quasiparticle (2qp) excitations grows, especially when  symmetries  of the nuclear ground state are internally broken.

In general, the QRPA equations can be derived  within nuclear density functional theory (DFT), as it provides a robust foundation for practical implementations \cite{Bender2003a}. Nuclear energy density functionals (EDFs) --- basic ingredients of DFT --- can be represented by around a dozen parameters optimized to selected experimental data \cite{Klupfel2009a,Kortelainen2010a}. Up to now, only a few EDFs have been calibrated by including excited-state properties obtained within the QRPA \cite{Yuksel2019a,Lin2025a}. Indeed,  a  comprehensive description of giant resonances and related observables,  would require a dataset consisting of a wide range of nuclei and nuclear response functions. However, such considerations are currently computationally prohibitive -- a full-fledged Bayesian model calibration~\cite{Phillips2021a} would require thousands to millions of model calculations, which remains beyond the reach of high-fidelity QRPA solvers.

Addressing the computational demands of modern models has motivated the development of emulators, or surrogate models.
% ~\cite{mcdonnell2015uncertainty, wesolowski2019exploring, melendez2019quantifying, surer2022, chan2023surrogate, svensson2023inference,frame2018eigenvector,furnstahl2020efficient,melendez2021fast,drischler2021toward,konig2020eigenvector,Zhang:2021jmi,melendez2022model,garcia2023wave,sarkar2021convergence,sarkar2022self,demol2020improved,djarv2022bayesian,yapa2022volume,franzke2022excited,bonilla2022training,anderson2022applications,giuliani2023bayes,bai2021generalizing,BUQEYE,ekstrom2019global,rose2024,anderson2024,Cook2024a,duguet2024colloquium, maldonado2025greedy, bakurov2024discovering,reed2024toward,armstrong2025emulators,somasundaram2025emulators,lay2024neural,lalit2024star}
 An emulator is a
fast model that approximates the output of complex, computationally intensive  models, referred  to as Full Order Model (FOM) in the following. Using FOM evaluations, the emulators are trained to connect selected inputs (i.e., the underlying model parameters), with selected outputs (i.e. calculated observables). See Refs.~\cite{Brunton2022a,Boehnlein2022a,Bonilla2022a,Duguet2024a} for some recent applications.

We highlight two general approaches for building emulators: data-driven and model-driven. Data-driven emulators usually attempt to connect inputs and outputs of a model directly, as exemplified by Gaussian Processes~\cite{Gramacy2020a} and neural networks. Emulators following a model-driven approach usually perform this task by finding a set of reduced coordinates for the model's latent variables (i.e., wave functions, densities), and constructing their dynamical equations by exploiting the original operators~\cite{Benner2015a}. 

Emulators based on the reduced basis method (RBM)~\cite{Quarteroni2015a,Hesthaven2016a} are examples that have followed the model-driven approach by identifying the reduced coordinates as amplitudes of a small tailored basis, and constructing the equations by projecting the operators into a small linear subspace. Since model-driven emulators are informed by the underlying system's equations --- which often include physics-driven constraints --- they require appreciably less training data to achieve similar predictive accuracy compared to their data-driven counterparts. However, this structural integration also makes them more cumbersome to construct~\cite{Giuliani2023a,Odell2024a}. These challenges have motivated developments to augment model-driven emulators to construct not only the low-dimensional latent space characterized by a set of reduced coordinates, but also the governing equations for those coordinates. This insight has enabled the creation of many successful emulators and model discovery tools in science and engineering broadly~\cite{Burohman2023a,Brunton2016a,Peherstorfer2016a,Benner2020a,Champion2019a,Loiseau2018a}, and within nuclear science and engineering applications in particular~\cite{Armstrong2025a,Somasundaram2025a,Cook2025a,
Bakurov2025,%Figueroa2024a,
Reed2024a,Xiao2024a,DiRonco2020a,Hardy2024a}. 

Based on these developments, in this work we create and benchmark two emulators for the QRPA calculations of electric dipole polarizability  and $\beta$-decay half-life $T_{1/2}$. The first emulator follows a hybrid data- and model-driven approach by identifying effective reduced coordinates to represent the strength function, and constructing the necessary equations by mirroring the structure of the QRPA eigensystem. From the emulated strength function, the system's observables can be calculated. The second emulator adopts the Parametric Matrix Model (PMM) \cite{Cook2025a} architecture in a purely data-driven setting: its trainable parametric matrices are optimized to predict the target observables directly, bypassing the intermediate construction of the strength function. We train both emulators using the FOM evaluations, and compare their performance in reproducing observables as a function of their  complexity. Emulators developed in this work are publicly available~\cite{GitHub2025a}. We have also developed online Yinteractive visualization tools to further illustrate key concepts of the work~\cite{smlrWebsite}.
 
 The paper is organized as follows. In Sec. \ref{sec:qrpa} we present the theoretical formalism behind  QRPA, while the corresponding emulators are described in Sec. \ref{sec:emulators}. In Sec. \ref{sec:results}, we present the emulation results for electric  dipole polarizability and $\beta$-decay half-lives. Finally, Sec. \ref{sec:conclusion} contains the summary of the paper. The glossary of acronyms and symbols  used  can be found in Appendix~\ref{sec: appendix}. 

\section{Quasiparticle Random Phase Approximation}\label{sec:qrpa}

We now present a summary of the QRPA formalism. In this work, the ground-state configuration of the nucleus is obtained by solving the Relativistic Hartree-Bogoliubov (RHB) equations in either a spherical or axially deformed geometry, with the mean-field generated self-consistently by the underlying EDF \cite{Niksic2014a}. This self-consistent solution yields the static generalized density matrix
$\mathcal{R}^0$. Subjected to an external field $\mathcal{F}(t)$, the system evolves according to the time-dependent generalized density:
\begin{equation}\label{eq:time_dependent_density}
\mathcal{R}(t) = \mathcal{R}^0 + \delta\mathcal{R}(t),
\end{equation}
with the harmonic ansatz in the time domain
\begin{equation}\label{harmonic_ansatz}
\delta\mathcal{R}(t)
= \delta\mathcal{R}(\omega)\,e^{-i\omega t} + \delta\mathcal{R}^\dagger(\omega)\,e^{i\omega t}.
\end{equation}
The linearization of the time-dependent RHB equations in $\delta \mathcal{R}$ leads to the QRPA eigenvalue problem. In the two-quasiparticle basis, the dimension of the QRPA matrix is given by
\begin{equation}\label{qrpa_dimensions}N_d=2n_{\rm 2qp},\end{equation}
where $n_{\rm 2qp}$ denotes the number of two-quasiparticle excitations allowed by selection rules \cite{Ring1984a}.

The linear response equation takes the form \cite{Paar2007a}:
\begin{equation}\label{eq:linear_response_equation}
\left[ \mathcal{M}(\boldsymbol{\alpha}) - \omega \mathcal{N} \right] \delta \mathcal{R}_{\boldsymbol{\alpha}}(\omega) = - \mathcal{F}(\boldsymbol{\alpha}),
\end{equation}
where the metric $\mathcal{N}$ is 
\begin{equation}\label{eq:norm matrix}
\mathcal{N}=
\begin{pmatrix}
\mathbb{1}_{n_{2qp}} & 0 \\
0 & -\mathbb1_{n_{2qp}}
\end{pmatrix}_{N_d\times N_d},
\end{equation}
and $\omega$ is the (generally complex) excitation energy, while $\boldsymbol{\alpha}$ denotes the set of $k$ parameters defining the interaction --- typically the coupling constants of the underlying EDF.
The QRPA Hermitian matrix $\mathcal{M}(\boldsymbol{\alpha})$ has the block structure~\cite{Ring2004a}
\begin{equation}\label{eq: QRPA Matrix}
\mathcal{M}(\boldsymbol{\alpha}) = \begin{pmatrix}
A(\boldsymbol{\alpha}) & B(\boldsymbol{\alpha}) \\
B^*(\boldsymbol{\alpha}) & A^*(\boldsymbol{\alpha})
\end{pmatrix}_{N_d\times N_d},
\end{equation}
where $A, B$ are complex matrices constructed from the residual two-body interaction, satisfying $A^\dag = A$, and $B^T = B$.
The external field vector $\mathcal{F}(\boldsymbol{\alpha}) = (F^{20}(\boldsymbol{\alpha}), F^{02}(\boldsymbol{\alpha}))$ represents the external field operator
in the quasiparticle basis. The QRPA amplitudes are of the form $\delta \mathcal{R}_{\boldsymbol{\alpha}}(\omega) = (x_{\boldsymbol{\alpha}}(\omega), y_{\boldsymbol{\alpha}}(\omega))$.

Once the QRPA amplitudes are computed, the strength function -- a physical observable representing the response of the system to the external field-- can be written  as:
\begin{equation}\label{eq:linear_response_strength}
    S(\boldsymbol{\alpha}; \omega) = -\frac{1}{\pi}\text{Im} \mathcal{F}(\boldsymbol{\alpha})^\dag \delta \mathcal{R}_{\boldsymbol{\alpha}}(\omega).
\end{equation}

While the QRPA amplitudes may, in principle, be obtained by the matrix inversion, the dimensionality of the quasiparticle space $n_{\rm 2qp}$ is often prohibitively large, rendering the direct approach computationally infeasible. Accordingly, alternative methods have been developed. One such method reformulates the linear response equation, Eq. (\ref{eq:linear_response_equation}), as a variational problem within the finite amplitude method (FAM) \cite{Nakatsukasa2007a, Kortelainen2015a}. Other techniques exploit symmetries and structure in the residual interaction to bypass explicit inversion in the large quasiparticle basis \cite{Ravlic2021a, Ravlic2024a}.

The QRPA linear response equation may also be cast as a non-Hermitian eigenvalue problem, commonly referred to as the matrix QRPA:
\begin{equation}\label{eq: non-Hermician Eq}
\mathcal{N}
\mathcal{M}(\boldsymbol{\alpha}) \delta \mathcal{R}_{\boldsymbol{\alpha}; i} = E_{\boldsymbol{\alpha}; i} \delta \mathcal{R}_{\boldsymbol{\alpha}; i}, \quad i = 1, \ldots , N_d
% 2n_{\rm 2qp},
\end{equation}
 where 
  $\delta \mathcal{R}_{\boldsymbol{\alpha}; i} = (x_{\boldsymbol{\alpha}; i}, y_{\boldsymbol{\alpha}; i})$
 are the $N_d$ eigenvectors corresponding to eigenvalues $E_{\boldsymbol{\alpha};i}$. In general, the spectrum of the QRPA matrix consists of complex-conjugate pairs $\{E_{\boldsymbol{\alpha};i}, -E_{\boldsymbol{\alpha};i}^* \}$. However, if the matrix $\mathcal{M(\boldsymbol{\alpha})}$ is positive definite (this condition is often satisfied when the QRPA is constructed on top of a self-consistent RHB minimum), then the eigenvalues of \eqref{eq: non-Hermician Eq}
 are real.
 To compute transition matrix elements between the ground state and excited states, we define the reduced transition probability:
\begin{equation}\label{eq:reduced_transition_probability}
\mathcal{B}_{\boldsymbol{\alpha}; i} = \left| \begin{pmatrix}
F^{20}(\boldsymbol{\alpha}) \\
F^{02}(\boldsymbol{\alpha})
\end{pmatrix}^\dag \begin{pmatrix}
x_{\boldsymbol{\alpha}; i} \\
y_{\boldsymbol{\alpha}; i}
\end{pmatrix} \right|^2.
\end{equation}
 In practice, the discrete strengths $\mathcal{B}_{\boldsymbol{\alpha}; i}$ are folded with a Lorentzian of width $\eta$ to produce a continuous strength function:

\begin{equation}\label{eq:strength_function}
S(\boldsymbol{\alpha}; \omega^\prime + i \eta) = \frac{\eta}{2 \pi} \sum \limits_{i=1}^{N_d/2} \frac{\mathcal{B}_{\boldsymbol{\alpha}; i}}{(\omega^\prime - E_{\boldsymbol{\alpha}; i})^2 + (\eta/2)^2},
\end{equation}
where  $\omega = \omega^\prime$.
Once the strength function $S(\omega)$ is obtained, other observables may be
computed. In this work, we focus on the electric dipole polarizability $\alpha_D$ and the $\beta$-decay half-life $T_{1/2}$, both of which are strongly sensitive to nuclear excitations.
Both observables are important. The dipole polarizability correlates closely with the neutron skin thickness, offering a stringent constraint on the density dependence of the symmetry energy.  The $\beta$-decay half-life, on the other hand,  is essential for a realistic description of charge-exchange nuclear excitations. We now summarize how these two observables are calculated from the QRPA strength functions.

\subsection{Dipole Polarizability $\alpha_D$}

Physically, the electric dipole polarizability determines the response of the nuclear system to an external electric field. This quantity is of significant interest in both nuclear structure physics and astrophysics due to its strong correlation with the symmetry energy of nuclear matter, as well as the neutron skin thickness \cite{Reinhard2010a,Piekarewicz2012a,Roca-Maza2018a}. It can be expressed through the QRPA strength function as \cite{Lipparini1989a}:
\begin{equation}\label{eq:alphaD}
    \alpha_D(\boldsymbol{\alpha}) = \frac{8 \pi}{9}e^2 \text{lim}_{\eta \to 0^+} \int d\omega^\prime \frac{S(\boldsymbol{\alpha};\omega^\prime + i\eta)}{\omega^\prime}.
\end{equation}
 In terms of all the $N_d$ eigenpairs, involving the discrete strengths $\mathcal{B}_{\boldsymbol{\alpha};i}$ and eigenvalues $E_{\boldsymbol{\alpha};i}$, the above equation can be rewritten as:
\begin{equation}\label{eq:alphaD_discrete}
    \alpha_D(\boldsymbol{\alpha}) = \frac{8 \pi}{9}e^2 \sum\limits_{i = 1}^{N_d/2} \frac{\mathcal{B}_{\boldsymbol{\alpha};i}}{E_{\boldsymbol{\alpha};i}}.
\end{equation}
In practical calculations of $\alpha_D$ in deformed nuclei with $N_d > 10^5$, contour integration techniques \cite{Hinohara2013a} eliminate the need to work directly with the strength function. 

\begin{figure*}[!ht]
\includegraphics[width=1\textwidth]{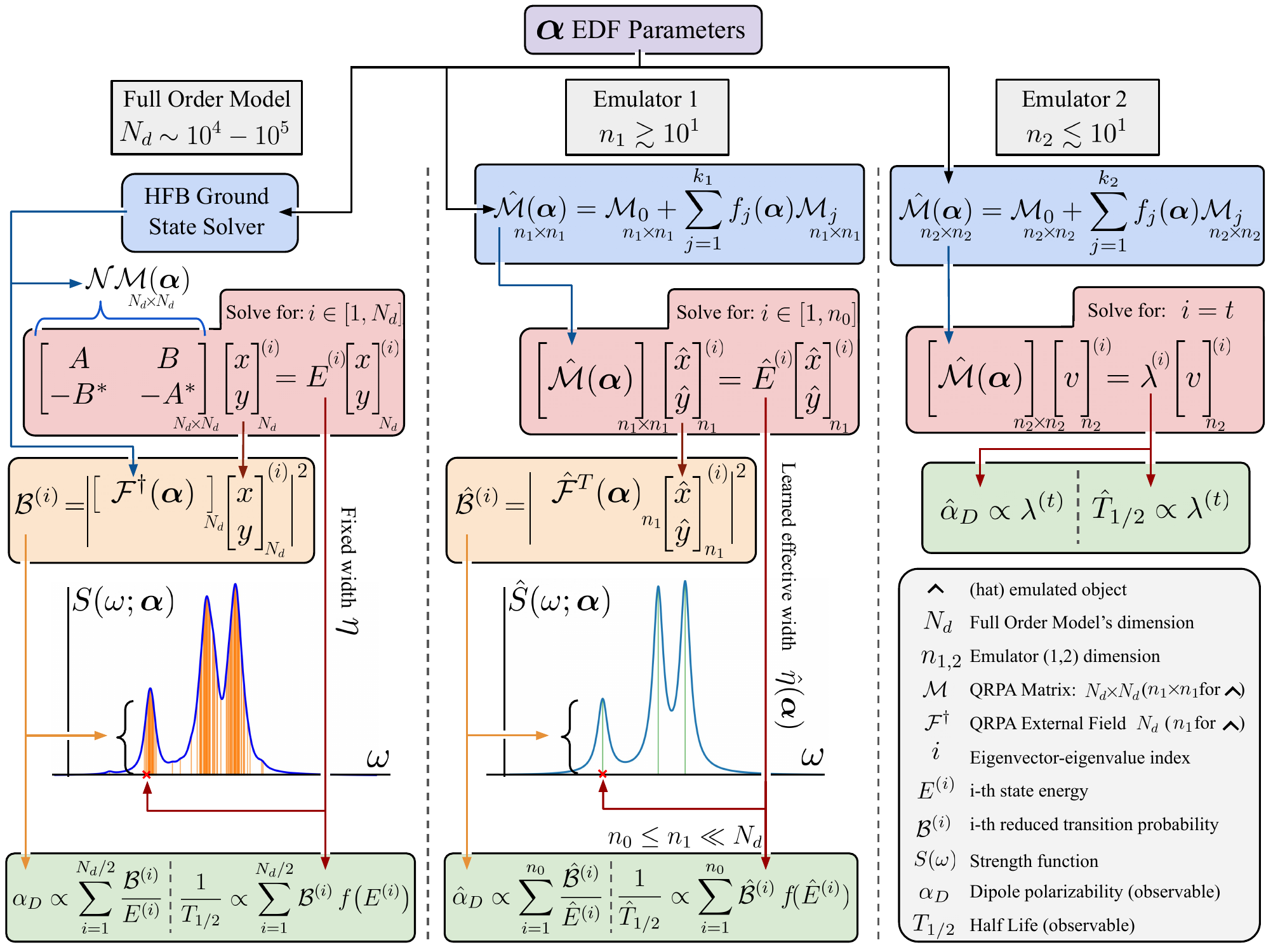}
\caption{Flowchart illustrating the computational algorithms for the full order model described in Sec.~\ref{subsec: FOM} and the two emulators described in Secs.~\ref{subsec:emulator 1} and~\ref{subsec:emulator 2}. The three approaches aim to connect the controlling EDF parameters $\boldsymbol{\alpha}$ (purple box at the top) with the two observables $\alpha_D$ and $T_{1/2}$ (green boxes at the bottom).  A detailed description of the diagram is given in the text.}    
    \label{fig:Flowchart}
\end{figure*}

\subsection{$\beta$-decay half-life $T_{1/2}$}

$\beta$-decay is a nuclear decay governed by the weak interaction. Within QRPA, we can compute the charge-exchange linear-response in the allowed Gamow-Teller (GT) channel to obtain the GT strength function $S_{GT}(\boldsymbol{\alpha};\omega)$ \cite{Marketin2016a, Ney2020a, Ravlic2025a}. Restricting to allowed transitions, the inverse half-life is given by:
\begin{equation}\label{eq: t1/2}
\begin{aligned}
    &\frac{1}{T_{1/2}(\boldsymbol{\alpha})}=\frac{g_A^2}{K} 
    \lim_{\eta \to 0^+} 
    \int \limits_{-\lambda_{np}}^{\Delta_{nH}} 
    d\omega \, f(\Delta_{np} - \omega, Z, A) 
    S_{GT}(\boldsymbol{\alpha}; \omega),
\end{aligned}
\end{equation}
for a nucleus with $Z$ protons and $A$ total nucleons, where $g_A = 1.2$ is the axial-vector coupling (for which we take the vacuum value) and $K = 6147$ s is the constant measured in superallowed $\beta$ decays \cite{Marketin2016a}. The 
integration in \eqref{eq: t1/2}
 is performed from $\lambda_{np} = \lambda_{n} - \lambda_p$, where $\lambda_{n(p)}$ is the neutron(proton) chemical potential, determined by solving the underlying RHB equations, up to the neutron-hydrogen atom energy difference $\Delta_{nH} = 0.782$ MeV, effectively enclosing the $Q_\beta$-value window. The phase-space factor $f(E_0, Z, A)$ takes into account the kinematics of outgoing leptons with the end-point energy $E_0 = \Delta_{np} - \omega$, where the neutron-proton mass difference is $\Delta_{np} = 1.293$ MeV. The function $f(E_0, Z, A)$ accounts for Coulomb distortion of electron wave functions through the Fermi function dependence \cite{Behrens1971a}. Having calculated the full QRPA spectrum, Eq.~\eqref{eq: t1/2} can be written as
\begin{equation}\label{eq:t12_discrete}
 \frac{1}{T_{1/2}(\boldsymbol{\alpha})}=\frac{g_A^2}{K} 
\sum \limits_{i = 1}^{N_d/2} f(\Delta_{np} - E_{\boldsymbol{\alpha};i}, Z, A) 
    \mathcal{B}_{\boldsymbol{\alpha};i},
\end{equation}
where the phase-space function imposes the condition $E_{\boldsymbol{\alpha};i} < \Delta_{nH}$.

\section{Full and Reduced Order Models}\label{sec:emulators}
In this section, we illustrate the methodology we developed to reduce the computational cost associated with QRPA calculations. The core principle behind various emulation strategies is that the parametric or time dynamics of high-dimensional objects can be approximated through a much smaller set of coordinates or degrees of freedom. Here, we are interested in characterizing the outcomes of the QRPA calculations, namely the strength function $S(\omega)$, Eq.~\eqref{eq:strength_function}, and the two observables $\mathcal{O}\equiv [\alpha_D,T_{1/2}]$
as we vary the  parameters $\boldsymbol{\alpha}$. 

Figure~\ref{fig:Flowchart} shows a diagrammatic comparison of how the FOM and the two emulators we developed connect the control variables $\boldsymbol{\alpha}$ to the response variables $S(\omega)$ and $\mathcal{O}$. In the following,  we use the ``hat" symbol to represent approximations to quantities of interest. For example, $\hat S(\omega)$  represents a low-dimensional approximation to the original strength function $S(\omega)$. We now proceed to describe these approaches and how the two emulators are trained by using a set of FOM evaluations.

 \subsection{Full Order Model}\label{subsec: FOM}
 The computational chart for the FOM is shown in the left column of Fig.~\ref{fig:Flowchart}. The FOM operates with matrices of dimensions $N_d\sim10^4-10^5$
 % The FOM shown in the left column of Fig.~\ref{fig:Flowchart} has an intrinsic dimension of $N_d\sim10^4-10^5$
 and connects $\boldsymbol{\alpha}$ with the observable $\mathcal{O}$ by first solving for the system's ground state through the RHB equations (blue box) from which the high-dimensional QRPA matrix $\mathcal{M}_{N_d \times N_d}(\boldsymbol{\alpha})$ is created.
 
The isovector dipole strength $S(\boldsymbol{\alpha};\omega)$ is computed with the axially-deformed FAM code of Ref.~\cite{Bjelcic2020a}, based on the relativistic point-coupling EDF DD-PC1~\cite{Niksic2008a} in the mean field channel, and the pairing interaction from Ref.~\cite{Tian2009a}. The DD-PC1 EDF contains twelve parameters: four scalar couplings ($a_S, b_S, c_S, d_S$); three vector couplings  $(a_V, b_V, d_V)$; two isovector couplings ($b_{TV}, d_{TV}$); one derivative coupling $\delta_S$, and two pairing strengths ($G_{p}, G_n$).

Since $\alpha_D$ -- and more generally the isovector giant dipole resonance (IVGDR) 
-- are most sensitive to the isovector sector of the EDF \cite{Reinhard2010a,Piekarewicz2012a}, we perform systematic calculations of the strength function by varying the two parameters: $b_{TV}$ and $d_{TV}$.

\begin{figure}
    \includegraphics[width=\linewidth]{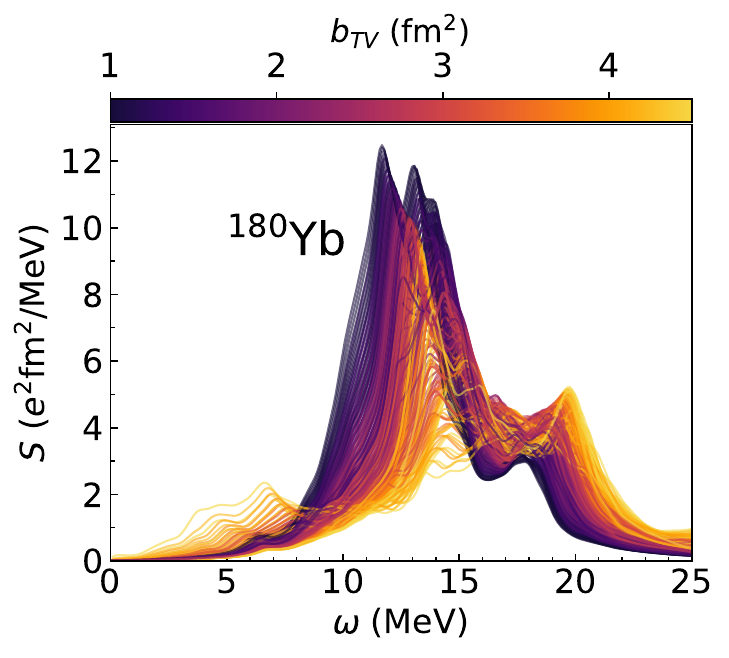}
    \caption{The total isovector electric dipole strength function in ${}^{180}$Yb calculated with the FAM QRPA. Starting from the DD-PC1 EDF parameterization, each line corresponds to an individual QRPA calculation at fixed parameters $b_{TV}$ and $d_{TV}$; each curve is colored according to the value of $b_{TV}$. A total of 225 calculations were performed on a regular grid in the parameter space spanning $b_{TV} \in [1.0,4.5]$ fm$^2$ and $d_{TV}  \in [0.1,2.0]$. (See Fig.~\ref{fig:YbNiPerformance}(a) for a graphical representation of the parameter distribution, and Fig.~\ref{fig:ManyCurves}(a) for the response of  $\alpha_D$ to changes in  $b_{TV}$ and $d_{TV}$.)}
    \label{fig:isovector_dipole_example}
\end{figure}

All FAM QRPA calculations are carried out for a deformed nucleus  ${}^{180}$Yb, in a stretched axially-deformed harmonic oscillator basis with $N_{osc} = 16$ shells. Both $K = 0$ and $K = 1$ projections of the dipole operator are included. The strength function is evaluated on the interval $\omega \in [0,40]$ MeV with a step size of 0.1 MeV and a Lorentzian smearing of $\eta = 1$ MeV. The dipole polarizability is calculated by placing the semi-circular contour around $\omega = 0$ in the complex energy plane with radius $\gamma = 0.1$ MeV, and performing the integral by Simpson integration using 10 discretized points, following the prescription from Ref. \cite{Hinohara2015a}. Each calculation is run on 40 CPUs with 10 GB of memory per CPU on the MSU High Performance Computing Cluster. Evaluating the full strength function requires $\mathcal{O}(10^4)$ s of wall-clock time. Figure~\ref{fig:isovector_dipole_example} displays the resulting strength distributions as $b_{TV}$ and $d_{TV}$ are varied.
%%%%%%%

\begin{figure}
    \includegraphics[width=\linewidth]{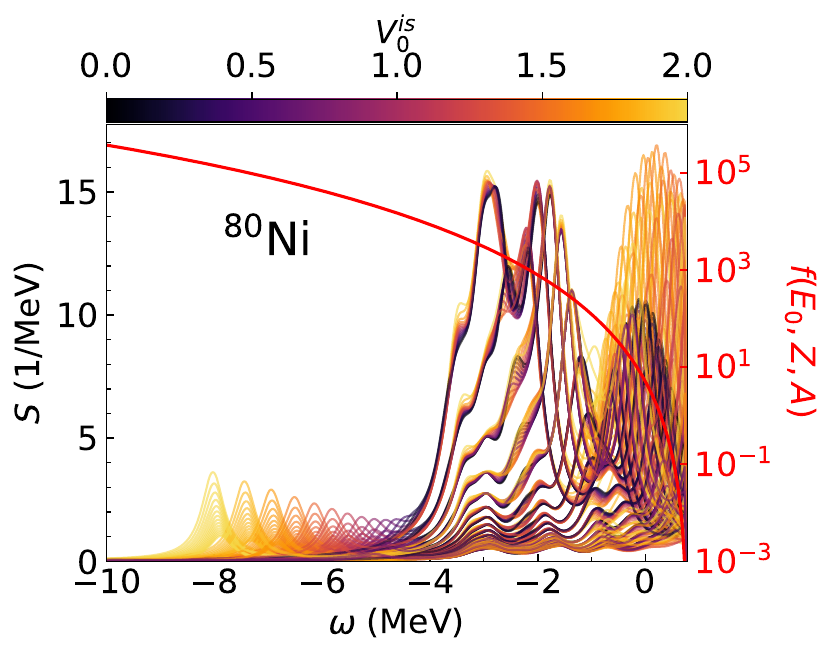}
    \caption{The Gamow-Teller strength function in ${}^{80}$Ni, within the $Q_\beta$-window, calculated with the linear response QRPA using DD-PC1 EDF parameterization at fixed parameters $g_0$ and $V_0^{is}$; each  curve is colored according to its value of $V_0^{is}$. The red solid line indicates the the phase-space factor $f(E_0,Z,A)$ at $E_0 = \Delta_{np} - \omega$.  A total of 225 calculations were performed on a regular grid in the parameter space spanning $V_0^{is} \in [0.0,2.0]$ and $g_0  \in [0.0,1.0]$. (See Fig.~\ref{fig:YbNiPerformance}(c) for a graphical representation of the parameter distribution, and Fig.~\ref{fig:ManyCurves}(b) for the response of  $T_{1/2}$ to changes in  $g_0$ and $V_0^{is}$.)}
    \label{fig:gamow_teller_example}
\end{figure}

As an example of charge-exchange response, we focus on studying the $\beta$-decay spectrum of the neutron-rich ${}^{80}$Ni nucleus, as the GT strength of nuclei close to the drip line shows considerable complexity within the $Q_\beta$-window. Since this nucleus is spherical in its ground state, we employ the spherical matrix QRPA solver from Ref. \cite{Vale2021a}, which guarantees that $N_d \approx 10^4$, allowing for direct diagonalization and obtaining the complete spectrum $\{E_{\boldsymbol{\alpha}; i}, \mathcal{B}_{\boldsymbol{\alpha};i } \}$. Therefore, in this case, we directly employ Eq.~\eqref{eq:t12_discrete} to calculate the half-lives. The ground-state configuration is obtained by solving the RHB equations in spherical harmonic oscillator basis with $N_{osc} = 20$. The strength function is sampled over two parameters, the isoscalar pairing $V_0^{is}$ and the Landau-Migdal coupling $g_0$. These correspond to time-odd terms, which need careful calibration to reproduce measured half-lives within the EDF framework \cite{Ravlic2025a, Ney2020a, Marketin2016a}. The matrix diagonalization and subsequent calculation of half-life is performed on one node utilizing 56 cores, requiring around $\mathcal{O}(10^2)$\,s. Figure~\ref{fig:gamow_teller_example} shows the calculated Gamow-Teller strength functions for ${}^{80}$Ni across the ranges of the two parameters  $g_0$ and $V_0^{is}$ we explore in this work. 

%%%%%%%
%%%%%%%%
 
 Using one of the previously described QRPA solvers, a spectral decomposition numerically yields all eigenpairs of the QRPA matrix, namely the $N_d$ eigenstates $[x,y]^{(i)}$ and eigenergies $E^{(i)}$ (red box). These eigenpairs are then used to construct the strength function $S(\omega)$ as a sum of $N_d$ Lorentzians, see Eq.~\eqref{eq:strength_function}. Each Lorentzian has a fixed width $\eta$, it is centered at energy $E^{(i)}$, and has an amplitude $\mathcal{B}^{(i)}$ obtained by calculating the overlap of the respective eigenstate $[x,y]^{(i)}$ with the external field $\mathcal{F}(\boldsymbol{\alpha})$ (orange box) for $i\in [1,N_d]$. The observables $\alpha_D$ and $T_{1/2}$ are then calculated as explicit functions of the amplitudes $\mathcal{B}^{(i)}$ through Eqs.~\eqref{eq:alphaD_discrete} and~\eqref{eq:t12_discrete}.

\subsection{Emulator 1}\label{subsec:emulator 1}

The computational chart for the first emulator (EM1) is shown in the middle column of Fig.~\ref{fig:Flowchart}. EM1 operates with matrices of dimension $n_1\gtrsim 10^1$ and connects the underlying parameters $\boldsymbol{\alpha}$ with the observables $\mathcal{O}$ by first reproducing the strength function $S(\omega)$ as an intermediate high-dimensional latent step of the calculations. To achieve that, we observe that although $S(\omega)$ is indeed a high dimensional object comprised of the superposition of $N_d$ Lorentzians, it can be well approximated by a much smaller number of $n_0\ll N_d$ Lorentzians (see the schematic strength functions $S(\omega)$ and $\hat S(\omega)$ shown in the bottom part of columns one and two of Fig.~\ref{fig:Flowchart}, respectively). This approximation $\hat S (\omega) \approx S (\omega)$ effectively compresses the original high dimension, providing a much smaller set of reduced coordinates --- the $n_0$ locations, $n_0$ amplitudes, and effective width $\hat \eta$ --- that need to be calculated for each parameter evaluation.

To compute the values of the identified reduced coordinates, EM1 mirrors the structure of the FOM, except that the reduced matrix $\mathcal{\hat M}(\boldsymbol{\alpha})$ is not derived from solving the RHB equations. Instead, akin to data-driven operator learning methods that fit low-dimensional representations of operators directly from FOM data \cite{Peherstorfer2016a, Benner2020a,Kovachki2023a}, and following the PMM structure~\cite{Cook2025a}, we represent this matrix as a linear combination of rank-$n_1$ matrices $\mathcal{M}_j$:
 \begin{equation}\label{eq: matrix hat}
\hat{\mathcal{M}}(\boldsymbol{\alpha}) = \mathcal{M}_0 + \sum \limits_{j=1}^{k_1} f_j(\boldsymbol{\alpha}) \mathcal{M}_j,
\end{equation}
with $k_1$ chosen functions $f_j$ that characterize the parametric dependence on $\boldsymbol{\alpha}$. In this work, we choose $\mathcal{M}_0$ as a real-valued diagonal matrix and $\mathcal{M}_j$ to be real-valued symmetric matrices. A subset $n_0\leq n_1$ of the obtained eigenvectors $[\hat x,\hat y]^{(i)}$  is used to construct the approximate reduced transitions $\mathcal{\hat{B}}^{(i)}$ through dot products with the approximate fixed external field $\mathcal{\hat F}$ learned from FOM evaluations:
\begin{equation}
    \mathcal{\hat{B}}^{(i)} = \left| \hat{\mathcal{F}}^T(\boldsymbol{\alpha}) \ \cdot \  [\hat x,\hat y]^{(i)} \right|^2, \quad i \in  [1,n_0].
\end{equation}
These approximate transition probabilities $\mathcal{\hat{B}}^{(i)}$ form the amplitudes of the $n_0$ Lorentzians that are centered at the obtained eigenenergies $\hat E^{(i)}$ and used to construct the approximate strength function:
\begin{equation}\label{eq: approx S}
    \hat{S}( \omega ;\boldsymbol{\alpha}) = \frac{\hat \eta}{2 \pi} \sum \limits_{i = 1}^{n_0} \frac{\mathcal{\hat{B}}^{(i)}}{(\omega - \hat{E}^{(i)})^2 + (\hat\eta/2)^2}.
\end{equation}
The effective Lorentzian width $\hat \eta(\boldsymbol{\alpha})$ is learned from FOM evaluations and changes with the controlling parameters $\boldsymbol{\alpha}$ to compensate for the fact that the strength function is now built from a greatly reduced  number of Lorentzians  $n_0\ll N_d$. The observables $\hat{\alpha}_D$ and $\hat{T}_{1/2}$ are then obtained from $\hat{\mathcal{B}}$ via Eqs.~\eqref{eq:alphaD_discrete} and \eqref{eq:t12_discrete}, respectively.

% \begin{figure*}
%     \centering
%     \includegraphics[width=\linewidth]{SpectrumEvolution.pdf}
%     \caption{Evolution of the eigenvalue spectrum of EM1 for the Gamow-Teller strength function of $^{80}$Ni as $V_0^{is}$ changes while keeping $g_0=0.5$ constant. The color coding follows the same values as the coloring in Figure~\ref{fig:gamow_teller_example}. The left (black curve) and right (yellow curve) plots correspond to the respective FOM strength functions $S(\omega)$ for the two extreme values of $V_0^{is}=0.0$ and $V_0^{is}=2.0$. The colored lines in the central plot represent the $n_0$ kept eigenvalues of the $n_1\times n_1$ matrix in Eq.~\eqref{eq: matrix hat} as $V_0^{is}$ changes ($n_1=16$, $n_0=14$). These eigenvalues represent the locations of the approximate energies $\hat E_i$ that are used to construct the approximate strength function in Eq.~\eqref{eq: approx S}, and their values for both spectra on the left and right plots are shown as green lines, similar to the three green lines shown in the approximate strength function in the middle column of Figure~\ref{fig:Flowchart}.  The gray line shows the lowest eigenvalue that is not kept to reconstruct the emulated strength function $\hat S(\omega)$, but which dynamics influences the position of the others. The highest non-kept eigenvalue is not visible within this range.  }
%     \label{fig:spectrumEvolution}
% \end{figure*}

\begin{figure}
    \centering
    % L B R T
\includegraphics[width=\linewidth,
  trim=63mm 50mm 130mm 40mm, clip]{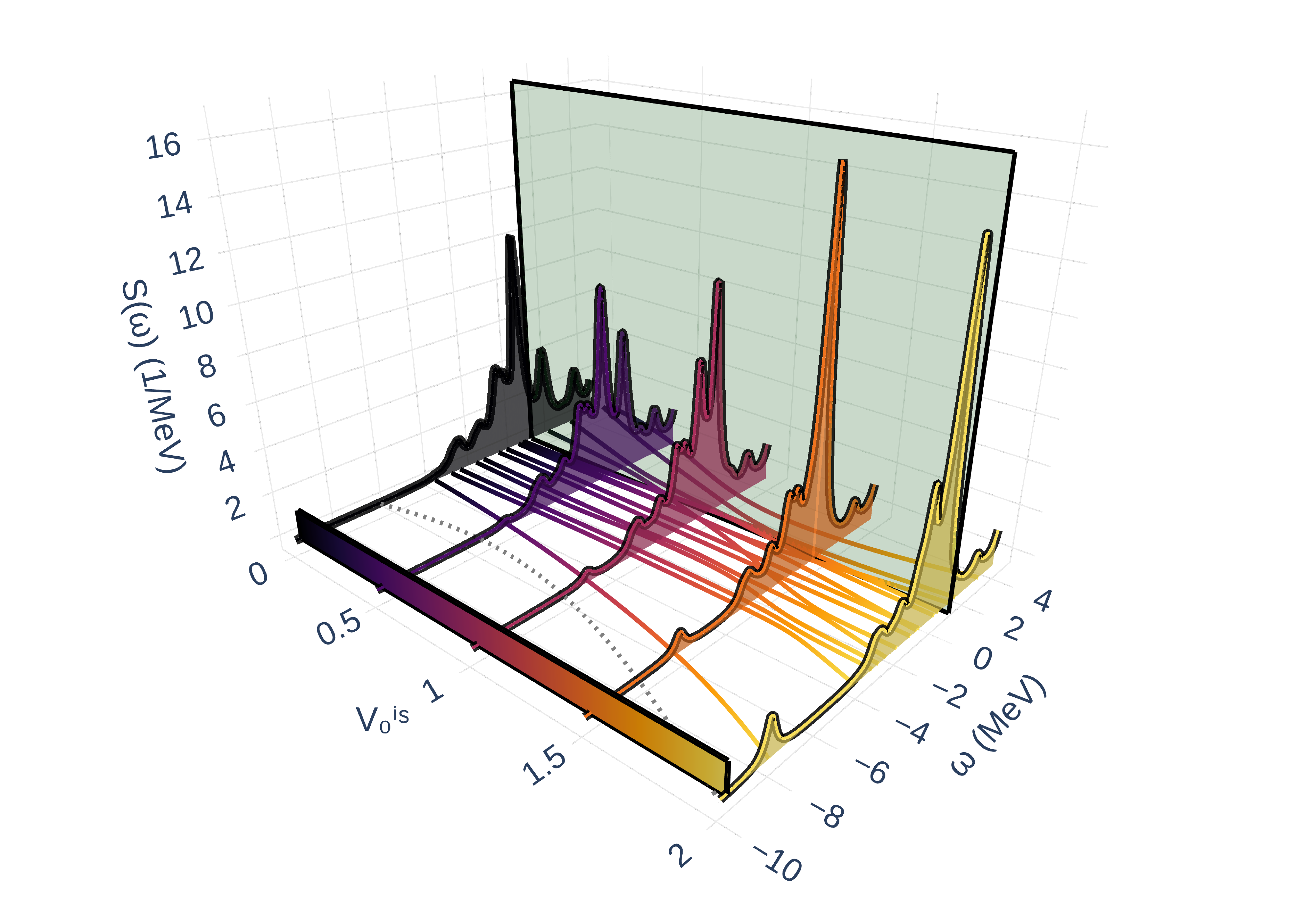}
    \caption{Evolution of the Gamow-Teller strength function in $^{80}$Ni as a function of  $V_0^{is}$  at $g_0=0.5$. The color coding follows the pattern of Fig.~\ref{fig:gamow_teller_example}. The  FOM strength functions $S(\omega)$ at five  values of $V_0^{is}=[0.0,0.5,1.0,1.5,2.0]$ are shown. The  tracks in the horizontal plane represent the subset of $n_0$ eigenvalues of matrix 
    $ \hat{\mathcal{M}}$. These eigenvalues represent the locations of the approximate energies $\hat E_i$ that are used to construct the emulated strength function in Eq.~\eqref{eq: approx S}. The values of $S(\hat E_i)$ are shown as vertical curves. The dotted line shows the lowest eigenvalue that is not retained in emulation; its behavior influences the overall dynamics. The vertical green plane shows the cutoff at $\Delta_{nH} = 0.782$ MeV, representing the limit of the $Q_\beta$-window. For an interactive 3D visualization of the figure see Ref.~\cite{smlrWebsite}.}
    \label{fig:3DSpectrum}
\end{figure}
EM1 is effectively constructing a low-dimensional algorithm 
that follows the original FOM structure as close as possible, with the numerical value of its hyperparameters  --- the matrix elements in Eq.~\eqref{eq: matrix hat} and the parameter dependence of $\hat \eta(\boldsymbol{\alpha})$ --- being tuned by comparing to the FOM evaluations. The specific form of ansatz in Eq. (\ref{eq: matrix hat}) depends on the physics of the problem. For instance, in the case of $\beta$-decay emulations, it is known that the external field vector does not depend on $g_0$ and $V_0^{is}$ parameters, and therefore we choose $\mathcal{F}(\boldsymbol{\alpha}) = \mathcal{F}_0$. Furthermore, since there is no direct relationship between $V_0^{is}$ and $g_0$, we assume the following form for the matrix for calculations involving $^{80}$Ni:
\begin{equation}\label{eq:PMM_t12}
    \mathcal{\hat M}(g_0, V_0^{is}) = \mathcal{M}_0 + (g_0 - \bar{g}_0)\mathcal{M}_1 + (V_{0}^{is} - \bar{V}_0^{is}) \mathcal{M}_2,
\end{equation}
where we define:
\begin{equation} \label{eq: central parameters}
    \begin{pmatrix}
    \bar{g}_0, \bar{V}_0^{is}
\end{pmatrix}
\equiv \boldsymbol{\alpha}_C
\end{equation}
% $\boldsymbol{\alpha}_C\equiv \begin{pmatrix}
%     \bar{g}_0, \bar{V}_0^{is}
% \end{pmatrix}$,
as the central parameters on the training dataset. On the other hand, for the dipole polarizability $\alpha_D$ emulation, selected parameters actually impact the external field matrix $\mathcal{F}(\boldsymbol{\alpha})$. Therefore, for calculations involving $^{180}$Yb, we choose:
\begin{align}\label{eq:PMM_alphaD}
\begin{split}
    \mathcal{\hat F}(b_{TV},d_{TV}) =& \mathcal{F}_0 
    + (b_{TV} - \bar{b}_{TV})\mathcal{F}_1 \\
    &+ (d_{TV} - \bar{d}_{TV})\mathcal{F}_2 ,
\end{split}
\end{align}
where the expansion is done around the central parameters $\boldsymbol{\alpha}_C$.  Furthermore, in DD-PC1, the $b_{TV}$ and $d_{TV}$ parameters appear in a combination  $b_{TV}e^{-d_{TV}x}$, where $x = \rho/\rho_{sat}$ is the ratio between vector density and the saturation density of symmetric nuclear matter~\cite{Niksic2008a}. Therefore, for $\alpha_D$ calculations, we choose $\mathcal{\hat M}$ as:
% \begin{align}
% \begin{split}
%     \mathcal{M}(b_{TV}, d_{TV}) &= \mathcal{M}_0 + (b_{TV} - \bar{b}_{TV})\mathcal{M}_1 + (d_{TV} - \bar{d}_{TV}) \mathcal{M}_2 \\
%     &+ (b_{TV} - \bar{b}_{TV})e^{-(d_{TV} - \bar{d}_{TV})x_1}\mathcal{M}_3,
% \end{split}
% \end{align}
\begin{align}\label{eq: matrix alphaD}
\begin{split}
    \mathcal{\hat M}(b_{TV}, d_{TV}) =& \mathcal{M}_0 +(b_{TV} - \bar{b}_{TV})\mathcal{M}_1 + \\
    & (d_{TV} - \bar{d}_{TV}) \mathcal{M}_2 +\\
    &(b_{TV} - \bar{b}_{TV})e^{-(d_{TV} - \bar{d}_{TV})x_1}\mathcal{M}_3,
\end{split}
\end{align}
where $x_1$ is an additional parameter to be learned from the data.

Figure~\ref{fig:3DSpectrum} shows as an example the dynamical evolution of the eigenvalues (approximate energies $\hat{E}^{(i)}$) of the matrix $\mathcal{\hat M}(g_0,V_0^{is})$ when calibrated to reproduce the Gamow-Teller strength function of $^{80}$Ni. For the fixed value of $\bar{g}_0=0.5$, the eigenvalues only depend on $V_0^{is}$ through the diagonal matrix $\mathcal{M}_0$ and the symmetric matrix $\mathcal{M}_2$. 

%This eigenvalue evolution can be understood in terms interacting repulsive particles in 1 dimension by interpreting the parameter $V_0^{is}$ as time,\PG{ a system known as the Pechukas-Yukawa Gas~\cite{pechukas1983distribution,yukawa1985new,haake2018level}.  }
% It is a system that has been extensively studied and known as the Pechukas-Yukawa Gas~\cite{pechukas1983distribution,yukawa1985new,haake2018level}. This interpretation can help us increase our insight on the inner workings of both EM1 as well as the full order model. Since the ``particles" repel each other through a ``force" proportional to the inverse of their distance cubed~\cite{haake2018level}, for large values of the parameter (time) $V_0^{is}$, the eigenvalues will escape from each other in opposite directions, effectively creating a spread out strength function $S(\omega)$, while for some intermediate parameter range the strength function $S(\omega)$ will be more collective, as shown in Figures~\ref{fig:gamow_teller_example} and~\ref{fig:3DSpectrum}. The small peak that can be seen in Figure~\ref{fig:gamow_teller_example} between -5 and -8 MeV as $V_0^{is}$ increases can also be seen in Figure~\ref{fig:3DSpectrum}, and this peak is responsible for a big contribution to the half life calculation because of the phase space factor $f(E_0,Z,A)$.

The final step to build EM1 is to model the effective width  $\hat{\eta}(\boldsymbol{\alpha})$. Knowing that $\hat{\eta}(\boldsymbol{\alpha})$ is positive,  we decided to use the following ansatz:
\begin{equation}
    \hat{\eta}(\boldsymbol{\alpha}) = \sqrt{\eta_0^2 + \left[ \sum\limits_{j = 1}^{k_0} \eta_j (\alpha_{j} - \bar{\alpha}_{j}) +\eta_{k_0+1} \right]^2},
\end{equation}
where $k_0$ is the number of varied EDF parameters, and $\eta_j$ for $j\in[0,k_0+1]$ are tunable parameters.

The total number of trainable parameters for EM1 for the dipole polarizability $\alpha_D$ calculations, with $k_0=2$ varied parameters $\boldsymbol{\alpha}$, $k_1=3$ parameter functions $f_j(\boldsymbol{\alpha})$ for the matrix parametric dependence in Eq.~\eqref{eq: matrix hat}, and the selected functional form for the approximated field $\mathcal{\hat F}(\boldsymbol{\alpha})$ in Eq.~\eqref{eq:PMM_alphaD}, is then given by:
% \begin{equation}\label{eq: Em 1 parameters}
%    N_{param}^{\text{Em1}} = \underbrace{3n_1}_{\mathcal{F}} + \underbrace{k+2}_{\eta} + \underbrace{n_1}_{\mathcal{M}_0} + \underbrace{\frac{1}{2}k_1 n_1(n_1+1)}_{\mathcal{M}_j} + \underbrace{1}_{x_1}.
% \end{equation}
\begin{equation}\label{eq:em1_num_param_alphaD}
    N_{param}^{\rm Em1} = \frac{3}{2}n_1^2 + \frac{11}{2}n_1 + 5.
\end{equation}
As an example, for a selected matrix size of $n_1=10$ we will have $N_{param}^{\text{Em1}} =210$ tunable parameters. For the $T_{1/2}$ emulator, and the matrix form (\ref{eq:PMM_t12}), the number of tunable parameters is given by
\begin{equation}\label{eq:em1_num_param_t12}
    N_{param}^{\rm Em1} = n_1^2 + 3n_1 + 5,
\end{equation}
corresponding to 135 for $n_1 = 10$.

Principles behind EM1 can be traced back to the \textit{random-matrix theory}, first introduced in nuclear physics to describe excitations in complex systems, where distances between neighboring energy levels are  small~\cite{Brody1981a}. In particular, it was applied in Ref.~\cite{Severyukhin2017a} to streamline computation of the coupling between complex configurations in RPA to explain the spreading widths of giant resonances. Both matrix elements of effective couplings as well as RPA eigenmodes can be formulated within random-matrix theory, by replacing them with random variables. Although such an approach seems promising in the case of giant resonances, where the density of the states is very large, it does not apply to the case where the structure of individual low-lying states are important. On the other hand, the ideas of EM1 — with matrix elements inferred from data rather than assumed to be random numbers — can apply to both.

\subsection{Emulator 2}\label{subsec:emulator 2}

\begin{table}[htb]
\centering
\renewcommand\theadfont{\bfseries}
\caption{Training and testing ranges (in square brackets) for each parameter, and the number $N_\text{FOM}$ of FOM evaluations used in the two cases studied. The FOM evaluations were distributed on a uniform regular grid in the  parameter spaces, with the training points selected inside a rectangle region and the testing points outside to benchmark extrapolation (see 
Fig.~\ref{fig:YbNiPerformance}).}
\begin{tabular}{@{} c| c c c c c @{}}
\toprule
\textbf{Case} &
$\boldsymbol{\alpha}$ &
\makecell[c]{Train \\ Range} &
\makecell[c]{Train \\ $N_\text{FOM}$} &
\makecell[c]{Test \\ Range} &
\makecell[c]{Test \\ $N_\text{FOM}$} \\
\midrule
\multirow{2}{*}{$^{180}$Yb $\alpha_D$} 
    & $d_\text{TV}$ & [0.5, 1.75] & \multirow{2}{*}{121} & [0.1, 2.0] & \multirow{2}{*}{104} \\
    &   $b_\text{TV}$(fm$^2$) & [1.5, 4.0]  &                        & [1.0, 4.5] &                       \\
\midrule
\multirow{2}{*}{$^{80}$Ni $T_{1/2}$} 
    & $V_0^{is}$   & [0.29, 1.71] &    \multirow{2}{*}{121}                    & [0.0, 2.0] &      \multirow{2}{*}{104}                 \\
    & $g_0$ & [0.14, 0.86] & & [0.0, 1.0] & \\
    % & $V_0^{is}$   & [0.29, 1.71] &                       & [0.0, 2.0] &                      \\
\bottomrule
\end{tabular}
\label{tab:case_ranges}
\end{table}

% The second emulator shown in the right column of Fig.~\ref{fig:Flowchart} has intrinsic dimension $n_2\lesssim 10^1$. 
The flowchart for the second emulator (EM2) is shown in the right column of Fig.~\ref{fig:Flowchart}. EM2 operates with matrices of dimension $n_2\lesssim10^1$ and bypasses building any intermediate latent space involving the strength function $S(\boldsymbol{{\alpha}};\omega)$. Instead, it  aims to directly connect the controlling parameters $\boldsymbol{\alpha}$ with the observables $\mathcal{O}$. To achieve that, EM2 uses a fully data-driven approach through PMMs~\cite{Cook2025a} to build a parametric matrix from data  with the same structure as Eq.~\eqref{eq: matrix hat} with rank-$n_2$ matrix and a total of $k_2$ functions of the controlling parameters $f_j (\boldsymbol{\alpha})$. The eigenvalues of this matrix are obtained numerically and a single selected target eigenvalue $\lambda^{(t)}$ is used to yield approximated observables. In our particular implementation, we select the innermost eigenvalue, as it shows most variability with the evolution of model parameters. The form of the matrix is selected as in Eq.~\eqref{eq: matrix alphaD} and \eqref{eq:PMM_t12}, for $\alpha_D$ and $T_{1/2}$, respectively. 
% For instance, in case of $\alpha_D$ emulator, the total number of trainable parameters is given by
For instance, for $\alpha_D$ emulations, with $k_0=2$ varied parameters $\boldsymbol{\alpha}$, $k_2=3$ parameter functions $f_j(\boldsymbol{\alpha})$ for the matrix parametric dependence in Eq.~\eqref{eq: matrix hat}, the total number of tunable parameters is given by:
\begin{equation}\label{eq: Em 2 parameters}
   N_{param}^{\text{EM2}} =  \frac{3}{2}n_2^2 + \frac{5}{2}n_2 + 1.
\end{equation}
For  $n_2=5$, one gets $N_{param}^{\text{EM2}} =51$ tunable parameters. For $T_{1/2}$ emulations, there are only $k_2=2$ parametric functions $f_j(\boldsymbol{\alpha})$ in Eq.~\eqref{eq: matrix hat}, and no term $x_1$, such that the total number of tunable parameters is
\begin{equation}
     N_{param}^{\text{EM2}} = n_2^2 + 2n_2 + 1.
\end{equation}
For  $n_2 = 5$, the number of tunable parameters is 36.

\subsection{Calibration}\label{subsec: training}

To accurately reproduce the observables, an emulator must be calibrated, which accounts for finding optimal numerical values for its hyperparameters. These hyperparameters are the elements of all the matrices involved in Eq.~\eqref{eq: matrix hat}, any free hyperparameters used to construct the functions $f_j(\boldsymbol{\alpha})$, and in the case of EM1 the hyperparameters used to build the effective width $\hat\eta(\boldsymbol{\alpha})$ and external field $\hat{\mathcal{F}}(\boldsymbol{\alpha})$. 
%Expressions~\eqref{eq: Em 1 parameters} and ~\eqref{eq: Em 2 parameters} show the total number of tunable parameters for each emulator, respectively.

We treat the learning procedure in a traditional machine learning optimization approach and build appropriate cost functions using a set of FOM evaluations for $m\in [1,N_\text{FOM}]$ parameters $\boldsymbol{\alpha}^{(m)}$ sampled from a chosen region in the parameter space. For the strength function, which plays a critical role in EM1, we build the cost function:
\begin{equation}\label{eq:relative_error_strength}
    \chi^2_{S} = \frac{1}{w_{S}}\sum \limits_{m = 1 }^{N_\text{FOM}} \frac{ \int d \omega\left| S(\omega;\boldsymbol{\alpha}^{(m)}) - \hat{S}(\omega;\boldsymbol{\alpha}^{(m)})\right|^2}{\int d \omega\left| S(\omega;\boldsymbol{\alpha}^{(m)}) \right|^2},
\end{equation}
where $w_S$ is a chosen weight to scale the cost function,
% the $N_\text{FOM}$ parameters $\boldsymbol{\alpha}^{(m)}$ sampled from a chosen region in the parameter space, 
$S(\omega;\boldsymbol{\alpha}^{(m)})$  is the FOM strength function, and  $\hat{S}(\omega;\boldsymbol{\alpha}^{(m)})$ is calculated with  EM1.

For the observables we build the cost function:
\begin{equation}
    \chi^2_{\mathcal{O}} = \frac{1}{w_{\mathcal{O}}}\sum \limits_{m = 1 }^{N_\text{FOM}} \left[ \mathcal{O}(\boldsymbol{\alpha}^{(m)}) - \mathcal{\hat{O}}(\boldsymbol{\alpha}^{(m)})\right]^2,
\end{equation}
where $w_\mathcal{O}$ is a chosen weight to make the cost function dimensionless, $\mathcal{O}$ and $\mathcal{\hat O}$ are either  $\alpha_D$ or $\ln T_{1/2}$ (as half-lives span orders of magnitude). These cost functions are then minimized with respect to the  emulator's hyperparameters. For the case of EM1, both cost functions are combined, $ \chi^2_{S}+ \chi^2_{\mathcal{O}}$, with the weights $w_S$ and $w_\mathcal{O}$ determining the relative importance between them. For both $\alpha_D$ and $T_{1/2}$ emulators we choose $w_S=1$. On the other hand, for the observables, in the case of the $\alpha_D$ emulator we select $w_{\alpha_D} = 0.5 \text{ fm}^{-6}$, while for $T_{1/2}$ emulator, $w_{T_{1/2}} = 1$. For EM2, the hyperparameters are only adjusted to minimize the corresponding observable cost function.

\begin{figure*}[!t] 
  \includegraphics[width=0.9\linewidth]{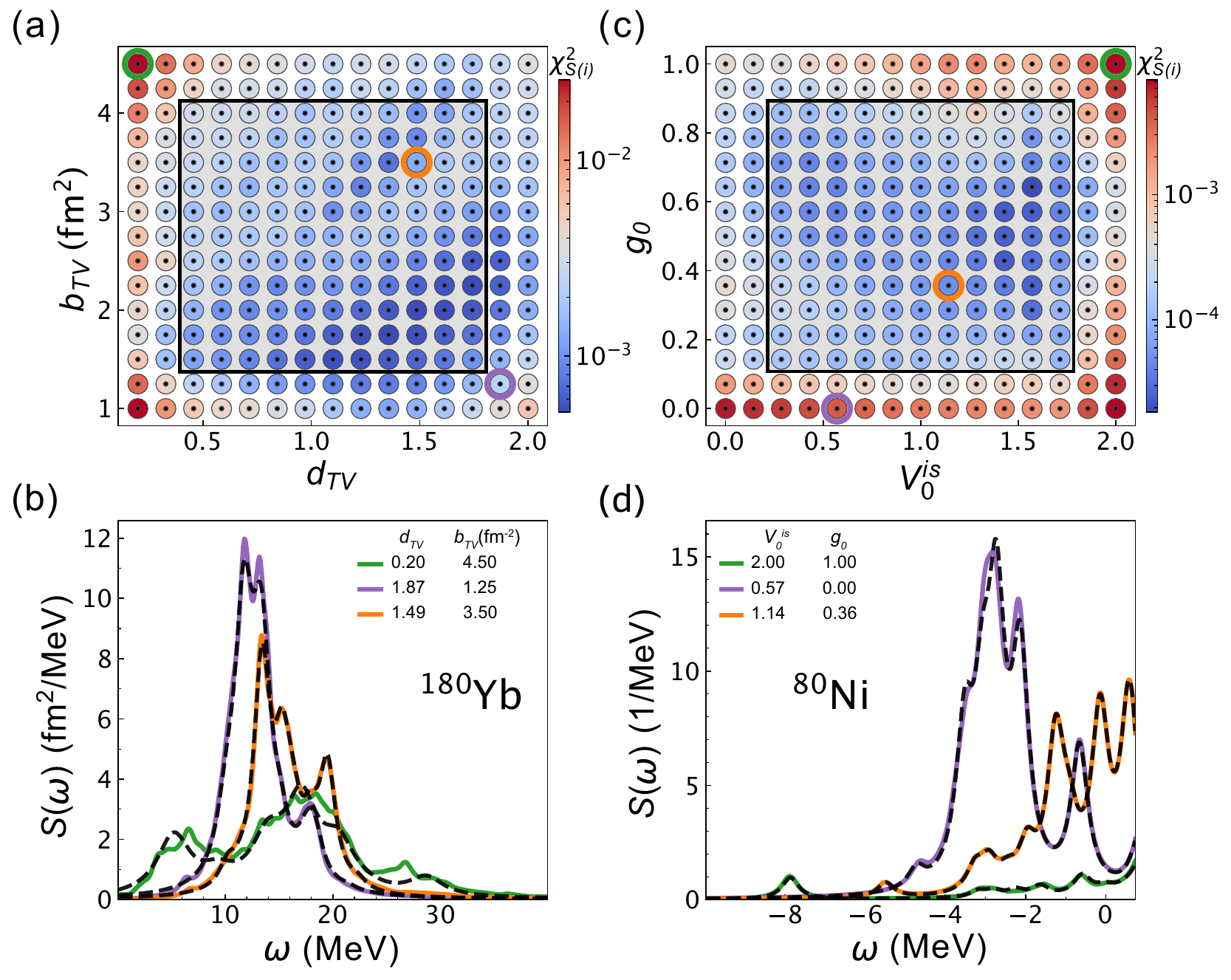}
  \caption{Performance of EM1 in reproducing isovector dipole strength in $^{180}$Yb (left panels), and the Gamow-Teller strength  in ${}^{80}$Ni (right panels). The emulator dimension was chosen as $n_1=13\,(n_0 = 8)$ for the isovector dipole strength, and $n_1=13\,(n_0=12)$ for the Gamow-Teller strength. The top panels show the relative error $\chi_{S(i)}^2$(\ref{eq:relative_error_strength_individual}) across the explored parameter range shown in Table~\ref{tab:case_ranges}. In both cases,  121 points inside the black rectangle were used for training, while  external 104 points form the test set. Bottom panels show three  strength functions
 as calculated in  QRPA (solid  lines) and EM1 
 % for $n_1 = 16$($n = 10$)
 (dashed lines) for the selected values of parameters (shown in  panels (a) and (c)  as thick circles matching the  colors of $S(\omega)$).}
  \label{fig:YbNiPerformance}
\end{figure*}

Since EM1 is aiming to reproduce both the high-dimensional strength function $S$ as an intermediate step, as well as the associated observable, each FOM evaluation is information-rich for its training. We could interpret that for an approximate strength function $\hat S$ that is being created by $n_0$ Lorentzians, each FOM calculations provides the optimal values of $n_0$ locations and $n_0$ amplitude peaks, for a total of $2n_0+1$ units of data taking into account the associated value of the observable. This allows us to train EM1 with far less FOM evaluations than the total number of hyperparameters expressed in Eqs.~\eqref{eq:em1_num_param_alphaD} and \eqref{eq:em1_num_param_t12}. In contrast, since EM2 is only calibrated using the end value of the observable, each FOM evaluation produces only a single number for the  cost function. In principle, this would suggest that one needs at least $N_{param}^\text{Em2}$ FOM evaluations to completely fit all the hyperparameters of EM2 to obtain good performance. Nevertheless, we have observed in this and previous works~\cite{Somasundaram2025a,Armstrong2025a} that a good performance can be achieved with significantly lower number of evaluations than those required by these arguments. A possible reason lies in only a small subset of the eigenvalues of the parametric matrix meaningfully contributing to the response of the observable to parameter changes. While the targeted eigenvalue $\lambda^{(t)}$, and those in its vicinity are well constrained by FOM data, those in more distant parts of the spectrum are not, yet they do not affect the emulator's performance.

\begin{figure*}
    \centering{\includegraphics[width=1\textwidth]{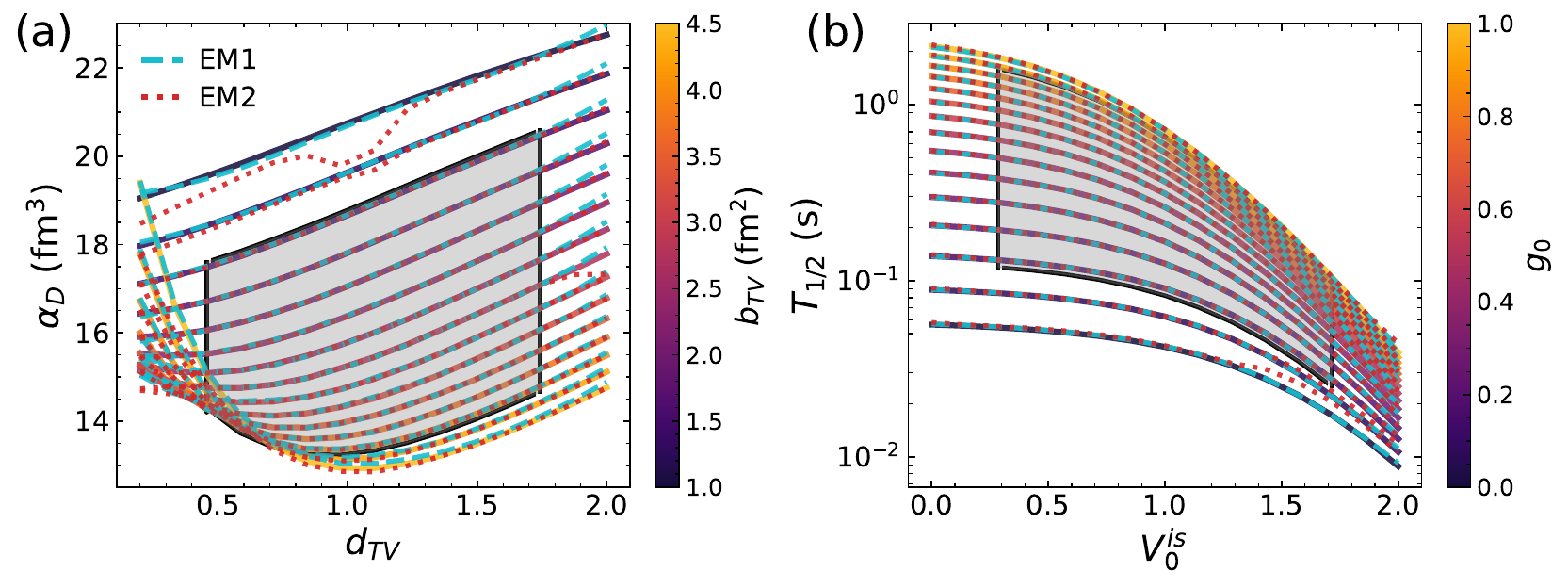}}
\caption{Performance of EM1 with dimension $n_1 = 13$ (cyan dashed line) and EM2  with dimension $n_2 = 9$ (red dotted line), in reproducing (a)  $\alpha_D$ of $^{180}$Yb, and (b) $T_{1/2}$ of $^{80}$Ni. The FOM calculations are shown by solid colored lines, with a color coding that matches that of Figs.~\ref{fig:isovector_dipole_example} and~\ref{fig:gamow_teller_example}. The calculations pertaining to the training set are shown by a gray area   enclosed  with a black boundary. Since $\alpha_D$ twists in a non-linear way near $d_{TV}\sim0.2-0.7$, some curves at low values of $b_{TV}$ cross into the training region even though these points were not used for training.}    
    \label{fig:ManyCurves}
\end{figure*}

Finally, in terms of the parameter initialization for the tuning, we note that the functions involved in expressions \eqref{eq:PMM_t12}, ~\eqref{eq:PMM_alphaD}, and ~\eqref{eq: matrix alphaD} are expanded in terms of the difference between the parameters $\boldsymbol{\alpha}$ and a central value $\boldsymbol{\alpha}_C$~\eqref{eq: central parameters}. This implies that the parametric functions involved are zero in the central parameters $f_j(\boldsymbol{\alpha}_C)=0$ for all cases, allowing for the initialization of the hyperparameter tuning to be done in an efficient way. For EM1, the parameters are initialized by fitting the energies and transition probabilities to reproduce the strength function of the  central point $\boldsymbol{\alpha}_C$, which directly determines $n_0$ out of the $n_1$ matrix elements of $\mathcal{M}_0$ (energies of the peaks), as well as the $n_0$ of the $n_1$ components of the vector $\mathcal{F}_0$ (height of the peaks). The remaining hyperparameters are initialized with small random values sampled from a uniform distribution. We have found that this initialization for EM1 achieves faster and more reliable convergence than if all hyperparameters were chosen at random. In contrast, hyperparameters of EM2 are initialized randomly in a small window around zero. To minimize the uncertainty due to optimizer finding different local minima based on random initialization, we train both types of emulators 10 times, sampling each time a different parameter set within a small window around zero. Subsequently, the training run which minimizes the overall training cost is selected.

The emulators based on both algorithms described in this section are implemented as Python scripts in \texttt{tensorflow} framework \cite{TensorFlow2016a}, and are publicly available in the repository \cite{GitHub2025a}.

\section{Results}\label{sec:results}

To study the performance of both  emulators, we divide the  parameter spaces into training and testing regions. This enables us to study the extrapolation ability of  emulators beyond their calibration domains. To this end, we train the emulators on the \textit{internal} region of the entire FOM dataset, and then extrapolate into the \textit{external} region. Table~\ref{tab:case_ranges} summarizes the ranges of the parameters $\boldsymbol{\alpha}$ used and the total number of FOM evaluations for training and testing.

%%%%%WN: finished here

We first analyze the performance of EM1 in reproducing the strength functions  in $^{180}$Yb  and $^{80}$Ni.  Figure~\ref{fig:YbNiPerformance}(a,c) shows the relative accuracy
$\chi^2_{S(i)}$, defined by rewriting Eq. \eqref{eq:relative_error_strength} as
\begin{equation}\label{eq:relative_error_strength_individual}
    \chi_S^2 = \sum \limits_{i = 1}^{N_{\text{FOM}}} \chi_{S(i)}^2.
\end{equation}
We employ $n_1 = 13$ in both cases.  For isovector dipole response
calculations  in $^{180}$Yb, $n_0 = 8$ eigenvectors and eigenvalues are retained, while for the Gamow-Teller response in $^{80}$Ni we take $n_0 = 12$. The error is 
very low in the internal training region (indicated by a black rectangle), between $0.01\%-0.2\%$, and remains small  in the testing region with most values of $\chi^2_{S}$ less than $1\%$ and a few points where $\chi^2_{S}$ remains below 2\%. In panels (b) and (d) we compare the emulated and FOM strength functions for three selected parameter values, one inside the training region and two points in the testing region, including also the worst performing value for both nuclei. Although the shape of the strength function changes significantly throughout the dataset, with a noticeable fragmentation on the edges of the testing region, the emulator  captures its behavior effectively.

\begin{figure}[h!]
    \includegraphics[width=\linewidth]{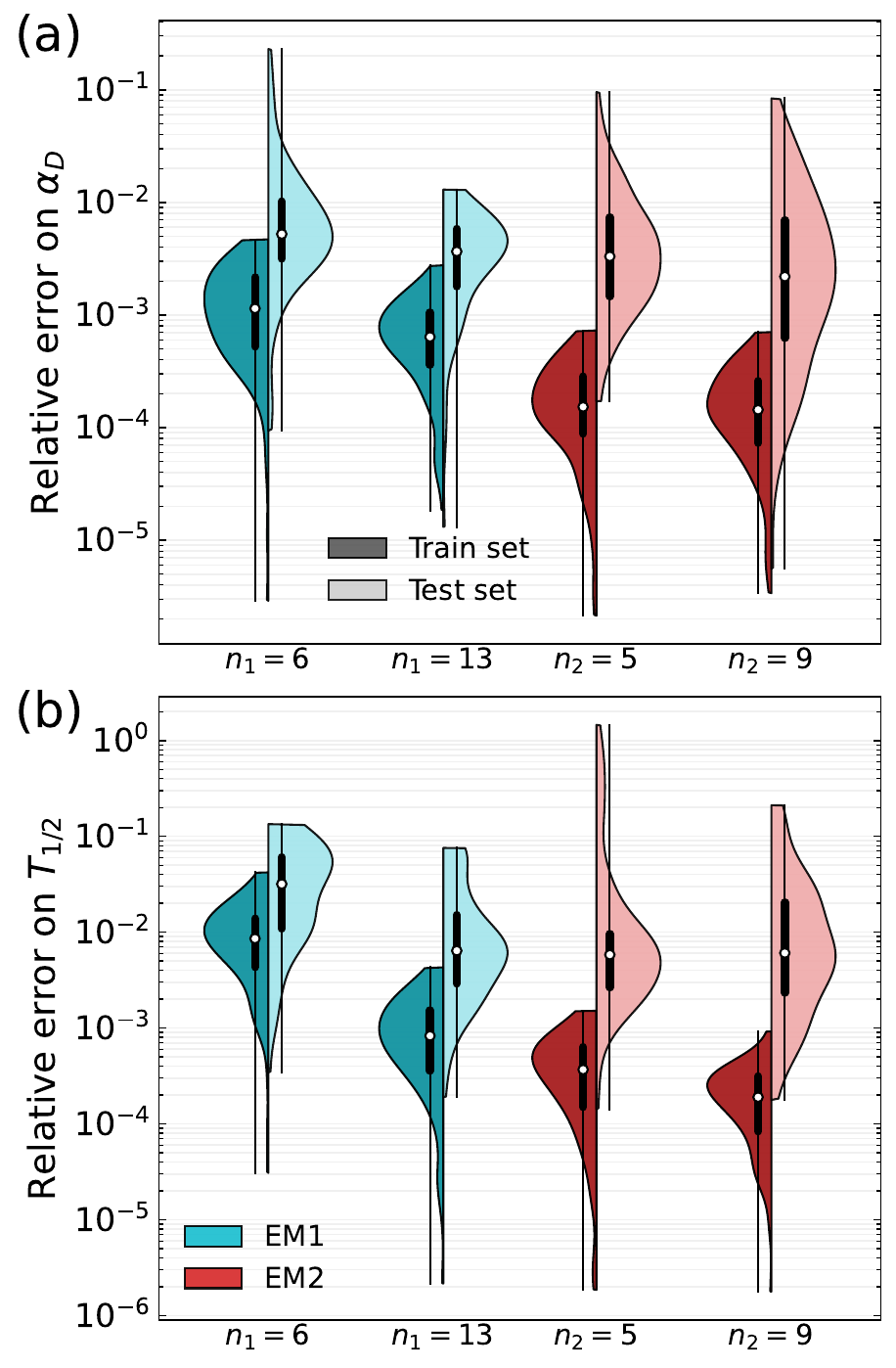}
    \caption{Violin plots~\cite{Hintze1998a} benchmarking the performance of EM1 and EM2 on the training and test sets for two choices of their dimensions $n$. Panels (a) and (b) show the relative error on  $\alpha_D$ in $^{180}$Yb and $T_{1/2}$ of $^{80}$Ni, respectively. For each emulator, we display two model dimensions for EM1 ($n_1=6$ and 13) and for EM2 ($n_2=5$ and 9). In each split violin, the left (darker) half corresponds to the training set and the right (lighter) half to the test set. The white dot marks the median, the thick black bar indicates the middle 50\% of the data, and thin segments show the full range of the relative error.}
    \label{fig:CombinedCAT}
\end{figure}

% Having observed that EM1 can accurately reconstruct the strength function $S(\omega)$ across the parameter set in both nuclei, 
Having established the ability of EM1 to reproduce the strength function, we now compare the performance of both emulators in reconstructing the two observables
of interest: $\alpha_D$ and $T_{1/2}$. In Fig.~\ref{fig:ManyCurves}, we inspect the overall predictions across the parameter range. Both observables present high sensitivity to changes in the parameters, with challenging features for the emulators to learn, such as $\alpha_D$ showing highly non-linear behavior near $d_{TV}\sim 0.2-0.7$ and $b_{TV}\sim3.5-4.5$ fm$^2$, and $T_{1/2}$ varying by orders of magnitude. Despite only being trained in the region shown by a black boundary, both emulators are able to  capture well the response of the observables in the test-extrapolation region. EM1 can reproduce the behavior of $\alpha_D$ with a discrepancy of only a few percent, while EM2 struggles more in the strong non-linear region, having a larger discrepancy of around 10\%. For $T_{1/2}$, both emulators reproduce the response, but EM2 shows some deviations of more than $50\%$ on a small region around $g_{0}\sim0$ and $V_0^{is} > 1.5$. Considering that EM2 only used observables as training data, its predictive power in the extrapolation region is remarkable and showcases its ability to reproduce system's dynamics through the eigensystems of parametric matrices.

% In Fig.~\ref{fig:CombinedCAT} we show violin plots~\cite{hintze1998violin} of the relative error distribution for both emulators to study how the choice of their dimensions, such as the matrix rank in Eq.~\eqref{eq: matrix hat}, impacts their accuracy. The emulator dimension did not appreciably affect computational times, with EM1 taking around 0.5 ms per evaluation, while EM2 takes around 0.2 ms, both showing appreciable speedup with respect to the FOM's computational time of around $10^3$ seconds.  
The emulator dimension did not appreciably affect computational times, with EM1 taking around twice as much time per evaluation as EM2, yet both showing appreciable speedups of $10^{6}-10^{7}$ with respect to the FOM's computational time.  For  $\alpha_D$, the median relative error for all emulators ranges between $0.06\%-0.1\%$ for the training set, and between $0.2\%-0.5\%$ for the testing  set, while for  $T_{1/2}$ it shows a bigger spread between $0.02\%-1\%$ for training and $0.6\%-3\%$ for testing. EM1 shows more consistency between its training and testing errors, with the testing error decreasing for both observables when the emulator dimension  is increased from 6 to 13. In contrast, EM2 shows a bigger discrepancy between its training and testing errors, a difference that becomes larger when its intrinsic dimension is increased from $n_2 = 5$ to 9, with training errors being reduced while testing errors remain almost unchanged. However, the relative error distribution significantly broadens. These two behaviors suggest that, for these values of dimensions we explored, EM1 better captures the intrinsic dynamics of the systems, with good generalization to extrapolating to unobserved parameters, while EM2 shows signs of overfitting, with almost two orders of magnitude spread between its training and testing errors for $T_{1/2}$ with $n_2=9$.

\begin{figure}
    \includegraphics[width=\linewidth]{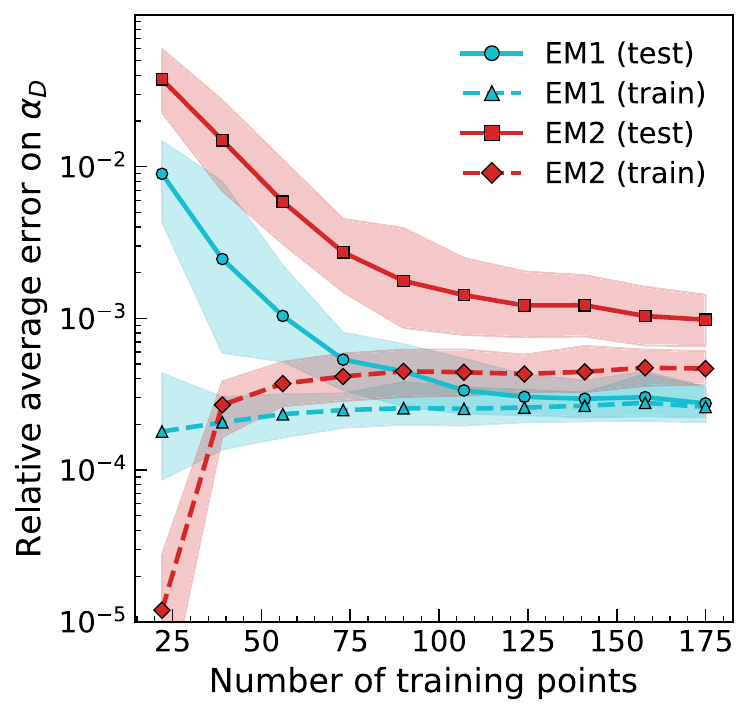}
    \caption{The relative average error on $\alpha_D$ on the training set (triangles: EM1, diamonds: EM2) and test set (circles: EM1, squares: EM2)  as a function of the number of training points. The comparison is made for both EM1 (cyan) with $n_1 = 8 \,(n_0 = 4)$ and EM2 (red) $n_2 = 5$. Both emulators are trained with the same data and randomly initialized 100 times, with mean values denoted by solid and dashed lines, and 95\% coverage intervals marked by the shaded areas. Similar results have been obtained   for $T_{1/2}$.}
    \label{fig:NumberOfTrain}
\end{figure}

Finally, we study how the performance of both emulators in reproducing $\alpha_D$ is affected by the amount of FOM training data.  The main results are shown in Fig.~\ref{fig:NumberOfTrain} for EM1 with $n_1 = 8 \,(n_0 = 4)$ and  EM2  with $n_2=5$. 
While Figs.~\ref{fig:YbNiPerformance},~\ref{fig:ManyCurves}, and~\ref{fig:CombinedCAT}, illustrate extrapolation capabilities, in Fig.~\ref{fig:NumberOfTrain} we follow a setup more likely to be used in practical calculations, where the emulator is purposely trained to cover the expected evaluation region in parameter space. For testing purposes we select 50 FOM calculations drawn randomly from the points of Fig. \ref{fig:YbNiPerformance}(a,c). From the remaining 175 points, we retain a varying fraction 
of training points 
ensuring that the $i$-th partition contains all the points of the previous one. This is done to minimize the impact on the results of the specific parameter locations of FOM evaluations, which is known to affect the quality of trained emulators~\cite{Bonilla2022a,Maldonado2025a}. In all training partitions all four corner points spanning the edges of the parameter ranges in Table~\ref{tab:case_ranges} are included, together with the central point $\boldsymbol{\alpha}_C$. 
% The solid and dashed lines in the figure shows the mean error while the bands encompass 95\% 
To take into account the variability on results associated with the initialization of their hyperparameters we perform the training 100 times, each time initializing matrix elements of matrices and vectors in EM1 and 2 with different random seeds. The solid and dashed lines in the figure shows the mean error while the bands encompass 95\% of the obtained curves.

The test accuracy of EM1 improves steeply at the beginning and then plateaus around 125 training points, while the test accuracy of EM2 continues to slightly improve as we use more data. The opposite trend can be observed for the relative error on the training set, as expected since more points are being included that the emulators must reproduce, yet their complexity (dimension) remains fixed. Based on the performance of the test set, we see that EM1 can achieve an accuracy of almost an order of magnitude better than EM2 for a fixed number of training points, and its performance in the train and test sets converges to the same value. We attribute these observations to the fact that EM1 is reproducing the strength function $S(\omega)$ as an intermediate step in the calculation, highlighting the impact that physics-informed decisions in the construction of data-informed emulators  can have on their final performance. The fact that EM2 is able to achieve sub-percent accuracy with only $\sim50$ training points is quite impressive, given that each FOM evaluation only gives one observable, which is used to train 51 free parameters. This observation further showcases PMMs~\cite{Cook2025a} as a flexible and robust machine learning approach capable of learning complicated mappings directly from the data.

\section{Conclusions}\label{sec:conclusion}

% 1. We developed two emulators for QRPA.
% 2. Structure and flexibility makes it good for usage depending on time and purpose (do you need a fast emulator or do you need accuracy+extrapolation?)
% 3. What's the result of its features?
% 4. What will this allow us to investigate? What are the next steps? What is the philosophical message with these emulators?

We have developed and benchmarked two emulators, EM1 and EM2, that are able to reproduce full QRPA strength-function calculations with sub-percent accuracy while offering  a speedup of 6-7 orders of magnitude compared to the the original full-order solver. As representative benchmarks, we examined two key nuclear observables: (i) the electric dipole polarizability of $^{180}$Yb as a function of the two isovector couplings ($b_{TV}, d_{TV}$), and (ii) the Gamow-Teller $\beta$-decay half-life  of $^{80}$Ni as a function of  the isoscalar pairing strength and the Landau-Migdal coupling$(V_0^{is},g_0$).

For EM1, we followed an approach inspired by data-driven operator learning methods that approximate low-dimensional representations of operators using  FOM evaluations \cite{Peherstorfer2016a,Benner2020a,Kovachki2023a}. We created a low-dimensional representation of  $S(\omega)$ through the sum of a few effective Lorentzians, and learned the necessary governing equations  as a function of the underlying parameters $\boldsymbol{\alpha}$ from FOM. The forms of these equations were inspired by the structure of the full QRPA pipeline, with the governing matrix $\mathcal{M}(\boldsymbol{\alpha})$ following the structure of the  PMM framework~\cite{Cook2025a}. 
%The dynamics of the eigenvalues of the approximated strength function $\hat{S}(\omega)$ in the case of linear dependence on the parameters $\boldsymbol{\alpha}$ resemble what is known as the Pechukas-Yukawa Gas~\cite{pechukas1983distribution,yukawa1985new,haake2018level}, a connection that can help beyond building the emulator but also in increasing the understanding of the underlying FOM. 
We built EM2 in a fully data-driven approach by using the PMM framework to directly connect the underlying parameters $\boldsymbol{\alpha}$ with the observables. EM1 showed better accuracy on the observables for test-set  parameters, reproducing strong non-linear features in $\alpha_D$ and the variation over two orders of magnitude in $\beta$-decay half-lives. EM2 only required data from the final QRPA observables to be trained, and usually showed better accuracy in the training region, while still being able to reproduce part of the response of the observables in the test set. 
% In addition, 
Besides the half-life and polarizability, EM1 was able to also provide a good reproduction of the strength function $S(\omega)$, which can be used to calculate other quantities of interest. The web-based tools deployed alongside this work~\cite{smlrWebsite} can run EM1 in real time to explore how $S(\omega)$ changes as the interaction parameters are tuned, yielding precise and rapid insight on parameter influence while avoiding repeated FOM runs. 
% EM2 only required data from the final QRPA observables to be trained, and usually showed better accuracy in the training region, while still being able to reproduce part of the response of the observables in the test set. 

By restructuring the training region to represent a more realistic  case, we have demonstrated that EM1 is able to achieve sub-percent accuracy on the test set being trained with only 22 data points, over which the strength function varies considerably. On the other hand, due to its simplicity, EM2 can be straightforwardly applied to learn observables directly, with the tradeoff of requiring more training data than EM1.
% , but possibly significantly lower number of samples compared to classical neural networks given its resemblance to the FOM structure.

% If higher accuracy in reproducing the observables is required, or the access to full strength function is needed, EM1 is the obvious best choice.
% We created visualization tools and made them available online~\cite{smlrWebsite} to run EM1 in real time, allowing users to explore the changes in the strength functions as the two parameters of the interaction are varied. For experts, this gives immediate insight as to what these parameters precisely control in the strength functions, avoiding costly FOM computations, for a streamlined research process.  

In the current work we focused on varying two of the controlling parameters $\boldsymbol{\alpha}$ of the EDF as a first step to test the viability of our outlined framework to create QRPA emulators. The high computational cost associated with QRPA has limited the calibration of nuclear EDFs informed by excited state data. By extending these methods to incorporate all the relevant EDF parameters we will be in position to offer a possible future path for the implementation of QRPA-based methods in Bayesian optimization of EDFs and subsequent uncertainty quantification.  The reduction in variability, paired with methods to estimate the errors of the emulators during the on-line (deployment) phase, will provide the robustness needed to incorporate the emulators into calibration and large studies pipelines. Since our  approach is agnostic to the specific physical system, it should be straightforward to extend  to problems across physics and chemistry areas that apply linear-response theory, thus opening the door to scalable Bayesian uncertainty quantification in broad applications for scientific discovery.

% the outlined framework can be expanded to account for the remaining parameters, a direction we will explore in a future work. The high computational cost required to perform QRPA calculations has limited the calibration of EDFs informed by excited states data in addition to ground-state observables. The two emulators we developed in this work offer a possible future path for implementation of QRPA-based methods in EDF Bayesian optimization and subsequent uncertainty quantification. Beyond the nuclear structure, our developed emulators can be applied in other areas of physics and chemistry where linear response theory is used to construct the excitation spectrum, enabling Bayesian studies to be implemented in broader areas. 

\section{Acknowledgments}
We thank Edgard Bonilla and Patrick Cook for valuable discussions. Computational resources were provided by the Institute for Cyber-Enabled Research at Michigan State University. This work was supported by the NSF Division of Physics through Award No. PHY-2050733, as part of the Research Experiences for Undergraduates at Michigan State University. Additional support was provided by the National Science Foundation CSSI program under award No.~OAC-2004601 (BAND Collaboration), and by the U.S. Department of Energy under Award No. DOE-DE-NA0004074 (NNSA, the Stewardship Science Academic Alliances program), and by the DOE Office of Science under Grants DE-SC0013365 and DE-SC0023175 (Office of Advanced Scientific Computing Research and Office of Nuclear Physics, Scientific Discovery through Advanced Computing).

% \cleardoublepage

\appendix

%========================
% Appendix: Acronyms & Symbols
%========================
\section{Acronyms and Symbols}\label{sec: appendix}

Tables~\ref{tab:acronyms} and~\ref{tab:symbols} summarize the acronyms and symbols used. 

%------------------------
% Acronyms
%------------------------
% \subsection*{Acronyms}
\begin{table}[ht!]  % top of a column
  \caption{Glossary of acronyms used in the  paper}
  \label{tab:acronyms}
\begin{tabular}{p{0.22\linewidth}p{0.73\linewidth}}
%\hline
%\textbf{Acronym} & \textbf{Meaning / Context} \\
\hline
DFT & Density Functional Theory \\
EDF & Energy Density Functional \\
%(parameterized functional determining the mean field). \\
EM1 & Emulator~1 (physics-informed surrogate) \\
EM2 & Emulator~2 (data-driven PMM surrogate) \\
FAM & Finite-Amplitude Method (linear-response solver) \\
FOM & Full Order Model (high-fidelity calculation) \\
GT & Gamow–Teller (charge-exchange) transition \\
HF(B) & Hartree–Fock(-Bogoliubov)  framework \\
PMM & Parametric Matrix Model \\
(Q)RPA & (Quasiparticle) Random Phase Approximation \\
RHB & Relativistic Hartree–Bogoliubov framework \\
TDDFT & Time-Dependent Density Functional Theory \\
\hline

\end{tabular}
\end{table}
% \vspace{10em}

%------------------------
% Symbols and Parameters
%------------------------
\begin{table}[ht!]  
  \caption{Symbols and parameters used }
  \label{tab:symbols}
\begin{tabular}{p{0.24\linewidth}p{0.73\linewidth}}
\hline
$\boldsymbol{\alpha}$ & Vector of controlling  parameters  \\
$\boldsymbol{\alpha_C}$ & Central parameter point  (training center)~\eqref{eq: central parameters} \\
$\alpha_D$ & Electric dipole polarizability~\eqref{eq:alphaD} \\
$\eta$ & Lorentzian folding width~\eqref{eq:strength_function} \\
$\mathcal{B}_{\alpha;i}$ & Reduced transition probability \eqref{eq:reduced_transition_probability} \\
$b_{TV},\,d_{TV}$ & Isovector EDF couplings\\
$E_{\alpha;i}$ & QRPA excitations \eqref{eq: non-Hermician Eq} \\
$\mathcal{F}(\boldsymbol{\alpha})$ & External field\\
$f(E_0,Z,A)$ &  Fermi function \\
$\mathcal{M}(\boldsymbol{\alpha})$ & QRPA  matrix~\eqref{eq: QRPA Matrix} \\
$\mathcal{N}$ & QRPA metric~\eqref{eq:norm matrix} \\
$N_d$ & Rank of the  QRPA matrix\\
$n_0,n_1,n_2$ & Emulator latent dimensions \\
$n_{2\text{qp}}$ & Number of two–quasiparticle states \\
$\mathcal{O}$  & Observables \\
$\hat{\mathcal{O}}$ & Emulated observables \\
$\delta \mathcal{R}(\omega)$ & Linear response amplitude~\eqref{harmonic_ansatz}\\
$S(\boldsymbol{\alpha};\omega)$ & Strength function~\eqref{eq:linear_response_strength} \\
$T_{1/2}$ & $\beta$-decay half-life~\eqref{eq: t1/2} \\
$V^{is}_0,\,g_0$ & Isoscalar pairing strength and Landau–Migdal coupling \\
\hline
\end{tabular}
\end{table}

\clearpage

\bibliography{bibl_edit2}% Produces the bibliography via BibTeX.

\begin{thebibliography}{94}%
\makeatletter
\providecommand \@ifxundefined [1]{%
 \@ifx{#1\undefined}
}%
\providecommand \@ifnum [1]{%
 \ifnum #1\expandafter \@firstoftwo
 \else \expandafter \@secondoftwo
 \fi
}%
\providecommand \@ifx [1]{%
 \ifx #1\expandafter \@firstoftwo
 \else \expandafter \@secondoftwo
 \fi
}%
\providecommand \natexlab [1]{#1}%
\providecommand \enquote  [1]{``#1''}%
\providecommand \bibnamefont  [1]{#1}%
\providecommand \bibfnamefont [1]{#1}%
\providecommand \citenamefont [1]{#1}%
\providecommand \href@noop [0]{\@secondoftwo}%
\providecommand \href [0]{\begingroup \@sanitize@url \@href}%
\providecommand \@href[1]{\@@startlink{#1}\@@href}%
\providecommand \@@href[1]{\endgroup#1\@@endlink}%
\providecommand \@sanitize@url [0]{\catcode `\\12\catcode `\$12\catcode
  `\&12\catcode `\#12\catcode `\^12\catcode `\_12\catcode `\%12\relax}%
\providecommand \@@startlink[1]{}%
\providecommand \@@endlink[0]{}%
\providecommand \url  [0]{\begingroup\@sanitize@url \@url }%
\providecommand \@url [1]{\endgroup\@href {#1}{\urlprefix }}%
\providecommand \urlprefix  [0]{URL }%
\providecommand \Eprint [0]{\href }%
\providecommand \doibase [0]{https://doi.org/}%
\providecommand \selectlanguage [0]{\@gobble}%
\providecommand \bibinfo  [0]{\@secondoftwo}%
\providecommand \bibfield  [0]{\@secondoftwo}%
\providecommand \translation [1]{[#1]}%
\providecommand \BibitemOpen [0]{}%
\providecommand \bibitemStop [0]{}%
\providecommand \bibitemNoStop [0]{.\EOS\space}%
\providecommand \EOS [0]{\spacefactor3000\relax}%
\providecommand \BibitemShut  [1]{\csname bibitem#1\endcsname}%
\let\auto@bib@innerbib\@empty
%</preamble>
\bibitem [{\citenamefont {Kubo}(1966)}]{Kubo1966a}%
  \BibitemOpen
  \bibfield  {author} {\bibinfo {author} {\bibfnamefont {R.}~\bibnamefont
  {Kubo}},\ }\bibfield  {title} {\bibinfo {title} {The fluctuation-dissipation
  theorem},\ }\href {https://doi.org/10.1088/0034-4885/29/1/306} {\bibfield
  {journal} {\bibinfo  {journal} {Rep. Prog. Phys.}\ }\textbf {\bibinfo
  {volume} {29}},\ \bibinfo {pages} {255} (\bibinfo {year} {1966})}\BibitemShut
  {NoStop}%
\bibitem [{\citenamefont {Onida}\ \emph {et~al.}(2002)\citenamefont {Onida},
  \citenamefont {Reining},\ and\ \citenamefont {Rubio}}]{Onida2002a}%
  \BibitemOpen
  \bibfield  {author} {\bibinfo {author} {\bibfnamefont {G.}~\bibnamefont
  {Onida}}, \bibinfo {author} {\bibfnamefont {L.}~\bibnamefont {Reining}},\
  and\ \bibinfo {author} {\bibfnamefont {A.}~\bibnamefont {Rubio}},\ }\bibfield
   {title} {\bibinfo {title} {Electronic excitations: density-functional versus
  many-body {Green's}-function approaches},\ }\href
  {https://doi.org/10.1103/RevModPhys.74.601} {\bibfield  {journal} {\bibinfo
  {journal} {Rev. Mod. Phys.}\ }\textbf {\bibinfo {volume} {74}},\ \bibinfo
  {pages} {601} (\bibinfo {year} {2002})}\BibitemShut {NoStop}%
\bibitem [{\citenamefont {Aryasetiawan}\ and\ \citenamefont
  {Gunnarsson}(1998)}]{Aryasetiawan1998a}%
  \BibitemOpen
  \bibfield  {author} {\bibinfo {author} {\bibfnamefont {F.}~\bibnamefont
  {Aryasetiawan}}\ and\ \bibinfo {author} {\bibfnamefont {O.}~\bibnamefont
  {Gunnarsson}},\ }\bibfield  {title} {\bibinfo {title} {The {GW} method},\
  }\href {https://doi.org/10.1088/0034-4885/61/3/002} {\bibfield  {journal}
  {\bibinfo  {journal} {Rep. Prog. Phys.}\ }\textbf {\bibinfo {volume} {61}},\
  \bibinfo {pages} {237} (\bibinfo {year} {1998})}\BibitemShut {NoStop}%
\bibitem [{\citenamefont {Osterfeld}(1992)}]{Osterfeld1992a}%
  \BibitemOpen
  \bibfield  {author} {\bibinfo {author} {\bibfnamefont {F.}~\bibnamefont
  {Osterfeld}},\ }\bibfield  {title} {\bibinfo {title} {Nuclear spin and
  isospin excitations},\ }\href {https://doi.org/10.1103/RevModPhys.64.491}
  {\bibfield  {journal} {\bibinfo  {journal} {Rev. Mod. Phys.}\ }\textbf
  {\bibinfo {volume} {64}},\ \bibinfo {pages} {491} (\bibinfo {year}
  {1992})}\BibitemShut {NoStop}%
\bibitem [{\citenamefont {Harakeh}\ and\ \citenamefont {van~der
  Woude}(2001)}]{Harakeh2001a}%
  \BibitemOpen
  \bibfield  {author} {\bibinfo {author} {\bibfnamefont {M.~N.}\ \bibnamefont
  {Harakeh}}\ and\ \bibinfo {author} {\bibfnamefont {A.}~\bibnamefont {van~der
  Woude}},\ }\href@noop {} {\emph {\bibinfo {title} {Giant Resonances:
  Fundamental High-Frequency Modes of Nuclear Excitation}}},\ Vol.~\bibinfo
  {volume} {24}\ (\bibinfo  {publisher} {Oxford Stud. Nucl. Phys.},\ \bibinfo
  {year} {2001})\BibitemShut {NoStop}%
\bibitem [{\citenamefont {Szabo}\ and\ \citenamefont
  {Ostlund}(1996)}]{Szabo1996a}%
  \BibitemOpen
  \bibfield  {author} {\bibinfo {author} {\bibfnamefont {A.}~\bibnamefont
  {Szabo}}\ and\ \bibinfo {author} {\bibfnamefont {N.~S.}\ \bibnamefont
  {Ostlund}},\ }\href@noop {} {\emph {\bibinfo {title} {Modern Quantum
  Chemistry: Introduction to Advanced Electronic Structure Theory}}}\ (\bibinfo
   {publisher} {Dover Publications},\ \bibinfo {address} {Mineola, New York},\
  \bibinfo {year} {1996})\BibitemShut {NoStop}%
\bibitem [{\citenamefont {Casida}\ and\ \citenamefont
  {Huix-Rotllant}(2012)}]{Casida2012a}%
  \BibitemOpen
  \bibfield  {author} {\bibinfo {author} {\bibfnamefont {M.~E.}\ \bibnamefont
  {Casida}}\ and\ \bibinfo {author} {\bibfnamefont {M.}~\bibnamefont
  {Huix-Rotllant}},\ }\bibfield  {title} {\bibinfo {title} {Progress in
  time-dependent density-functional theory},\ }\href
  {https://doi.org/10.1146/annurev-physchem-032511-143803} {\bibfield
  {journal} {\bibinfo  {journal} {Annu. Rev. Phys. Chem.}\ }\textbf {\bibinfo
  {volume} {63}},\ \bibinfo {pages} {287} (\bibinfo {year} {2012})}\BibitemShut
  {NoStop}%
\bibitem [{\citenamefont {Molinelli}\ \emph {et~al.}(2013)\citenamefont
  {Molinelli}, \citenamefont {Korkut}, \citenamefont {Wang}, \citenamefont
  {Miller}, \citenamefont {Gauthier}, \citenamefont {Jing}, \citenamefont
  {Kaushik}, \citenamefont {He}, \citenamefont {Mills}, \citenamefont {Solit},
  \citenamefont {Pratilas}, \citenamefont {Weigt}, \citenamefont {Braunstein},
  \citenamefont {Pagnani}, \citenamefont {Zecchina},\ and\ \citenamefont
  {Sander}}]{Molinelli2013a}%
  \BibitemOpen
  \bibfield  {author} {\bibinfo {author} {\bibfnamefont {E.~J.}\ \bibnamefont
  {Molinelli}}, \bibinfo {author} {\bibfnamefont {A.}~\bibnamefont {Korkut}},
  \bibinfo {author} {\bibfnamefont {W.}~\bibnamefont {Wang}}, \bibinfo {author}
  {\bibfnamefont {M.~L.}\ \bibnamefont {Miller}}, \bibinfo {author}
  {\bibfnamefont {N.~P.}\ \bibnamefont {Gauthier}}, \bibinfo {author}
  {\bibfnamefont {X.}~\bibnamefont {Jing}}, \bibinfo {author} {\bibfnamefont
  {P.}~\bibnamefont {Kaushik}}, \bibinfo {author} {\bibfnamefont
  {Q.}~\bibnamefont {He}}, \bibinfo {author} {\bibfnamefont {G.}~\bibnamefont
  {Mills}}, \bibinfo {author} {\bibfnamefont {D.~B.}\ \bibnamefont {Solit}},
  \bibinfo {author} {\bibfnamefont {C.~A.}\ \bibnamefont {Pratilas}}, \bibinfo
  {author} {\bibfnamefont {M.}~\bibnamefont {Weigt}}, \bibinfo {author}
  {\bibfnamefont {A.}~\bibnamefont {Braunstein}}, \bibinfo {author}
  {\bibfnamefont {A.}~\bibnamefont {Pagnani}}, \bibinfo {author} {\bibfnamefont
  {R.}~\bibnamefont {Zecchina}},\ and\ \bibinfo {author} {\bibfnamefont
  {C.}~\bibnamefont {Sander}},\ }\bibfield  {title} {\bibinfo {title}
  {Perturbation biology: Inferring signaling networks in cellular systems},\
  }\href {https://doi.org/10.1371/journal.pcbi.1003290} {\bibfield  {journal}
  {\bibinfo  {journal} {PLoS Comput. Biol.}\ }\textbf {\bibinfo {volume} {9}},\
  \bibinfo {pages} {e1003290} (\bibinfo {year} {2013})}\BibitemShut {NoStop}%
\bibitem [{\citenamefont {Lucarini}(2018)}]{Lucarini2018a}%
  \BibitemOpen
  \bibfield  {author} {\bibinfo {author} {\bibfnamefont {V.}~\bibnamefont
  {Lucarini}},\ }\bibfield  {title} {\bibinfo {title} {Revising and extending
  the linear response theory for statistical mechanical systems: Evaluating
  observables as predictors and predictands},\ }\href
  {https://doi.org/10.1007/s10955-018-2151-5} {\bibfield  {journal} {\bibinfo
  {journal} {J. Stat. Phys.}\ }\textbf {\bibinfo {volume} {173}},\ \bibinfo
  {pages} {1698} (\bibinfo {year} {2018})}\BibitemShut {NoStop}%
\bibitem [{\citenamefont {Ruelle}(2009)}]{Ruelle2009a}%
  \BibitemOpen
  \bibfield  {author} {\bibinfo {author} {\bibfnamefont {D.}~\bibnamefont
  {Ruelle}},\ }\bibfield  {title} {\bibinfo {title} {A review of linear
  response theory for general differentiable dynamical systems},\ }\href
  {https://doi.org/10.1088/0951-7715/22/4/009} {\bibfield  {journal} {\bibinfo
  {journal} {Nonlinearity}\ }\textbf {\bibinfo {volume} {22}},\ \bibinfo
  {pages} {855} (\bibinfo {year} {2009})}\BibitemShut {NoStop}%
\bibitem [{\citenamefont {Guarnieri}\ \emph {et~al.}(2024)\citenamefont
  {Guarnieri}, \citenamefont {Eisert},\ and\ \citenamefont
  {Miller}}]{Guarnieri2024a}%
  \BibitemOpen
  \bibfield  {author} {\bibinfo {author} {\bibfnamefont {G.}~\bibnamefont
  {Guarnieri}}, \bibinfo {author} {\bibfnamefont {J.}~\bibnamefont {Eisert}},\
  and\ \bibinfo {author} {\bibfnamefont {H.~J.~D.}\ \bibnamefont {Miller}},\
  }\bibfield  {title} {\bibinfo {title} {Generalized linear response theory for
  the full quantum work statistics},\ }\href
  {https://doi.org/10.1103/PhysRevLett.133.070405} {\bibfield  {journal}
  {\bibinfo  {journal} {Phys. Rev. Lett.}\ }\textbf {\bibinfo {volume} {133}},\
  \bibinfo {pages} {070405} (\bibinfo {year} {2024})}\BibitemShut {NoStop}%
\bibitem [{\citenamefont {Puertas}\ \emph {et~al.}(2021)\citenamefont
  {Puertas}, \citenamefont {Trinidad-Segovia}, \citenamefont
  {S{\'a}nchez-Granero}, \citenamefont {Clara-Rahora},\ and\ \citenamefont
  {de~las Nieves}}]{Puertas2021a}%
  \BibitemOpen
  \bibfield  {author} {\bibinfo {author} {\bibfnamefont {A.~M.}\ \bibnamefont
  {Puertas}}, \bibinfo {author} {\bibfnamefont {J.~E.}\ \bibnamefont
  {Trinidad-Segovia}}, \bibinfo {author} {\bibfnamefont {M.~A.}\ \bibnamefont
  {S{\'a}nchez-Granero}}, \bibinfo {author} {\bibfnamefont {J.}~\bibnamefont
  {Clara-Rahora}},\ and\ \bibinfo {author} {\bibfnamefont {F.~J.}\ \bibnamefont
  {de~las Nieves}},\ }\bibfield  {title} {\bibinfo {title} {Linear response
  theory in stock markets},\ }\href
  {https://doi.org/10.1038/s41598-021-02263-6} {\bibfield  {journal} {\bibinfo
  {journal} {Sci. Rep.}\ }\textbf {\bibinfo {volume} {11}},\ \bibinfo {pages}
  {23076} (\bibinfo {year} {2021})}\BibitemShut {NoStop}%
\bibitem [{\citenamefont {Pines}\ and\ \citenamefont
  {Bohm}(1952)}]{Pines1952a}%
  \BibitemOpen
  \bibfield  {author} {\bibinfo {author} {\bibfnamefont {D.}~\bibnamefont
  {Pines}}\ and\ \bibinfo {author} {\bibfnamefont {D.}~\bibnamefont {Bohm}},\
  }\bibfield  {title} {\bibinfo {title} {A collective description of electron
  interactions: {II. Collective} vs individual particle aspects of the
  interactions},\ }\href {https://doi.org/10.1103/PhysRev.85.338} {\bibfield
  {journal} {\bibinfo  {journal} {Phys. Rev.}\ }\textbf {\bibinfo {volume}
  {85}},\ \bibinfo {pages} {338} (\bibinfo {year} {1952})}\BibitemShut
  {NoStop}%
\bibitem [{\citenamefont {Bohm}\ and\ \citenamefont {Pines}(1951)}]{Bohm1951a}%
  \BibitemOpen
  \bibfield  {author} {\bibinfo {author} {\bibfnamefont {D.}~\bibnamefont
  {Bohm}}\ and\ \bibinfo {author} {\bibfnamefont {D.}~\bibnamefont {Pines}},\
  }\bibfield  {title} {\bibinfo {title} {A collective description of electron
  interactions. {I. Magnetic} interactions},\ }\href
  {https://doi.org/10.1103/PhysRev.82.625} {\bibfield  {journal} {\bibinfo
  {journal} {Phys. Rev.}\ }\textbf {\bibinfo {volume} {82}},\ \bibinfo {pages}
  {625} (\bibinfo {year} {1951})}\BibitemShut {NoStop}%
\bibitem [{\citenamefont {Bohm}\ and\ \citenamefont {Pines}(1953)}]{Bohm1953a}%
  \BibitemOpen
  \bibfield  {author} {\bibinfo {author} {\bibfnamefont {D.}~\bibnamefont
  {Bohm}}\ and\ \bibinfo {author} {\bibfnamefont {D.}~\bibnamefont {Pines}},\
  }\bibfield  {title} {\bibinfo {title} {A collective description of electron
  interactions: {III. Coulomb} interactions in a degenerate electron gas},\
  }\href {https://doi.org/10.1103/PhysRev.92.609} {\bibfield  {journal}
  {\bibinfo  {journal} {Phys. Rev.}\ }\textbf {\bibinfo {volume} {92}},\
  \bibinfo {pages} {609} (\bibinfo {year} {1953})}\BibitemShut {NoStop}%
\bibitem [{\citenamefont {Gell-Mann}\ and\ \citenamefont
  {Brueckner}(1957)}]{Gell-Mann1957a}%
  \BibitemOpen
  \bibfield  {author} {\bibinfo {author} {\bibfnamefont {M.}~\bibnamefont
  {Gell-Mann}}\ and\ \bibinfo {author} {\bibfnamefont {K.~A.}\ \bibnamefont
  {Brueckner}},\ }\bibfield  {title} {\bibinfo {title} {Correlation energy of
  an electron gas at high density},\ }\href
  {https://doi.org/10.1103/PhysRev.106.364} {\bibfield  {journal} {\bibinfo
  {journal} {Phys. Rev.}\ }\textbf {\bibinfo {volume} {106}},\ \bibinfo {pages}
  {364} (\bibinfo {year} {1957})}\BibitemShut {NoStop}%
\bibitem [{\citenamefont {Chen}\ \emph {et~al.}(2017)\citenamefont {Chen},
  \citenamefont {Voora}, \citenamefont {Agee}, \citenamefont {Balasubramani},\
  and\ \citenamefont {Furche}}]{Chen2017a}%
  \BibitemOpen
  \bibfield  {author} {\bibinfo {author} {\bibfnamefont {G.~P.}\ \bibnamefont
  {Chen}}, \bibinfo {author} {\bibfnamefont {V.~K.}\ \bibnamefont {Voora}},
  \bibinfo {author} {\bibfnamefont {M.~M.}\ \bibnamefont {Agee}}, \bibinfo
  {author} {\bibfnamefont {S.~G.}\ \bibnamefont {Balasubramani}},\ and\
  \bibinfo {author} {\bibfnamefont {F.}~\bibnamefont {Furche}},\ }\bibfield
  {title} {\bibinfo {title} {Random-phase approximation methods},\ }\href
  {https://doi.org/10.1146/annurev-physchem-040215-112308} {\bibfield
  {journal} {\bibinfo  {journal} {Annu. Rev. Phys. Chem.}\ }\textbf {\bibinfo
  {volume} {68}},\ \bibinfo {pages} {421} (\bibinfo {year} {2017})}\BibitemShut
  {NoStop}%
\bibitem [{\citenamefont {Heßelmann}\ and\ \citenamefont
  {Görling}(2011)}]{Hesselmann2011a}%
  \BibitemOpen
  \bibfield  {author} {\bibinfo {author} {\bibfnamefont {A.}~\bibnamefont
  {Heßelmann}}\ and\ \bibinfo {author} {\bibfnamefont {A.}~\bibnamefont
  {Görling}},\ }\bibfield  {title} {\bibinfo {title} {Random-phase
  approximation correlation methods for molecules and solids},\ }\href
  {https://doi.org/10.1080/00268976.2011.614282} {\bibfield  {journal}
  {\bibinfo  {journal} {Mol. Phys.}\ }\textbf {\bibinfo {volume} {109}},\
  \bibinfo {pages} {2473} (\bibinfo {year} {2011})}\BibitemShut {NoStop}%
\bibitem [{\citenamefont {Yu}\ \emph {et~al.}(2021)\citenamefont {Yu},
  \citenamefont {Nguyen}, \citenamefont {Tsai}, \citenamefont {Hernandez},\
  and\ \citenamefont {Furche}}]{Yu2021a}%
  \BibitemOpen
  \bibfield  {author} {\bibinfo {author} {\bibfnamefont {J.~M.}\ \bibnamefont
  {Yu}}, \bibinfo {author} {\bibfnamefont {B.~D.}\ \bibnamefont {Nguyen}},
  \bibinfo {author} {\bibfnamefont {J.}~\bibnamefont {Tsai}}, \bibinfo {author}
  {\bibfnamefont {D.~J.}\ \bibnamefont {Hernandez}},\ and\ \bibinfo {author}
  {\bibfnamefont {F.}~\bibnamefont {Furche}},\ }\bibfield  {title} {\bibinfo
  {title} {Self-consistent random phase approximation methods},\ }\href
  {https://doi.org/10.1063/5.0056565} {\bibfield  {journal} {\bibinfo
  {journal} {J. Chem. Phys.}\ }\textbf {\bibinfo {volume} {155}},\ \bibinfo
  {pages} {040902} (\bibinfo {year} {2021})}\BibitemShut {NoStop}%
\bibitem [{\citenamefont {Thouless}(1961)}]{Thouless1961a}%
  \BibitemOpen
  \bibfield  {author} {\bibinfo {author} {\bibfnamefont {D.~J.}\ \bibnamefont
  {Thouless}},\ }\bibfield  {title} {\bibinfo {title} {Vibrational states of
  nuclei in the random phase approximation},\ }\href
  {https://doi.org/10.1016/0029-5582(61)90364-9} {\bibfield  {journal}
  {\bibinfo  {journal} {Nucl. Phys.}\ }\textbf {\bibinfo {volume} {22}},\
  \bibinfo {pages} {78} (\bibinfo {year} {1961})}\BibitemShut {NoStop}%
\bibitem [{\citenamefont {Fetter}\ and\ \citenamefont
  {Walecka}(1971)}]{Fetter1971a}%
  \BibitemOpen
  \bibfield  {author} {\bibinfo {author} {\bibfnamefont {A.~L.}\ \bibnamefont
  {Fetter}}\ and\ \bibinfo {author} {\bibfnamefont {J.~D.}\ \bibnamefont
  {Walecka}},\ }\href@noop {} {\emph {\bibinfo {title} {Quantum Theory of Many
  Particle Systems}}}\ (\bibinfo  {publisher} {McGraw-Hill},\ \bibinfo
  {address} {New York},\ \bibinfo {year} {1971})\BibitemShut {NoStop}%
\bibitem [{\citenamefont {Nakatsukasa}\ \emph {et~al.}(2016)\citenamefont
  {Nakatsukasa}, \citenamefont {Matsuyanagi}, \citenamefont {Matsuo},\ and\
  \citenamefont {Yabana}}]{Nakatsukasa2016a}%
  \BibitemOpen
  \bibfield  {author} {\bibinfo {author} {\bibfnamefont {T.}~\bibnamefont
  {Nakatsukasa}}, \bibinfo {author} {\bibfnamefont {K.}~\bibnamefont
  {Matsuyanagi}}, \bibinfo {author} {\bibfnamefont {M.}~\bibnamefont
  {Matsuo}},\ and\ \bibinfo {author} {\bibfnamefont {K.}~\bibnamefont
  {Yabana}},\ }\bibfield  {title} {\bibinfo {title} {Time-dependent
  density-functional description of nuclear dynamics},\ }\href
  {https://doi.org/10.1103/RevModPhys.88.045004} {\bibfield  {journal}
  {\bibinfo  {journal} {Rev. Mod. Phys.}\ }\textbf {\bibinfo {volume} {88}},\
  \bibinfo {pages} {045004} (\bibinfo {year} {2016})}\BibitemShut {NoStop}%
\bibitem [{\citenamefont {Ullrich}(2012)}]{Ullrich2012a}%
  \BibitemOpen
  \bibfield  {author} {\bibinfo {author} {\bibfnamefont {C.~A.}\ \bibnamefont
  {Ullrich}},\ }\href
  {https://doi.org/10.1093/acprof:oso/9780199563029.001.0001} {\emph {\bibinfo
  {title} {Time-Dependent Density-Functional Theory: Concepts and
  Applications}}}\ (\bibinfo  {publisher} {Oxford University Press},\ \bibinfo
  {address} {Oxford},\ \bibinfo {year} {2012})\BibitemShut {NoStop}%
\bibitem [{\citenamefont {Ring}\ and\ \citenamefont
  {Schuck}(2004)}]{Ring2004a}%
  \BibitemOpen
  \bibfield  {author} {\bibinfo {author} {\bibfnamefont {P.}~\bibnamefont
  {Ring}}\ and\ \bibinfo {author} {\bibfnamefont {P.}~\bibnamefont {Schuck}},\
  }\href@noop {} {\emph {\bibinfo {title} {The Nuclear Many-Body Problem}}}\
  (\bibinfo  {publisher} {Springer},\ \bibinfo {address} {New York},\ \bibinfo
  {year} {2004})\BibitemShut {NoStop}%
\bibitem [{\citenamefont {Rowe}(2010)}]{Rowe2010a}%
  \BibitemOpen
  \bibfield  {author} {\bibinfo {author} {\bibfnamefont {D.~J.}\ \bibnamefont
  {Rowe}},\ }\href@noop {} {\emph {\bibinfo {title} {Nuclear Collective Motion:
  Models and Theory}}}\ (\bibinfo  {publisher} {World Scientific},\ \bibinfo
  {year} {2010})\BibitemShut {NoStop}%
\bibitem [{\citenamefont {Scuseria}\ \emph {et~al.}(2008)\citenamefont
  {Scuseria}, \citenamefont {Henderson},\ and\ \citenamefont
  {Sorensen}}]{Scuseria2008a}%
  \BibitemOpen
  \bibfield  {author} {\bibinfo {author} {\bibfnamefont {G.~E.}\ \bibnamefont
  {Scuseria}}, \bibinfo {author} {\bibfnamefont {T.~M.}\ \bibnamefont
  {Henderson}},\ and\ \bibinfo {author} {\bibfnamefont {D.~C.}\ \bibnamefont
  {Sorensen}},\ }\bibfield  {title} {\bibinfo {title} {The ground state
  correlation energy of the random phase approximation from a ring coupled
  cluster doubles approach},\ }\href {https://doi.org/10.1063/1.3043729}
  {\bibfield  {journal} {\bibinfo  {journal} {J. Chem. Phys.}\ }\textbf
  {\bibinfo {volume} {129}},\ \bibinfo {pages} {231101} (\bibinfo {year}
  {2008})}\BibitemShut {NoStop}%
\bibitem [{\citenamefont {Rowe}(1968)}]{Rowe1968a}%
  \BibitemOpen
  \bibfield  {author} {\bibinfo {author} {\bibfnamefont {D.~J.}\ \bibnamefont
  {Rowe}},\ }\bibfield  {title} {\bibinfo {title} {Equations-of-motion method
  and the extended shell model},\ }\href
  {https://doi.org/10.1103/RevModPhys.40.153} {\bibfield  {journal} {\bibinfo
  {journal} {Rev. Mod. Phys.}\ }\textbf {\bibinfo {volume} {40}},\ \bibinfo
  {pages} {153} (\bibinfo {year} {1968})}\BibitemShut {NoStop}%
\bibitem [{\citenamefont {Bogolyubov}(1959)}]{Bogolyubov1959a}%
  \BibitemOpen
  \bibfield  {author} {\bibinfo {author} {\bibfnamefont {N.~N.}\ \bibnamefont
  {Bogolyubov}},\ }\bibfield  {title} {\bibinfo {title} {The compensation
  principle and the self-consistent field method},\ }\href
  {https://doi.org/10.1070/PU1959v002n02ABEH003122} {\bibfield  {journal}
  {\bibinfo  {journal} {Phys.-Usp.}\ }\textbf {\bibinfo {volume} {2}},\
  \bibinfo {pages} {236} (\bibinfo {year} {1959})}\BibitemShut {NoStop}%
\bibitem [{\citenamefont {Baranger}(1960)}]{Baranger1960a}%
  \BibitemOpen
  \bibfield  {author} {\bibinfo {author} {\bibfnamefont {M.}~\bibnamefont
  {Baranger}},\ }\bibfield  {title} {\bibinfo {title} {Extension of the shell
  model for heavy spherical nuclei},\ }\href
  {https://doi.org/10.1103/PhysRev.120.957} {\bibfield  {journal} {\bibinfo
  {journal} {Phys. Rev.}\ }\textbf {\bibinfo {volume} {120}},\ \bibinfo {pages}
  {957} (\bibinfo {year} {1960})}\BibitemShut {NoStop}%
\bibitem [{\citenamefont {Paar}\ \emph {et~al.}(2007)\citenamefont {Paar},
  \citenamefont {Vretenar}, \citenamefont {Khan},\ and\ \citenamefont
  {Col{\`o}}}]{Paar2007a}%
  \BibitemOpen
  \bibfield  {author} {\bibinfo {author} {\bibfnamefont {N.}~\bibnamefont
  {Paar}}, \bibinfo {author} {\bibfnamefont {D.}~\bibnamefont {Vretenar}},
  \bibinfo {author} {\bibfnamefont {E.}~\bibnamefont {Khan}},\ and\ \bibinfo
  {author} {\bibfnamefont {G.}~\bibnamefont {Col{\`o}}},\ }\bibfield  {title}
  {\bibinfo {title} {Exotic modes of excitation in atomic nuclei far from
  stability},\ }\href {https://doi.org/10.1088/0034-4885/70/5/R02} {\bibfield
  {journal} {\bibinfo  {journal} {Rep. Prog. Phys.}\ }\textbf {\bibinfo
  {volume} {70}},\ \bibinfo {pages} {691} (\bibinfo {year} {2007})}\BibitemShut
  {NoStop}%
\bibitem [{\citenamefont {Roca-Maza}\ and\ \citenamefont
  {Paar}(2018)}]{Roca-Maza2018a}%
  \BibitemOpen
  \bibfield  {author} {\bibinfo {author} {\bibfnamefont {X.}~\bibnamefont
  {Roca-Maza}}\ and\ \bibinfo {author} {\bibfnamefont {N.}~\bibnamefont
  {Paar}},\ }\bibfield  {title} {\bibinfo {title} {Nuclear equation of state
  from ground and collective excited state properties of nuclei},\ }\href
  {https://doi.org/10.1016/j.ppnp.2018.04.001} {\bibfield  {journal} {\bibinfo
  {journal} {Prog. Part. Nucl. Phys.}\ }\textbf {\bibinfo {volume} {101}},\
  \bibinfo {pages} {96} (\bibinfo {year} {2018})}\BibitemShut {NoStop}%
\bibitem [{\citenamefont {Lund}\ \emph {et~al.}(2023)\citenamefont {Lund},
  \citenamefont {Engel}, \citenamefont {McLaughlin}, \citenamefont {Mumpower},
  \citenamefont {Ney},\ and\ \citenamefont {Surman}}]{Lund2023a}%
  \BibitemOpen
  \bibfield  {author} {\bibinfo {author} {\bibfnamefont {K.~A.}\ \bibnamefont
  {Lund}}, \bibinfo {author} {\bibfnamefont {J.}~\bibnamefont {Engel}},
  \bibinfo {author} {\bibfnamefont {G.~C.}\ \bibnamefont {McLaughlin}},
  \bibinfo {author} {\bibfnamefont {M.~R.}\ \bibnamefont {Mumpower}}, \bibinfo
  {author} {\bibfnamefont {E.~M.}\ \bibnamefont {Ney}},\ and\ \bibinfo {author}
  {\bibfnamefont {R.}~\bibnamefont {Surman}},\ }\bibfield  {title} {\bibinfo
  {title} {The influence of $\beta$-decay rates on r-process observables},\
  }\href {https://doi.org/10.3847/1538-4357/acaf56} {\bibfield  {journal}
  {\bibinfo  {journal} {Astrophys. J.}\ }\textbf {\bibinfo {volume} {944}},\
  \bibinfo {pages} {144} (\bibinfo {year} {2023})}\BibitemShut {NoStop}%
\bibitem [{\citenamefont {Langanke}\ \emph {et~al.}(2021)\citenamefont
  {Langanke}, \citenamefont {Mart{\'i}nez-Pinedo},\ and\ \citenamefont
  {Zegers}}]{Langanke2021a}%
  \BibitemOpen
  \bibfield  {author} {\bibinfo {author} {\bibfnamefont {K.}~\bibnamefont
  {Langanke}}, \bibinfo {author} {\bibfnamefont {G.}~\bibnamefont
  {Mart{\'i}nez-Pinedo}},\ and\ \bibinfo {author} {\bibfnamefont {R.~G.~T.}\
  \bibnamefont {Zegers}},\ }\bibfield  {title} {\bibinfo {title} {Electron
  capture in stars},\ }\href {https://doi.org/10.1088/1361-6633/abf207}
  {\bibfield  {journal} {\bibinfo  {journal} {Rep. Prog. Phys.}\ }\textbf
  {\bibinfo {volume} {84}},\ \bibinfo {pages} {066301} (\bibinfo {year}
  {2021})}\BibitemShut {NoStop}%
\bibitem [{\citenamefont {Suzuki}(2022)}]{Suzuki2022a}%
  \BibitemOpen
  \bibfield  {author} {\bibinfo {author} {\bibfnamefont {T.}~\bibnamefont
  {Suzuki}},\ }\bibfield  {title} {\bibinfo {title} {Nuclear weak rates and
  nuclear weak processes in stars},\ }\href
  {https://doi.org/10.1016/j.ppnp.2022.103974} {\bibfield  {journal} {\bibinfo
  {journal} {Prog. Part. Nucl. Phys.}\ }\textbf {\bibinfo {volume} {126}},\
  \bibinfo {pages} {103974} (\bibinfo {year} {2022})}\BibitemShut {NoStop}%
\bibitem [{\citenamefont {Engel}\ and\ \citenamefont
  {Men{\'e}ndez}(2017)}]{Engel2017a}%
  \BibitemOpen
  \bibfield  {author} {\bibinfo {author} {\bibfnamefont {J.}~\bibnamefont
  {Engel}}\ and\ \bibinfo {author} {\bibfnamefont {J.}~\bibnamefont
  {Men{\'e}ndez}},\ }\bibfield  {title} {\bibinfo {title} {Status and future of
  nuclear matrix elements for neutrinoless double-beta decay: a review},\
  }\href {https://doi.org/10.1088/1361-6633/aa5bc5} {\bibfield  {journal}
  {\bibinfo  {journal} {Rep. Prog. Phys.}\ }\textbf {\bibinfo {volume} {80}},\
  \bibinfo {pages} {046301} (\bibinfo {year} {2017})}\BibitemShut {NoStop}%
\bibitem [{\citenamefont {Hinohara}\ \emph {et~al.}(2013)\citenamefont
  {Hinohara}, \citenamefont {Kortelainen},\ and\ \citenamefont
  {Nazarewicz}}]{Hinohara2013a}%
  \BibitemOpen
  \bibfield  {author} {\bibinfo {author} {\bibfnamefont {N.}~\bibnamefont
  {Hinohara}}, \bibinfo {author} {\bibfnamefont {M.}~\bibnamefont
  {Kortelainen}},\ and\ \bibinfo {author} {\bibfnamefont {W.}~\bibnamefont
  {Nazarewicz}},\ }\bibfield  {title} {\bibinfo {title} {Low-energy collective
  modes of deformed superfluid nuclei within the finite-amplitude method},\
  }\href {https://doi.org/10.1103/PhysRevC.87.064309} {\bibfield  {journal}
  {\bibinfo  {journal} {Phys. Rev. C}\ }\textbf {\bibinfo {volume} {87}},\
  \bibinfo {pages} {064309} (\bibinfo {year} {2013})}\BibitemShut {NoStop}%
\bibitem [{\citenamefont {Kortelainen}\ \emph {et~al.}(2015)\citenamefont
  {Kortelainen}, \citenamefont {Hinohara},\ and\ \citenamefont
  {Nazarewicz}}]{Kortelainen2015a}%
  \BibitemOpen
  \bibfield  {author} {\bibinfo {author} {\bibfnamefont {M.}~\bibnamefont
  {Kortelainen}}, \bibinfo {author} {\bibfnamefont {N.}~\bibnamefont
  {Hinohara}},\ and\ \bibinfo {author} {\bibfnamefont {W.}~\bibnamefont
  {Nazarewicz}},\ }\bibfield  {title} {\bibinfo {title} {Multipole modes in
  deformed nuclei within the finite amplitude method},\ }\href
  {https://doi.org/10.1103/PhysRevC.92.051302} {\bibfield  {journal} {\bibinfo
  {journal} {Phys. Rev. C}\ }\textbf {\bibinfo {volume} {92}},\ \bibinfo
  {pages} {051302} (\bibinfo {year} {2015})}\BibitemShut {NoStop}%
\bibitem [{\citenamefont {Marketin}\ \emph {et~al.}(2016)\citenamefont
  {Marketin}, \citenamefont {Huther},\ and\ \citenamefont
  {Mart{\'i}nez-Pinedo}}]{Marketin2016a}%
  \BibitemOpen
  \bibfield  {author} {\bibinfo {author} {\bibfnamefont {T.}~\bibnamefont
  {Marketin}}, \bibinfo {author} {\bibfnamefont {L.}~\bibnamefont {Huther}},\
  and\ \bibinfo {author} {\bibfnamefont {G.}~\bibnamefont
  {Mart{\'i}nez-Pinedo}},\ }\bibfield  {title} {\bibinfo {title} {Large-scale
  evaluation of $\beta$-decay rates of $r$-process nuclei with the inclusion of
  first-forbidden transitions},\ }\href
  {https://doi.org/10.1103/PhysRevC.93.025805} {\bibfield  {journal} {\bibinfo
  {journal} {Phys. Rev. C}\ }\textbf {\bibinfo {volume} {93}},\ \bibinfo
  {pages} {025805} (\bibinfo {year} {2016})}\BibitemShut {NoStop}%
\bibitem [{\citenamefont {Ney}\ \emph {et~al.}(2020)\citenamefont {Ney},
  \citenamefont {Engel}, \citenamefont {Li},\ and\ \citenamefont
  {Schunck}}]{Ney2020a}%
  \BibitemOpen
  \bibfield  {author} {\bibinfo {author} {\bibfnamefont {E.~M.}\ \bibnamefont
  {Ney}}, \bibinfo {author} {\bibfnamefont {J.}~\bibnamefont {Engel}}, \bibinfo
  {author} {\bibfnamefont {T.}~\bibnamefont {Li}},\ and\ \bibinfo {author}
  {\bibfnamefont {N.}~\bibnamefont {Schunck}},\ }\bibfield  {title} {\bibinfo
  {title} {Global description of $\beta^-$ decay with the axially deformed
  {Skyrme} finite-amplitude method: Extension to odd-mass and odd-odd nuclei},\
  }\href {https://doi.org/10.1103/PhysRevC.102.034326} {\bibfield  {journal}
  {\bibinfo  {journal} {Phys. Rev. C}\ }\textbf {\bibinfo {volume} {102}},\
  \bibinfo {pages} {034326} (\bibinfo {year} {2020})}\BibitemShut {NoStop}%
\bibitem [{\citenamefont {Horowitz}\ \emph {et~al.}(2019)\citenamefont
  {Horowitz} \emph {et~al.}}]{Horowitz2019a}%
  \BibitemOpen
  \bibfield  {author} {\bibinfo {author} {\bibfnamefont {C.~J.}\ \bibnamefont
  {Horowitz}} \emph {et~al.},\ }\bibfield  {title} {\bibinfo {title} {r-process
  nucleosynthesis: connecting rare-isotope beam facilities with the cosmos},\
  }\href {https://doi.org/10.1088/1361-6471/ab0849} {\bibfield  {journal}
  {\bibinfo  {journal} {J. Phys. G}\ }\textbf {\bibinfo {volume} {46}},\
  \bibinfo {pages} {083001} (\bibinfo {year} {2019})}\BibitemShut {NoStop}%
\bibitem [{\citenamefont {Suhonen}(2007)}]{Suhonen2007a}%
  \BibitemOpen
  \bibfield  {author} {\bibinfo {author} {\bibfnamefont {J.}~\bibnamefont
  {Suhonen}},\ }\href {https://doi.org/10.1007/978-3-540-48861-3} {\emph
  {\bibinfo {title} {From Nucleons to Nucleus: Concepts of Microscopic Nuclear
  Theory}}}\ (\bibinfo  {publisher} {Springer},\ \bibinfo {address} {Berlin,
  Heidelberg},\ \bibinfo {year} {2007})\BibitemShut {NoStop}%
\bibitem [{\citenamefont {Nakatsukasa}\ \emph {et~al.}(2007)\citenamefont
  {Nakatsukasa}, \citenamefont {Inakura},\ and\ \citenamefont
  {Yabana}}]{Nakatsukasa2007a}%
  \BibitemOpen
  \bibfield  {author} {\bibinfo {author} {\bibfnamefont {T.}~\bibnamefont
  {Nakatsukasa}}, \bibinfo {author} {\bibfnamefont {T.}~\bibnamefont
  {Inakura}},\ and\ \bibinfo {author} {\bibfnamefont {K.}~\bibnamefont
  {Yabana}},\ }\bibfield  {title} {\bibinfo {title} {Finite amplitude method
  for the solution of the random-phase approximation},\ }\href
  {https://doi.org/10.1103/PhysRevC.76.024318} {\bibfield  {journal} {\bibinfo
  {journal} {Phys. Rev. C}\ }\textbf {\bibinfo {volume} {76}},\ \bibinfo
  {pages} {024318} (\bibinfo {year} {2007})}\BibitemShut {NoStop}%
\bibitem [{\citenamefont {Bender}\ \emph {et~al.}(2003)\citenamefont {Bender},
  \citenamefont {Heenen},\ and\ \citenamefont {Reinhard}}]{Bender2003a}%
  \BibitemOpen
  \bibfield  {author} {\bibinfo {author} {\bibfnamefont {M.}~\bibnamefont
  {Bender}}, \bibinfo {author} {\bibfnamefont {P.-H.}\ \bibnamefont {Heenen}},\
  and\ \bibinfo {author} {\bibfnamefont {P.-G.}\ \bibnamefont {Reinhard}},\
  }\bibfield  {title} {\bibinfo {title} {Self-consistent mean-field models for
  nuclear structure},\ }\href {https://doi.org/10.1103/RevModPhys.75.121}
  {\bibfield  {journal} {\bibinfo  {journal} {Rev. Mod. Phys.}\ }\textbf
  {\bibinfo {volume} {75}},\ \bibinfo {pages} {121} (\bibinfo {year}
  {2003})}\BibitemShut {NoStop}%
\bibitem [{\citenamefont {Kl{\"u}pfel}\ \emph {et~al.}(2009)\citenamefont
  {Kl{\"u}pfel}, \citenamefont {Reinhard}, \citenamefont {B{\"u}rvenich},\ and\
  \citenamefont {Maruhn}}]{Klupfel2009a}%
  \BibitemOpen
  \bibfield  {author} {\bibinfo {author} {\bibfnamefont {P.}~\bibnamefont
  {Kl{\"u}pfel}}, \bibinfo {author} {\bibfnamefont {P.-G.}\ \bibnamefont
  {Reinhard}}, \bibinfo {author} {\bibfnamefont {T.~J.}\ \bibnamefont
  {B{\"u}rvenich}},\ and\ \bibinfo {author} {\bibfnamefont {J.~A.}\
  \bibnamefont {Maruhn}},\ }\bibfield  {title} {\bibinfo {title} {Variations on
  a theme by {Skyrme: A} systematic study of adjustments of model parameters},\
  }\href {https://doi.org/10.1103/PhysRevC.79.034310} {\bibfield  {journal}
  {\bibinfo  {journal} {Phys. Rev. C}\ }\textbf {\bibinfo {volume} {79}},\
  \bibinfo {pages} {034310} (\bibinfo {year} {2009})},\ \Eprint
  {https://arxiv.org/abs/0804.3385} {arXiv:0804.3385 [nucl-th]} \BibitemShut
  {NoStop}%
\bibitem [{\citenamefont {Kortelainen}\ \emph {et~al.}(2010)\citenamefont
  {Kortelainen}, \citenamefont {Lesinski}, \citenamefont {Mor{\'e}},
  \citenamefont {Nazarewicz}, \citenamefont {Sarich}, \citenamefont {Schunck},
  \citenamefont {Stoitsov},\ and\ \citenamefont {Wild}}]{Kortelainen2010a}%
  \BibitemOpen
  \bibfield  {author} {\bibinfo {author} {\bibfnamefont {M.}~\bibnamefont
  {Kortelainen}}, \bibinfo {author} {\bibfnamefont {T.}~\bibnamefont
  {Lesinski}}, \bibinfo {author} {\bibfnamefont {J.}~\bibnamefont {Mor{\'e}}},
  \bibinfo {author} {\bibfnamefont {W.}~\bibnamefont {Nazarewicz}}, \bibinfo
  {author} {\bibfnamefont {J.}~\bibnamefont {Sarich}}, \bibinfo {author}
  {\bibfnamefont {N.}~\bibnamefont {Schunck}}, \bibinfo {author} {\bibfnamefont
  {M.~V.}\ \bibnamefont {Stoitsov}},\ and\ \bibinfo {author} {\bibfnamefont
  {S.}~\bibnamefont {Wild}},\ }\bibfield  {title} {\bibinfo {title} {Nuclear
  energy density optimization},\ }\href
  {https://doi.org/10.1103/PhysRevC.82.024313} {\bibfield  {journal} {\bibinfo
  {journal} {Phys. Rev. C}\ }\textbf {\bibinfo {volume} {82}},\ \bibinfo
  {pages} {024313} (\bibinfo {year} {2010})}\BibitemShut {NoStop}%
\bibitem [{\citenamefont {Y{\"u}ksel}\ \emph {et~al.}(2019)\citenamefont
  {Y{\"u}ksel}, \citenamefont {Marketin},\ and\ \citenamefont
  {Paar}}]{Yuksel2019a}%
  \BibitemOpen
  \bibfield  {author} {\bibinfo {author} {\bibfnamefont {E.}~\bibnamefont
  {Y{\"u}ksel}}, \bibinfo {author} {\bibfnamefont {T.}~\bibnamefont
  {Marketin}},\ and\ \bibinfo {author} {\bibfnamefont {N.}~\bibnamefont
  {Paar}},\ }\bibfield  {title} {\bibinfo {title} {Optimizing the relativistic
  energy density functional with nuclear ground-state and collective excitation
  properties},\ }\href {https://doi.org/10.1103/PhysRevC.99.034318} {\bibfield
  {journal} {\bibinfo  {journal} {Phys. Rev. C}\ }\textbf {\bibinfo {volume}
  {99}},\ \bibinfo {pages} {034318} (\bibinfo {year} {2019})}\BibitemShut
  {NoStop}%
\bibitem [{\citenamefont {Lin}\ \emph {et~al.}(2025)\citenamefont {Lin},
  \citenamefont {Col{\`o}}, \citenamefont {Stiner},\ and\ \citenamefont
  {Stinson}}]{Lin2025a}%
  \BibitemOpen
  \bibfield  {author} {\bibinfo {author} {\bibfnamefont {Z.}~\bibnamefont
  {Lin}}, \bibinfo {author} {\bibfnamefont {G.}~\bibnamefont {Col{\`o}}},
  \bibinfo {author} {\bibfnamefont {A.~W.}\ \bibnamefont {Stiner}},\ and\
  \bibinfo {author} {\bibfnamefont {A.}~\bibnamefont {Stinson}},\ }\bibfield
  {title} {\bibinfo {title} {Bayesian inference of the {Landau} parameter
  $g'_0$ from joint {Gamow-Teller} measurements},\ }\href
  {https://doi.org/10.48550/arXiv.2506.05564} {\bibfield  {journal} {\bibinfo
  {journal} {arXiv preprint arXiv:2506.05564}\ } (\bibinfo {year}
  {2025})}\BibitemShut {NoStop}%
\bibitem [{\citenamefont {Phillips}\ \emph {et~al.}(2021)\citenamefont
  {Phillips}, \citenamefont {Furnstahl}, \citenamefont {Heinz}, \citenamefont
  {Maiti}, \citenamefont {Nazarewicz}, \citenamefont {Nunes}, \citenamefont
  {Plumlee}, \citenamefont {Pratola}, \citenamefont {Pratt}, \citenamefont
  {Viens} \emph {et~al.}}]{Phillips2021a}%
  \BibitemOpen
  \bibfield  {author} {\bibinfo {author} {\bibfnamefont {D.~R.}\ \bibnamefont
  {Phillips}}, \bibinfo {author} {\bibfnamefont {R.~J.}\ \bibnamefont
  {Furnstahl}}, \bibinfo {author} {\bibfnamefont {U.}~\bibnamefont {Heinz}},
  \bibinfo {author} {\bibfnamefont {T.}~\bibnamefont {Maiti}}, \bibinfo
  {author} {\bibfnamefont {W.}~\bibnamefont {Nazarewicz}}, \bibinfo {author}
  {\bibfnamefont {F.~M.}\ \bibnamefont {Nunes}}, \bibinfo {author}
  {\bibfnamefont {M.}~\bibnamefont {Plumlee}}, \bibinfo {author} {\bibfnamefont
  {M.~T.}\ \bibnamefont {Pratola}}, \bibinfo {author} {\bibfnamefont
  {S.}~\bibnamefont {Pratt}}, \bibinfo {author} {\bibfnamefont {F.~G.}\
  \bibnamefont {Viens}}, \emph {et~al.},\ }\bibfield  {title} {\bibinfo {title}
  {Get on the {BAND wagon: A Bayesian }framework for quantifying model
  uncertainties in nuclear dynamics},\ }\href
  {https://doi.org/10.1088/1361-6471/abf1df} {\bibfield  {journal} {\bibinfo
  {journal} {J. Phys. G: Nucl. Part. Phys.}\ }\textbf {\bibinfo {volume}
  {48}},\ \bibinfo {pages} {072001} (\bibinfo {year} {2021})}\BibitemShut
  {NoStop}%
\bibitem [{\citenamefont {Brunton}\ and\ \citenamefont
  {Kutz}(2022)}]{Brunton2022a}%
  \BibitemOpen
  \bibfield  {author} {\bibinfo {author} {\bibfnamefont {S.~L.}\ \bibnamefont
  {Brunton}}\ and\ \bibinfo {author} {\bibfnamefont {J.~N.}\ \bibnamefont
  {Kutz}},\ }\href@noop {} {\emph {\bibinfo {title} {Data-driven Science and
  Engineering: Machine Learning, Dynamical Systems, and Control}}}\ (\bibinfo
  {publisher} {Cambridge University Press},\ \bibinfo {year}
  {2022})\BibitemShut {NoStop}%
\bibitem [{\citenamefont {Boehnlein}\ \emph {et~al.}(2022)\citenamefont
  {Boehnlein}, \citenamefont {Diefenthaler}, \citenamefont {Sato},
  \citenamefont {Schram}, \citenamefont {Ziegler}, \citenamefont {Fanelli},
  \citenamefont {Hjorth-Jensen}, \citenamefont {Horn}, \citenamefont {Kuchera},
  \citenamefont {Lee}, \citenamefont {Nazarewicz}, \citenamefont {Ostroumov},
  \citenamefont {Orginos}, \citenamefont {Poon}, \citenamefont {Wang},
  \citenamefont {Scheinker}, \citenamefont {Smith},\ and\ \citenamefont
  {Pang}}]{Boehnlein2022a}%
  \BibitemOpen
  \bibfield  {author} {\bibinfo {author} {\bibfnamefont {A.}~\bibnamefont
  {Boehnlein}}, \bibinfo {author} {\bibfnamefont {M.}~\bibnamefont
  {Diefenthaler}}, \bibinfo {author} {\bibfnamefont {N.}~\bibnamefont {Sato}},
  \bibinfo {author} {\bibfnamefont {M.}~\bibnamefont {Schram}}, \bibinfo
  {author} {\bibfnamefont {V.}~\bibnamefont {Ziegler}}, \bibinfo {author}
  {\bibfnamefont {C.}~\bibnamefont {Fanelli}}, \bibinfo {author} {\bibfnamefont
  {M.}~\bibnamefont {Hjorth-Jensen}}, \bibinfo {author} {\bibfnamefont
  {T.}~\bibnamefont {Horn}}, \bibinfo {author} {\bibfnamefont {M.~P.}\
  \bibnamefont {Kuchera}}, \bibinfo {author} {\bibfnamefont {D.}~\bibnamefont
  {Lee}}, \bibinfo {author} {\bibfnamefont {W.}~\bibnamefont {Nazarewicz}},
  \bibinfo {author} {\bibfnamefont {P.}~\bibnamefont {Ostroumov}}, \bibinfo
  {author} {\bibfnamefont {K.}~\bibnamefont {Orginos}}, \bibinfo {author}
  {\bibfnamefont {A.}~\bibnamefont {Poon}}, \bibinfo {author} {\bibfnamefont
  {X.-N.}\ \bibnamefont {Wang}}, \bibinfo {author} {\bibfnamefont
  {A.}~\bibnamefont {Scheinker}}, \bibinfo {author} {\bibfnamefont {M.~S.}\
  \bibnamefont {Smith}},\ and\ \bibinfo {author} {\bibfnamefont {L.-G.}\
  \bibnamefont {Pang}},\ }\bibfield  {title} {\bibinfo {title} {Colloquium:
  Machine learning in nuclear physics},\ }\href
  {https://doi.org/10.1103/RevModPhys.94.031003} {\bibfield  {journal}
  {\bibinfo  {journal} {Rev. Mod. Phys.}\ }\textbf {\bibinfo {volume} {94}},\
  \bibinfo {pages} {031003} (\bibinfo {year} {2022})}\BibitemShut {NoStop}%
\bibitem [{\citenamefont {Bonilla}\ \emph {et~al.}(2022)\citenamefont
  {Bonilla}, \citenamefont {Giuliani}, \citenamefont {Godbey},\ and\
  \citenamefont {Lee}}]{Bonilla2022a}%
  \BibitemOpen
  \bibfield  {author} {\bibinfo {author} {\bibfnamefont {E.}~\bibnamefont
  {Bonilla}}, \bibinfo {author} {\bibfnamefont {P.}~\bibnamefont {Giuliani}},
  \bibinfo {author} {\bibfnamefont {K.}~\bibnamefont {Godbey}},\ and\ \bibinfo
  {author} {\bibfnamefont {D.}~\bibnamefont {Lee}},\ }\bibfield  {title}
  {\bibinfo {title} {Training and projecting: A reduced basis method emulator
  for many-body physics},\ }\href {https://doi.org/10.1103/PhysRevC.106.054322}
  {\bibfield  {journal} {\bibinfo  {journal} {Phys. Rev. C}\ }\textbf {\bibinfo
  {volume} {106}},\ \bibinfo {pages} {054322} (\bibinfo {year}
  {2022})}\BibitemShut {NoStop}%
\bibitem [{\citenamefont {Duguet}\ \emph {et~al.}(2024)\citenamefont {Duguet},
  \citenamefont {Ekstr{\"o}m}, \citenamefont {Furnstahl}, \citenamefont
  {K{\"o}nig},\ and\ \citenamefont {Lee}}]{Duguet2024a}%
  \BibitemOpen
  \bibfield  {author} {\bibinfo {author} {\bibfnamefont {T.}~\bibnamefont
  {Duguet}}, \bibinfo {author} {\bibfnamefont {A.}~\bibnamefont {Ekstr{\"o}m}},
  \bibinfo {author} {\bibfnamefont {R.~J.}\ \bibnamefont {Furnstahl}}, \bibinfo
  {author} {\bibfnamefont {S.}~\bibnamefont {K{\"o}nig}},\ and\ \bibinfo
  {author} {\bibfnamefont {D.}~\bibnamefont {Lee}},\ }\bibfield  {title}
  {\bibinfo {title} {Colloquium: Eigenvector continuation and projection-based
  emulators},\ }\href {https://doi.org/10.1103/RevModPhys.96.031002} {\bibfield
   {journal} {\bibinfo  {journal} {Rev. Mod. Phys.}\ }\textbf {\bibinfo
  {volume} {96}},\ \bibinfo {pages} {031002} (\bibinfo {year}
  {2024})}\BibitemShut {NoStop}%
\bibitem [{\citenamefont {Gramacy}(2020)}]{Gramacy2020a}%
  \BibitemOpen
  \bibfield  {author} {\bibinfo {author} {\bibfnamefont {R.~B.}\ \bibnamefont
  {Gramacy}},\ }\href {https://doi.org/10.1201/9780367815493} {\emph {\bibinfo
  {title} {Surrogates: Gaussian Process Modeling, Design, and Optimization for
  the Applied Sciences}}}\ (\bibinfo  {publisher} {Chapman and Hall/CRC},\
  \bibinfo {year} {2020})\BibitemShut {NoStop}%
\bibitem [{\citenamefont {Benner}\ \emph {et~al.}(2015)\citenamefont {Benner},
  \citenamefont {Gugercin},\ and\ \citenamefont {Willcox}}]{Benner2015a}%
  \BibitemOpen
  \bibfield  {author} {\bibinfo {author} {\bibfnamefont {P.}~\bibnamefont
  {Benner}}, \bibinfo {author} {\bibfnamefont {S.}~\bibnamefont {Gugercin}},\
  and\ \bibinfo {author} {\bibfnamefont {K.}~\bibnamefont {Willcox}},\
  }\bibfield  {title} {\bibinfo {title} {A survey of projection-based model
  reduction methods for parametric dynamical systems},\ }\href
  {https://doi.org/10.1137/130932715} {\bibfield  {journal} {\bibinfo
  {journal} {SIAM Rev.}\ }\textbf {\bibinfo {volume} {57}},\ \bibinfo {pages}
  {483} (\bibinfo {year} {2015})}\BibitemShut {NoStop}%
\bibitem [{\citenamefont {Quarteroni}\ \emph {et~al.}(2015)\citenamefont
  {Quarteroni}, \citenamefont {Manzoni},\ and\ \citenamefont
  {Negri}}]{Quarteroni2015a}%
  \BibitemOpen
  \bibfield  {author} {\bibinfo {author} {\bibfnamefont {A.}~\bibnamefont
  {Quarteroni}}, \bibinfo {author} {\bibfnamefont {A.}~\bibnamefont
  {Manzoni}},\ and\ \bibinfo {author} {\bibfnamefont {F.}~\bibnamefont
  {Negri}},\ }\href {https://doi.org/10.1007/978-3-319-15431-2} {\emph
  {\bibinfo {title} {Reduced Basis Methods for Partial Differential Equations:
  An Introduction}}}\ (\bibinfo  {publisher} {Springer},\ \bibinfo {address}
  {New York},\ \bibinfo {year} {2015})\BibitemShut {NoStop}%
\bibitem [{\citenamefont {Hesthaven}\ \emph {et~al.}(2016)\citenamefont
  {Hesthaven}, \citenamefont {Rozza}, \citenamefont {Stamm} \emph
  {et~al.}}]{Hesthaven2016a}%
  \BibitemOpen
  \bibfield  {author} {\bibinfo {author} {\bibfnamefont {J.~S.}\ \bibnamefont
  {Hesthaven}}, \bibinfo {author} {\bibfnamefont {G.}~\bibnamefont {Rozza}},
  \bibinfo {author} {\bibfnamefont {B.}~\bibnamefont {Stamm}}, \emph {et~al.},\
  }\href {https://doi.org/10.1007/978-3-319-22470-1} {\emph {\bibinfo {title}
  {Certified Reduced Basis Methods for Parametrized Partial Differential
  Equations}}}\ (\bibinfo  {publisher} {Springer},\ \bibinfo {address} {New
  York},\ \bibinfo {year} {2016})\BibitemShut {NoStop}%
\bibitem [{\citenamefont {Giuliani}\ \emph {et~al.}(2023)\citenamefont
  {Giuliani}, \citenamefont {Godbey}, \citenamefont {Bonilla}, \citenamefont
  {Viens},\ and\ \citenamefont {Piekarewicz}}]{Giuliani2023a}%
  \BibitemOpen
  \bibfield  {author} {\bibinfo {author} {\bibfnamefont {P.}~\bibnamefont
  {Giuliani}}, \bibinfo {author} {\bibfnamefont {K.}~\bibnamefont {Godbey}},
  \bibinfo {author} {\bibfnamefont {E.}~\bibnamefont {Bonilla}}, \bibinfo
  {author} {\bibfnamefont {F.}~\bibnamefont {Viens}},\ and\ \bibinfo {author}
  {\bibfnamefont {J.}~\bibnamefont {Piekarewicz}},\ }\bibfield  {title}
  {\bibinfo {title} {Bayes goes fast: Uncertainty quantification for a
  covariant energy density functional emulated by the reduced basis method},\
  }\href {https://doi.org/10.3389/fphy.2022.1054524} {\bibfield  {journal}
  {\bibinfo  {journal} {Front. Phys.}\ }\textbf {\bibinfo {volume} {10}},\
  \bibinfo {pages} {1054524} (\bibinfo {year} {2023})}\BibitemShut {NoStop}%
\bibitem [{\citenamefont {Odell}\ \emph {et~al.}(2024)\citenamefont {Odell},
  \citenamefont {Giuliani}, \citenamefont {Beyer}, \citenamefont
  {Catacora-Rios}, \citenamefont {Chan}, \citenamefont {Bonilla}, \citenamefont
  {Furnstahl}, \citenamefont {Godbey},\ and\ \citenamefont
  {Nunes}}]{Odell2024a}%
  \BibitemOpen
  \bibfield  {author} {\bibinfo {author} {\bibfnamefont {D.}~\bibnamefont
  {Odell}}, \bibinfo {author} {\bibfnamefont {P.}~\bibnamefont {Giuliani}},
  \bibinfo {author} {\bibfnamefont {K.}~\bibnamefont {Beyer}}, \bibinfo
  {author} {\bibfnamefont {M.}~\bibnamefont {Catacora-Rios}}, \bibinfo {author}
  {\bibfnamefont {M.~Y.-H.}\ \bibnamefont {Chan}}, \bibinfo {author}
  {\bibfnamefont {E.}~\bibnamefont {Bonilla}}, \bibinfo {author} {\bibfnamefont
  {R.~J.}\ \bibnamefont {Furnstahl}}, \bibinfo {author} {\bibfnamefont
  {K.}~\bibnamefont {Godbey}},\ and\ \bibinfo {author} {\bibfnamefont {F.~M.}\
  \bibnamefont {Nunes}},\ }\bibfield  {title} {\bibinfo {title} {Rose: A
  reduced-order scattering emulator for optical models},\ }\href
  {https://doi.org/10.1103/PhysRevC.109.044612} {\bibfield  {journal} {\bibinfo
   {journal} {Phys. Rev. C}\ }\textbf {\bibinfo {volume} {109}},\ \bibinfo
  {pages} {044612} (\bibinfo {year} {2024})}\BibitemShut {NoStop}%
\bibitem [{\citenamefont {Burohman}\ \emph {et~al.}(2023)\citenamefont
  {Burohman}, \citenamefont {Besselink}, \citenamefont {Scherpen},\ and\
  \citenamefont {Camlibel}}]{Burohman2023a}%
  \BibitemOpen
  \bibfield  {author} {\bibinfo {author} {\bibfnamefont {A.~M.}\ \bibnamefont
  {Burohman}}, \bibinfo {author} {\bibfnamefont {B.}~\bibnamefont {Besselink}},
  \bibinfo {author} {\bibfnamefont {J.~M.~A.}\ \bibnamefont {Scherpen}},\ and\
  \bibinfo {author} {\bibfnamefont {M.~K.}\ \bibnamefont {Camlibel}},\
  }\bibfield  {title} {\bibinfo {title} {From data to reduced-order models via
  generalized balanced truncation},\ }\href
  {https://doi.org/10.1016/j.sysconle.2024.105965} {\bibfield  {journal}
  {\bibinfo  {journal} {IEEE Trans. Autom. Control}\ }\textbf {\bibinfo
  {volume} {68}},\ \bibinfo {pages} {6160} (\bibinfo {year}
  {2023})}\BibitemShut {NoStop}%
\bibitem [{\citenamefont {Brunton}\ \emph {et~al.}(2016)\citenamefont
  {Brunton}, \citenamefont {Proctor},\ and\ \citenamefont
  {Kutz}}]{Brunton2016a}%
  \BibitemOpen
  \bibfield  {author} {\bibinfo {author} {\bibfnamefont {S.~L.}\ \bibnamefont
  {Brunton}}, \bibinfo {author} {\bibfnamefont {J.~L.}\ \bibnamefont
  {Proctor}},\ and\ \bibinfo {author} {\bibfnamefont {J.~N.}\ \bibnamefont
  {Kutz}},\ }\bibfield  {title} {\bibinfo {title} {Discovering governing
  equations from data by sparse identification of nonlinear dynamical
  systems},\ }\href {https://doi.org/10.1073/pnas.1517384113} {\bibfield
  {journal} {\bibinfo  {journal} {Proc. Natl. Acad. Sci. USA}\ }\textbf
  {\bibinfo {volume} {113}},\ \bibinfo {pages} {3932} (\bibinfo {year}
  {2016})}\BibitemShut {NoStop}%
\bibitem [{\citenamefont {Peherstorfer}\ and\ \citenamefont
  {Willcox}(2016)}]{Peherstorfer2016a}%
  \BibitemOpen
  \bibfield  {author} {\bibinfo {author} {\bibfnamefont {B.}~\bibnamefont
  {Peherstorfer}}\ and\ \bibinfo {author} {\bibfnamefont {K.}~\bibnamefont
  {Willcox}},\ }\bibfield  {title} {\bibinfo {title} {Data-driven operator
  inference for nonintrusive projection-based model reduction},\ }\href
  {https://doi.org/10.1016/j.cma.2016.03.025} {\bibfield  {journal} {\bibinfo
  {journal} {Comput. Methods Appl. Mech. Eng.}\ }\textbf {\bibinfo {volume}
  {306}},\ \bibinfo {pages} {196} (\bibinfo {year} {2016})}\BibitemShut
  {NoStop}%
\bibitem [{\citenamefont {Benner}\ \emph {et~al.}(2020)\citenamefont {Benner},
  \citenamefont {Goyal}, \citenamefont {Kramer}, \citenamefont {Peherstorfer},\
  and\ \citenamefont {Willcox}}]{Benner2020a}%
  \BibitemOpen
  \bibfield  {author} {\bibinfo {author} {\bibfnamefont {P.}~\bibnamefont
  {Benner}}, \bibinfo {author} {\bibfnamefont {P.}~\bibnamefont {Goyal}},
  \bibinfo {author} {\bibfnamefont {B.}~\bibnamefont {Kramer}}, \bibinfo
  {author} {\bibfnamefont {B.}~\bibnamefont {Peherstorfer}},\ and\ \bibinfo
  {author} {\bibfnamefont {K.}~\bibnamefont {Willcox}},\ }\bibfield  {title}
  {\bibinfo {title} {Operator inference for non-intrusive model reduction of
  systems with non-polynomial nonlinear terms},\ }\href
  {https://doi.org/10.1016/j.cma.2020.113433} {\bibfield  {journal} {\bibinfo
  {journal} {Comput. Methods Appl. Mech. Eng.}\ }\textbf {\bibinfo {volume}
  {372}},\ \bibinfo {pages} {113433} (\bibinfo {year} {2020})}\BibitemShut
  {NoStop}%
\bibitem [{\citenamefont {Champion}\ \emph {et~al.}(2019)\citenamefont
  {Champion}, \citenamefont {Lusch}, \citenamefont {Kutz},\ and\ \citenamefont
  {Brunton}}]{Champion2019a}%
  \BibitemOpen
  \bibfield  {author} {\bibinfo {author} {\bibfnamefont {K.}~\bibnamefont
  {Champion}}, \bibinfo {author} {\bibfnamefont {B.}~\bibnamefont {Lusch}},
  \bibinfo {author} {\bibfnamefont {J.~N.}\ \bibnamefont {Kutz}},\ and\
  \bibinfo {author} {\bibfnamefont {S.~L.}\ \bibnamefont {Brunton}},\
  }\bibfield  {title} {\bibinfo {title} {Data-driven discovery of coordinates
  and governing equations},\ }\href {https://doi.org/10.1073/pnas.1906995116}
  {\bibfield  {journal} {\bibinfo  {journal} {Proc. Natl. Acad. Sci. USA}\
  }\textbf {\bibinfo {volume} {116}},\ \bibinfo {pages} {22445} (\bibinfo
  {year} {2019})}\BibitemShut {NoStop}%
\bibitem [{\citenamefont {Loiseau}\ and\ \citenamefont
  {Brunton}(2018)}]{Loiseau2018a}%
  \BibitemOpen
  \bibfield  {author} {\bibinfo {author} {\bibfnamefont {J.-C.}\ \bibnamefont
  {Loiseau}}\ and\ \bibinfo {author} {\bibfnamefont {S.~L.}\ \bibnamefont
  {Brunton}},\ }\bibfield  {title} {\bibinfo {title} {Constrained sparse
  galerkin regression},\ }\href {https://doi.org/10.1017/jfm.2017.823}
  {\bibfield  {journal} {\bibinfo  {journal} {J. Fluid Mech.}\ }\textbf
  {\bibinfo {volume} {838}},\ \bibinfo {pages} {42} (\bibinfo {year}
  {2018})}\BibitemShut {NoStop}%
\bibitem [{\citenamefont {Armstrong}\ \emph {et~al.}(2025)\citenamefont
  {Armstrong}, \citenamefont {Giuliani}, \citenamefont {Godbey}, \citenamefont
  {Somasundaram},\ and\ \citenamefont {Tews}}]{Armstrong2025a}%
  \BibitemOpen
  \bibfield  {author} {\bibinfo {author} {\bibfnamefont {C.~L.}\ \bibnamefont
  {Armstrong}}, \bibinfo {author} {\bibfnamefont {P.}~\bibnamefont {Giuliani}},
  \bibinfo {author} {\bibfnamefont {K.}~\bibnamefont {Godbey}}, \bibinfo
  {author} {\bibfnamefont {R.}~\bibnamefont {Somasundaram}},\ and\ \bibinfo
  {author} {\bibfnamefont {I.}~\bibnamefont {Tews}},\ }\bibfield  {title}
  {\bibinfo {title} {Emulators for scarce and noisy data: {Application to
  auxiliary-field diffusion Monte Carlo} for neutron matter},\ }\href
  {https://doi.org/10.1103/9928-wyjm} {\bibfield  {journal} {\bibinfo
  {journal} {Phys. Rev. Lett.}\ }\textbf {\bibinfo {volume} {135}},\ \bibinfo
  {pages} {142501} (\bibinfo {year} {2025})}\BibitemShut {NoStop}%
\bibitem [{\citenamefont {Somasundaram}\ \emph {et~al.}(2025)\citenamefont
  {Somasundaram}, \citenamefont {Armstrong}, \citenamefont {Giuliani},
  \citenamefont {Godbey}, \citenamefont {Gandolfi},\ and\ \citenamefont
  {Tews}}]{Somasundaram2025a}%
  \BibitemOpen
  \bibfield  {author} {\bibinfo {author} {\bibfnamefont {R.}~\bibnamefont
  {Somasundaram}}, \bibinfo {author} {\bibfnamefont {C.~L.}\ \bibnamefont
  {Armstrong}}, \bibinfo {author} {\bibfnamefont {P.}~\bibnamefont {Giuliani}},
  \bibinfo {author} {\bibfnamefont {K.}~\bibnamefont {Godbey}}, \bibinfo
  {author} {\bibfnamefont {S.}~\bibnamefont {Gandolfi}},\ and\ \bibinfo
  {author} {\bibfnamefont {I.}~\bibnamefont {Tews}},\ }\bibfield  {title}
  {\bibinfo {title} {Emulators for scarce and noisy data: {Application to
  auxiliary-field diffusion Monte Carlo} for the deuteron},\ }\href
  {https://doi.org/10.1016/j.physletb.2025.139558} {\bibfield  {journal}
  {\bibinfo  {journal} {Phys. Lett. B}\ ,\ \bibinfo {pages} {139558}} (\bibinfo
  {year} {2025})}\BibitemShut {NoStop}%
\bibitem [{\citenamefont {Cook}\ \emph {et~al.}(2025)\citenamefont {Cook},
  \citenamefont {Jammooa}, \citenamefont {Hjorth-Jensen}, \citenamefont {Lee},\
  and\ \citenamefont {Lee}}]{Cook2025a}%
  \BibitemOpen
  \bibfield  {author} {\bibinfo {author} {\bibfnamefont {P.}~\bibnamefont
  {Cook}}, \bibinfo {author} {\bibfnamefont {D.}~\bibnamefont {Jammooa}},
  \bibinfo {author} {\bibfnamefont {M.}~\bibnamefont {Hjorth-Jensen}}, \bibinfo
  {author} {\bibfnamefont {D.~D.}\ \bibnamefont {Lee}},\ and\ \bibinfo {author}
  {\bibfnamefont {D.}~\bibnamefont {Lee}},\ }\bibfield  {title} {\bibinfo
  {title} {Parametric matrix models},\ }\href
  {https://doi.org/10.1038/s41467-025-61362-4} {\bibfield  {journal} {\bibinfo
  {journal} {Nat. Commun.}\ }\textbf {\bibinfo {volume} {16}},\ \bibinfo
  {pages} {5929} (\bibinfo {year} {2025})}\BibitemShut {NoStop}%
\bibitem [{\citenamefont {Bakurov}\ \emph {et~al.}(2025)\citenamefont
  {Bakurov}, \citenamefont {Giuliani}, \citenamefont {Godbey}, \citenamefont
  {Haut}, \citenamefont {Banzhaf},\ and\ \citenamefont
  {Nazarewicz}}]{Bakurov2025}%
  \BibitemOpen
  \bibfield  {author} {\bibinfo {author} {\bibfnamefont {I.}~\bibnamefont
  {Bakurov}}, \bibinfo {author} {\bibfnamefont {P.}~\bibnamefont {Giuliani}},
  \bibinfo {author} {\bibfnamefont {K.}~\bibnamefont {Godbey}}, \bibinfo
  {author} {\bibfnamefont {N.}~\bibnamefont {Haut}}, \bibinfo {author}
  {\bibfnamefont {W.}~\bibnamefont {Banzhaf}},\ and\ \bibinfo {author}
  {\bibfnamefont {W.}~\bibnamefont {Nazarewicz}},\ }\bibfield  {title}
  {\bibinfo {title} {Genetic programming for the nuclear many-body problem: a
  guide},\ }\href {https://doi.org/10.1088/1361-6471/ae0b98} {\bibfield
  {journal} {\bibinfo  {journal} {J. Phys. G}\ }\textbf {\bibinfo {volume}
  {52}},\ \bibinfo {pages} {102001} (\bibinfo {year} {2025})}\BibitemShut
  {NoStop}%
\bibitem [{\citenamefont {Reed}\ \emph {et~al.}(2024)\citenamefont {Reed},
  \citenamefont {Somasundaram}, \citenamefont {De}, \citenamefont {Armstrong},
  \citenamefont {Giuliani}, \citenamefont {Capano}, \citenamefont {Brown},\
  and\ \citenamefont {Tews}}]{Reed2024a}%
  \BibitemOpen
  \bibfield  {author} {\bibinfo {author} {\bibfnamefont {B.~T.}\ \bibnamefont
  {Reed}}, \bibinfo {author} {\bibfnamefont {R.}~\bibnamefont {Somasundaram}},
  \bibinfo {author} {\bibfnamefont {S.}~\bibnamefont {De}}, \bibinfo {author}
  {\bibfnamefont {C.~L.}\ \bibnamefont {Armstrong}}, \bibinfo {author}
  {\bibfnamefont {P.}~\bibnamefont {Giuliani}}, \bibinfo {author}
  {\bibfnamefont {C.}~\bibnamefont {Capano}}, \bibinfo {author} {\bibfnamefont
  {D.~A.}\ \bibnamefont {Brown}},\ and\ \bibinfo {author} {\bibfnamefont
  {I.}~\bibnamefont {Tews}},\ }\bibfield  {title} {\bibinfo {title} {Toward
  accelerated nuclear-physics parameter estimation from binary neutron star
  mergers: Emulators for the tolman--oppenheimer--volkoff equations},\ }\href
  {https://doi.org/10.3847/1538-4357/ad737c} {\bibfield  {journal} {\bibinfo
  {journal} {Astrophys. J.}\ }\textbf {\bibinfo {volume} {974}},\ \bibinfo
  {pages} {285} (\bibinfo {year} {2024})}\BibitemShut {NoStop}%
\bibitem [{\citenamefont {Xiao}\ \emph {et~al.}(2024)\citenamefont {Xiao},
  \citenamefont {Liu}, \citenamefont {Zu}, \citenamefont {Chai}, \citenamefont
  {He},\ and\ \citenamefont {Zhang}}]{Xiao2024a}%
  \BibitemOpen
  \bibfield  {author} {\bibinfo {author} {\bibfnamefont {W.}~\bibnamefont
  {Xiao}}, \bibinfo {author} {\bibfnamefont {X.}~\bibnamefont {Liu}}, \bibinfo
  {author} {\bibfnamefont {J.}~\bibnamefont {Zu}}, \bibinfo {author}
  {\bibfnamefont {X.}~\bibnamefont {Chai}}, \bibinfo {author} {\bibfnamefont
  {H.}~\bibnamefont {He}},\ and\ \bibinfo {author} {\bibfnamefont
  {T.}~\bibnamefont {Zhang}},\ }\bibfield  {title} {\bibinfo {title} {Operator
  inference driven data assimilation for high-fidelity neutron transport},\
  }\href {https://doi.org/10.1016/j.cma.2024.117214} {\bibfield  {journal}
  {\bibinfo  {journal} {Comput. Methods Appl. Mech. Eng.}\ }\textbf {\bibinfo
  {volume} {430}},\ \bibinfo {pages} {117214} (\bibinfo {year}
  {2024})}\BibitemShut {NoStop}%
\bibitem [{\citenamefont {Di~Ronco}\ \emph {et~al.}(2020)\citenamefont
  {Di~Ronco}, \citenamefont {Introini}, \citenamefont {Cervi}, \citenamefont
  {Lorenzi}, \citenamefont {Jeong}, \citenamefont {Seo}, \citenamefont {Bang},
  \citenamefont {Giacobbo},\ and\ \citenamefont {Cammi}}]{DiRonco2020a}%
  \BibitemOpen
  \bibfield  {author} {\bibinfo {author} {\bibfnamefont {A.}~\bibnamefont
  {Di~Ronco}}, \bibinfo {author} {\bibfnamefont {C.}~\bibnamefont {Introini}},
  \bibinfo {author} {\bibfnamefont {E.}~\bibnamefont {Cervi}}, \bibinfo
  {author} {\bibfnamefont {S.}~\bibnamefont {Lorenzi}}, \bibinfo {author}
  {\bibfnamefont {Y.~S.}\ \bibnamefont {Jeong}}, \bibinfo {author}
  {\bibfnamefont {S.~B.}\ \bibnamefont {Seo}}, \bibinfo {author} {\bibfnamefont
  {I.~C.}\ \bibnamefont {Bang}}, \bibinfo {author} {\bibfnamefont
  {F.}~\bibnamefont {Giacobbo}},\ and\ \bibinfo {author} {\bibfnamefont
  {A.}~\bibnamefont {Cammi}},\ }\bibfield  {title} {\bibinfo {title} {Dynamic
  mode decomposition for the stability analysis of the molten salt fast reactor
  core},\ }\href {https://doi.org/10.1016/j.nucengdes.2020.110529} {\bibfield
  {journal} {\bibinfo  {journal} {Nucl. Eng. Des.}\ }\textbf {\bibinfo {volume}
  {362}},\ \bibinfo {pages} {110529} (\bibinfo {year} {2020})}\BibitemShut
  {NoStop}%
\bibitem [{\citenamefont {Hardy}\ and\ \citenamefont
  {Morel}(2024)}]{Hardy2024a}%
  \BibitemOpen
  \bibfield  {author} {\bibinfo {author} {\bibfnamefont {Z.~K.}\ \bibnamefont
  {Hardy}}\ and\ \bibinfo {author} {\bibfnamefont {J.~E.}\ \bibnamefont
  {Morel}},\ }\bibfield  {title} {\bibinfo {title} {Proper orthogonal
  decomposition mode coefficient interpolation: A non-intrusive reduced-order
  model for parametric reactor kinetics},\ }\href
  {https://doi.org/10.1080/00295639.2023.2218581} {\bibfield  {journal}
  {\bibinfo  {journal} {Nucl. Sci. Eng.}\ }\textbf {\bibinfo {volume} {198}},\
  \bibinfo {pages} {832} (\bibinfo {year} {2024})}\BibitemShut {NoStop}%
\bibitem [{\citenamefont {Jin}\ \emph {et~al.}(2025)\citenamefont {Jin},
  \citenamefont {Ravli{\'c}}, \citenamefont {Godbey}, \citenamefont
  {Giuliani},\ and\ \citenamefont {Nazarewicz}}]{GitHub2025a}%
  \BibitemOpen
  \bibfield  {author} {\bibinfo {author} {\bibfnamefont {L.}~\bibnamefont
  {Jin}}, \bibinfo {author} {\bibfnamefont {A.}~\bibnamefont {Ravli{\'c}}},
  \bibinfo {author} {\bibfnamefont {K.}~\bibnamefont {Godbey}}, \bibinfo
  {author} {\bibfnamefont {P.}~\bibnamefont {Giuliani}},\ and\ \bibinfo
  {author} {\bibfnamefont {W.}~\bibnamefont {Nazarewicz}},\ }\href@noop {}
  {\bibinfo {title} {{Surrogate Models for Linear Response GitHub
  Repository}}},\ \bibinfo {howpublished} {\url{https://github.com/ascsn/SMLR}}
  (\bibinfo {year} {2025})\BibitemShut {NoStop}%
\bibitem [{sml()}]{smlrWebsite}%
  \BibitemOpen
  \href@noop {} {\bibinfo {title} {{Interactive Visualization Repository}}},\
  \bibinfo {howpublished}
  {\href{https://smlr.ascsn.net}{\nolinkurl{smlr.ascsn.net}}}\BibitemShut
  {NoStop}%
\bibitem [{\citenamefont {Nik{\v{s}}i{\'c}}\ \emph {et~al.}(2014)\citenamefont
  {Nik{\v{s}}i{\'c}}, \citenamefont {Paar}, \citenamefont {Vretenar},\ and\
  \citenamefont {Ring}}]{Niksic2014a}%
  \BibitemOpen
  \bibfield  {author} {\bibinfo {author} {\bibfnamefont {T.}~\bibnamefont
  {Nik{\v{s}}i{\'c}}}, \bibinfo {author} {\bibfnamefont {N.}~\bibnamefont
  {Paar}}, \bibinfo {author} {\bibfnamefont {D.}~\bibnamefont {Vretenar}},\
  and\ \bibinfo {author} {\bibfnamefont {P.}~\bibnamefont {Ring}},\ }\bibfield
  {title} {\bibinfo {title} {{DIRHB—A }relativistic self-consistent
  mean-field framework for atomic nuclei},\ }\href
  {https://doi.org/10.1016/j.cpc.2014.02.027} {\bibfield  {journal} {\bibinfo
  {journal} {Comput. Phys. Commun.}\ }\textbf {\bibinfo {volume} {185}},\
  \bibinfo {pages} {1808} (\bibinfo {year} {2014})}\BibitemShut {NoStop}%
\bibitem [{\citenamefont {Ring}\ \emph {et~al.}(1984)\citenamefont {Ring},
  \citenamefont {Robledo}, \citenamefont {Egido},\ and\ \citenamefont
  {Faber}}]{Ring1984a}%
  \BibitemOpen
  \bibfield  {author} {\bibinfo {author} {\bibfnamefont {P.}~\bibnamefont
  {Ring}}, \bibinfo {author} {\bibfnamefont {L.~M.}\ \bibnamefont {Robledo}},
  \bibinfo {author} {\bibfnamefont {J.~L.}\ \bibnamefont {Egido}},\ and\
  \bibinfo {author} {\bibfnamefont {M.}~\bibnamefont {Faber}},\ }\bibfield
  {title} {\bibinfo {title} {Microscopic theory of the isovector dipole
  resonance at high angular momenta},\ }\href
  {https://doi.org/10.1016/0375-9474(84)90393-2} {\bibfield  {journal}
  {\bibinfo  {journal} {Nucl. Phys. A}\ }\textbf {\bibinfo {volume} {419}},\
  \bibinfo {pages} {261} (\bibinfo {year} {1984})}\BibitemShut {NoStop}%
\bibitem [{\citenamefont {Ravli{\'c}}\ \emph {et~al.}(2021)\citenamefont
  {Ravli{\'c}}, \citenamefont {Niu}, \citenamefont {Nik{\v{s}}i{\'c}},
  \citenamefont {Paar},\ and\ \citenamefont {Ring}}]{Ravlic2021a}%
  \BibitemOpen
  \bibfield  {author} {\bibinfo {author} {\bibfnamefont {A.}~\bibnamefont
  {Ravli{\'c}}}, \bibinfo {author} {\bibfnamefont {Y.~F.}\ \bibnamefont {Niu}},
  \bibinfo {author} {\bibfnamefont {T.}~\bibnamefont {Nik{\v{s}}i{\'c}}},
  \bibinfo {author} {\bibfnamefont {N.}~\bibnamefont {Paar}},\ and\ \bibinfo
  {author} {\bibfnamefont {P.}~\bibnamefont {Ring}},\ }\bibfield  {title}
  {\bibinfo {title} {Finite-temperature linear response theory based on
  relativistic hartree–bogoliubov model with point-coupling interaction},\
  }\href {https://doi.org/10.1103/PhysRevC.104.064302} {\bibfield  {journal}
  {\bibinfo  {journal} {Phys. Rev. C}\ }\textbf {\bibinfo {volume} {104}},\
  \bibinfo {pages} {064302} (\bibinfo {year} {2021})}\BibitemShut {NoStop}%
\bibitem [{\citenamefont {Ravli{\'c}}\ \emph {et~al.}(2024)\citenamefont
  {Ravli{\'c}}, \citenamefont {Nik{\v{s}}i{\'c}}, \citenamefont {Niu},
  \citenamefont {Ring},\ and\ \citenamefont {Paar}}]{Ravlic2024a}%
  \BibitemOpen
  \bibfield  {author} {\bibinfo {author} {\bibfnamefont {A.}~\bibnamefont
  {Ravli{\'c}}}, \bibinfo {author} {\bibfnamefont {T.}~\bibnamefont
  {Nik{\v{s}}i{\'c}}}, \bibinfo {author} {\bibfnamefont {Y.~F.}\ \bibnamefont
  {Niu}}, \bibinfo {author} {\bibfnamefont {P.}~\bibnamefont {Ring}},\ and\
  \bibinfo {author} {\bibfnamefont {N.}~\bibnamefont {Paar}},\ }\bibfield
  {title} {\bibinfo {title} {Axially deformed relativistic quasiparticle
  random-phase approximation based on point-coupling interactions},\ }\href
  {https://doi.org/10.1103/PhysRevC.110.024323} {\bibfield  {journal} {\bibinfo
   {journal} {Phys. Rev. C}\ }\textbf {\bibinfo {volume} {110}},\ \bibinfo
  {pages} {024323} (\bibinfo {year} {2024})}\BibitemShut {NoStop}%
\bibitem [{\citenamefont {Reinhard}\ and\ \citenamefont
  {Nazarewicz}(2010)}]{Reinhard2010a}%
  \BibitemOpen
  \bibfield  {author} {\bibinfo {author} {\bibfnamefont {P.-G.}\ \bibnamefont
  {Reinhard}}\ and\ \bibinfo {author} {\bibfnamefont {W.}~\bibnamefont
  {Nazarewicz}},\ }\bibfield  {title} {\bibinfo {title} {Information content of
  a new observable: The case of the nuclear neutron skin},\ }\href
  {https://doi.org/10.1103/PhysRevC.81.051303} {\bibfield  {journal} {\bibinfo
  {journal} {Phys. Rev. C}\ }\textbf {\bibinfo {volume} {81}},\ \bibinfo
  {pages} {051303} (\bibinfo {year} {2010})}\BibitemShut {NoStop}%
\bibitem [{\citenamefont {Piekarewicz}\ \emph {et~al.}(2012)\citenamefont
  {Piekarewicz}, \citenamefont {Agrawal}, \citenamefont {Col{\`o}},
  \citenamefont {Nazarewicz}, \citenamefont {Paar}, \citenamefont {Reinhard},
  \citenamefont {Roca-Maza},\ and\ \citenamefont
  {Vretenar}}]{Piekarewicz2012a}%
  \BibitemOpen
  \bibfield  {author} {\bibinfo {author} {\bibfnamefont {J.}~\bibnamefont
  {Piekarewicz}}, \bibinfo {author} {\bibfnamefont {B.~K.}\ \bibnamefont
  {Agrawal}}, \bibinfo {author} {\bibfnamefont {G.}~\bibnamefont {Col{\`o}}},
  \bibinfo {author} {\bibfnamefont {W.}~\bibnamefont {Nazarewicz}}, \bibinfo
  {author} {\bibfnamefont {N.}~\bibnamefont {Paar}}, \bibinfo {author}
  {\bibfnamefont {P.-G.}\ \bibnamefont {Reinhard}}, \bibinfo {author}
  {\bibfnamefont {X.}~\bibnamefont {Roca-Maza}},\ and\ \bibinfo {author}
  {\bibfnamefont {D.}~\bibnamefont {Vretenar}},\ }\bibfield  {title} {\bibinfo
  {title} {Electric dipole polarizability and the neutron skin},\ }\href
  {https://doi.org/10.1103/PhysRevC.85.041302} {\bibfield  {journal} {\bibinfo
  {journal} {Phys. Rev. C}\ }\textbf {\bibinfo {volume} {85}},\ \bibinfo
  {pages} {041302} (\bibinfo {year} {2012})}\BibitemShut {NoStop}%
\bibitem [{\citenamefont {Lipparini}\ and\ \citenamefont
  {Stringari}(1989)}]{Lipparini1989a}%
  \BibitemOpen
  \bibfield  {author} {\bibinfo {author} {\bibfnamefont {E.}~\bibnamefont
  {Lipparini}}\ and\ \bibinfo {author} {\bibfnamefont {S.}~\bibnamefont
  {Stringari}},\ }\bibfield  {title} {\bibinfo {title} {Sum rules and giant
  resonances in nuclei},\ }\href {https://doi.org/10.1016/0370-1573(89)90029-X}
  {\bibfield  {journal} {\bibinfo  {journal} {Phys. Rep.}\ }\textbf {\bibinfo
  {volume} {175}},\ \bibinfo {pages} {103} (\bibinfo {year}
  {1989})}\BibitemShut {NoStop}%
\bibitem [{\citenamefont {Ravli{\'c}}\ and\ \citenamefont
  {Nazarewicz}(2025)}]{Ravlic2025a}%
  \BibitemOpen
  \bibfield  {author} {\bibinfo {author} {\bibfnamefont {A.}~\bibnamefont
  {Ravli{\'c}}}\ and\ \bibinfo {author} {\bibfnamefont {W.}~\bibnamefont
  {Nazarewicz}},\ }\bibfield  {title} {\bibinfo {title} {Weak decays in
  superheavy nuclei},\ }\href {https://doi.org/10.1103/PhysRevC.111.L051305}
  {\bibfield  {journal} {\bibinfo  {journal} {Phys. Rev. C}\ }\textbf {\bibinfo
  {volume} {111}},\ \bibinfo {pages} {L051305} (\bibinfo {year}
  {2025})}\BibitemShut {NoStop}%
\bibitem [{\citenamefont {Behrens}\ and\ \citenamefont
  {B{\"u}hring}(1971)}]{Behrens1971a}%
  \BibitemOpen
  \bibfield  {author} {\bibinfo {author} {\bibfnamefont {H.}~\bibnamefont
  {Behrens}}\ and\ \bibinfo {author} {\bibfnamefont {W.}~\bibnamefont
  {B{\"u}hring}},\ }\bibfield  {title} {\bibinfo {title} {Nuclear beta decay},\
  }\href {https://doi.org/10.1016/0375-9474(71)90489-1} {\bibfield  {journal}
  {\bibinfo  {journal} {Nucl. Phys. A}\ }\textbf {\bibinfo {volume} {162}},\
  \bibinfo {pages} {111} (\bibinfo {year} {1971})}\BibitemShut {NoStop}%
\bibitem [{\citenamefont {Bjel{\v{c}}i{\'c}}\ and\ \citenamefont
  {Nik{\v{s}}i{\'c}}(2020)}]{Bjelcic2020a}%
  \BibitemOpen
  \bibfield  {author} {\bibinfo {author} {\bibfnamefont {A.}~\bibnamefont
  {Bjel{\v{c}}i{\'c}}}\ and\ \bibinfo {author} {\bibfnamefont {T.}~\bibnamefont
  {Nik{\v{s}}i{\'c}}},\ }\bibfield  {title} {\bibinfo {title} {Implementation
  of the quasiparticle finite amplitude method within the relativistic
  self-consistent mean-field framework: The program {DIRQFAM}},\ }\href
  {https://doi.org/10.1016/j.cpc.2020.107184} {\bibfield  {journal} {\bibinfo
  {journal} {Comput. Phys. Commun.}\ }\textbf {\bibinfo {volume} {253}},\
  \bibinfo {pages} {107184} (\bibinfo {year} {2020})}\BibitemShut {NoStop}%
\bibitem [{\citenamefont {Nik{\v{s}}i{\'c}}\ \emph {et~al.}(2008)\citenamefont
  {Nik{\v{s}}i{\'c}}, \citenamefont {Vretenar},\ and\ \citenamefont
  {Ring}}]{Niksic2008a}%
  \BibitemOpen
  \bibfield  {author} {\bibinfo {author} {\bibfnamefont {T.}~\bibnamefont
  {Nik{\v{s}}i{\'c}}}, \bibinfo {author} {\bibfnamefont {D.}~\bibnamefont
  {Vretenar}},\ and\ \bibinfo {author} {\bibfnamefont {P.}~\bibnamefont
  {Ring}},\ }\bibfield  {title} {\bibinfo {title} {Relativistic nuclear energy
  density functionals: Adjusting parameters to binding energies},\ }\href
  {https://doi.org/10.1103/PhysRevC.78.034318} {\bibfield  {journal} {\bibinfo
  {journal} {Phys. Rev. C}\ }\textbf {\bibinfo {volume} {78}},\ \bibinfo
  {pages} {034318} (\bibinfo {year} {2008})}\BibitemShut {NoStop}%
\bibitem [{\citenamefont {Tian}\ \emph {et~al.}(2009)\citenamefont {Tian},
  \citenamefont {Ma},\ and\ \citenamefont {Ring}}]{Tian2009a}%
  \BibitemOpen
  \bibfield  {author} {\bibinfo {author} {\bibfnamefont {Y.}~\bibnamefont
  {Tian}}, \bibinfo {author} {\bibfnamefont {Z.-Y.}\ \bibnamefont {Ma}},\ and\
  \bibinfo {author} {\bibfnamefont {P.}~\bibnamefont {Ring}},\ }\bibfield
  {title} {\bibinfo {title} {Axially deformed relativistic
  {Hartree–Bogoliubov }theory with a separable pairing force},\ }\href
  {https://doi.org/10.1103/PhysRevC.80.024313} {\bibfield  {journal} {\bibinfo
  {journal} {Phys. Rev. C}\ }\textbf {\bibinfo {volume} {80}},\ \bibinfo
  {pages} {024313} (\bibinfo {year} {2009})}\BibitemShut {NoStop}%
\bibitem [{\citenamefont {Hinohara}\ \emph {et~al.}(2015)\citenamefont
  {Hinohara}, \citenamefont {Kortelainen}, \citenamefont {Nazarewicz},\ and\
  \citenamefont {Olsen}}]{Hinohara2015a}%
  \BibitemOpen
  \bibfield  {author} {\bibinfo {author} {\bibfnamefont {N.}~\bibnamefont
  {Hinohara}}, \bibinfo {author} {\bibfnamefont {M.}~\bibnamefont
  {Kortelainen}}, \bibinfo {author} {\bibfnamefont {W.}~\bibnamefont
  {Nazarewicz}},\ and\ \bibinfo {author} {\bibfnamefont {E.}~\bibnamefont
  {Olsen}},\ }\bibfield  {title} {\bibinfo {title} {Complex-energy approach to
  sum rules within nuclear density functional theory},\ }\href
  {https://doi.org/10.1103/PhysRevC.91.044323} {\bibfield  {journal} {\bibinfo
  {journal} {Phys. Rev. C}\ }\textbf {\bibinfo {volume} {91}},\ \bibinfo
  {pages} {044323} (\bibinfo {year} {2015})}\BibitemShut {NoStop}%
\bibitem [{\citenamefont {Vale}\ \emph {et~al.}(2021)\citenamefont {Vale},
  \citenamefont {Niu},\ and\ \citenamefont {Paar}}]{Vale2021a}%
  \BibitemOpen
  \bibfield  {author} {\bibinfo {author} {\bibfnamefont {D.}~\bibnamefont
  {Vale}}, \bibinfo {author} {\bibfnamefont {Y.~F.}\ \bibnamefont {Niu}},\ and\
  \bibinfo {author} {\bibfnamefont {N.}~\bibnamefont {Paar}},\ }\bibfield
  {title} {\bibinfo {title} {Nuclear charge-exchange excitations based on a
  relativistic density-dependent point-coupling model},\ }\href
  {https://doi.org/10.1103/PhysRevC.103.064307} {\bibfield  {journal} {\bibinfo
   {journal} {Phys. Rev. C}\ }\textbf {\bibinfo {volume} {103}},\ \bibinfo
  {pages} {064307} (\bibinfo {year} {2021})}\BibitemShut {NoStop}%
\bibitem [{\citenamefont {Kovachki}\ \emph {et~al.}(2023)\citenamefont
  {Kovachki}, \citenamefont {Li}, \citenamefont {Liu}, \citenamefont
  {Azizzadenesheli}, \citenamefont {Bhattacharya}, \citenamefont {Stuart},\
  and\ \citenamefont {Anandkumar}}]{Kovachki2023a}%
  \BibitemOpen
  \bibfield  {author} {\bibinfo {author} {\bibfnamefont {N.}~\bibnamefont
  {Kovachki}}, \bibinfo {author} {\bibfnamefont {Z.}~\bibnamefont {Li}},
  \bibinfo {author} {\bibfnamefont {B.}~\bibnamefont {Liu}}, \bibinfo {author}
  {\bibfnamefont {K.}~\bibnamefont {Azizzadenesheli}}, \bibinfo {author}
  {\bibfnamefont {K.}~\bibnamefont {Bhattacharya}}, \bibinfo {author}
  {\bibfnamefont {A.}~\bibnamefont {Stuart}},\ and\ \bibinfo {author}
  {\bibfnamefont {A.}~\bibnamefont {Anandkumar}},\ }\bibfield  {title}
  {\bibinfo {title} {Neural operator: Learning maps between function spaces
  with applications to pdes},\ }\href
  {https://www.jmlr.org/papers/v24/21-1524.html} {\bibfield  {journal}
  {\bibinfo  {journal} {J. Mach. Learn. Res.}\ }\textbf {\bibinfo {volume}
  {24}},\ \bibinfo {pages} {1} (\bibinfo {year} {2023})}\BibitemShut {NoStop}%
\bibitem [{\citenamefont {Brody}\ \emph {et~al.}(1981)\citenamefont {Brody},
  \citenamefont {Flores}, \citenamefont {French}, \citenamefont {Mello},
  \citenamefont {Pandey},\ and\ \citenamefont {Wong}}]{Brody1981a}%
  \BibitemOpen
  \bibfield  {author} {\bibinfo {author} {\bibfnamefont {T.~A.}\ \bibnamefont
  {Brody}}, \bibinfo {author} {\bibfnamefont {J.}~\bibnamefont {Flores}},
  \bibinfo {author} {\bibfnamefont {J.~B.}\ \bibnamefont {French}}, \bibinfo
  {author} {\bibfnamefont {P.~A.}\ \bibnamefont {Mello}}, \bibinfo {author}
  {\bibfnamefont {A.}~\bibnamefont {Pandey}},\ and\ \bibinfo {author}
  {\bibfnamefont {S.~S.~M.}\ \bibnamefont {Wong}},\ }\bibfield  {title}
  {\bibinfo {title} {Random-matrix physics: Spectrum and strength
  fluctuations},\ }\href {https://doi.org/10.1103/RevModPhys.53.385} {\bibfield
   {journal} {\bibinfo  {journal} {Rev. Mod. Phys.}\ }\textbf {\bibinfo
  {volume} {53}},\ \bibinfo {pages} {385} (\bibinfo {year} {1981})}\BibitemShut
  {NoStop}%
\bibitem [{\citenamefont {Severyukhin}\ \emph {et~al.}(2017)\citenamefont
  {Severyukhin}, \citenamefont {\AA{}berg}, \citenamefont {Arsenyev},\ and\
  \citenamefont {Nazmitdinov}}]{Severyukhin2017a}%
  \BibitemOpen
  \bibfield  {author} {\bibinfo {author} {\bibfnamefont {A.~P.}\ \bibnamefont
  {Severyukhin}}, \bibinfo {author} {\bibfnamefont {S.}~\bibnamefont
  {\AA{}berg}}, \bibinfo {author} {\bibfnamefont {N.~N.}\ \bibnamefont
  {Arsenyev}},\ and\ \bibinfo {author} {\bibfnamefont {R.~G.}\ \bibnamefont
  {Nazmitdinov}},\ }\bibfield  {title} {\bibinfo {title} {Spreading widths of
  giant resonances in spherical nuclei: Damped transient response},\ }\href
  {https://doi.org/10.1103/PhysRevC.95.061305} {\bibfield  {journal} {\bibinfo
  {journal} {Phys. Rev. C}\ }\textbf {\bibinfo {volume} {95}},\ \bibinfo
  {pages} {061305} (\bibinfo {year} {2017})}\BibitemShut {NoStop}%
\bibitem [{\citenamefont {Abadi}\ \emph {et~al.}(2016)\citenamefont {Abadi},
  \citenamefont {Barham}, \citenamefont {Chen}, \citenamefont {Chen},
  \citenamefont {Davis}, \citenamefont {Dean}, \citenamefont {Devin},
  \citenamefont {Ghemawat}, \citenamefont {Irving}, \citenamefont {Isard},
  \citenamefont {Kudlur}, \citenamefont {Levenberg}, \citenamefont {Monga},
  \citenamefont {Moore}, \citenamefont {Murray}, \citenamefont {Steiner},
  \citenamefont {Tucker}, \citenamefont {Vasudevan}, \citenamefont {Warden},
  \citenamefont {Wicke}, \citenamefont {Yu},\ and\ \citenamefont
  {Zheng}}]{TensorFlow2016a}%
  \BibitemOpen
  \bibfield  {author} {\bibinfo {author} {\bibfnamefont {M.}~\bibnamefont
  {Abadi}}, \bibinfo {author} {\bibfnamefont {P.}~\bibnamefont {Barham}},
  \bibinfo {author} {\bibfnamefont {J.}~\bibnamefont {Chen}}, \bibinfo {author}
  {\bibfnamefont {Z.}~\bibnamefont {Chen}}, \bibinfo {author} {\bibfnamefont
  {A.}~\bibnamefont {Davis}}, \bibinfo {author} {\bibfnamefont
  {J.}~\bibnamefont {Dean}}, \bibinfo {author} {\bibfnamefont {M.}~\bibnamefont
  {Devin}}, \bibinfo {author} {\bibfnamefont {S.}~\bibnamefont {Ghemawat}},
  \bibinfo {author} {\bibfnamefont {G.}~\bibnamefont {Irving}}, \bibinfo
  {author} {\bibfnamefont {M.}~\bibnamefont {Isard}}, \bibinfo {author}
  {\bibfnamefont {M.}~\bibnamefont {Kudlur}}, \bibinfo {author} {\bibfnamefont
  {J.}~\bibnamefont {Levenberg}}, \bibinfo {author} {\bibfnamefont
  {R.}~\bibnamefont {Monga}}, \bibinfo {author} {\bibfnamefont
  {S.}~\bibnamefont {Moore}}, \bibinfo {author} {\bibfnamefont {D.~G.}\
  \bibnamefont {Murray}}, \bibinfo {author} {\bibfnamefont {B.}~\bibnamefont
  {Steiner}}, \bibinfo {author} {\bibfnamefont {P.}~\bibnamefont {Tucker}},
  \bibinfo {author} {\bibfnamefont {V.}~\bibnamefont {Vasudevan}}, \bibinfo
  {author} {\bibfnamefont {P.}~\bibnamefont {Warden}}, \bibinfo {author}
  {\bibfnamefont {M.}~\bibnamefont {Wicke}}, \bibinfo {author} {\bibfnamefont
  {Y.}~\bibnamefont {Yu}},\ and\ \bibinfo {author} {\bibfnamefont
  {X.}~\bibnamefont {Zheng}},\ }\bibfield  {title} {\bibinfo {title}
  {Tensorflow: A system for large-scale machine learning},\ }in\ \href
  {https://www.usenix.org/conference/osdi16/technical-sessions/presentation/abadi}
  {\emph {\bibinfo {booktitle} {Proc. 12th USENIX Symp. Oper. Syst. Des.
  Implement. (OSDI 16)}}}\ (\bibinfo {address} {Savannah, GA},\ \bibinfo {year}
  {2016})\ pp.\ \bibinfo {pages} {265--283}\BibitemShut {NoStop}%
\bibitem [{\citenamefont {Hintze}\ and\ \citenamefont
  {Nelson}(1998)}]{Hintze1998a}%
  \BibitemOpen
  \bibfield  {author} {\bibinfo {author} {\bibfnamefont {J.~L.}\ \bibnamefont
  {Hintze}}\ and\ \bibinfo {author} {\bibfnamefont {R.~D.}\ \bibnamefont
  {Nelson}},\ }\bibfield  {title} {\bibinfo {title} {Violin plots: A box
  plot–density trace synergism},\ }\href
  {https://doi.org/10.1080/00031305.1998.10480559} {\bibfield  {journal}
  {\bibinfo  {journal} {Am. Stat.}\ }\textbf {\bibinfo {volume} {52}},\
  \bibinfo {pages} {181} (\bibinfo {year} {1998})}\BibitemShut {NoStop}%
\bibitem [{\citenamefont {Maldonado}\ \emph {et~al.}(2025)\citenamefont
  {Maldonado}, \citenamefont {Drischler}, \citenamefont {Furnstahl},\ and\
  \citenamefont {Mlinari\ifmmode~\acute{c}\else \'{c}\fi{}}}]{Maldonado2025a}%
  \BibitemOpen
  \bibfield  {author} {\bibinfo {author} {\bibfnamefont {J.~M.}\ \bibnamefont
  {Maldonado}}, \bibinfo {author} {\bibfnamefont {C.}~\bibnamefont
  {Drischler}}, \bibinfo {author} {\bibfnamefont {R.~J.}\ \bibnamefont
  {Furnstahl}},\ and\ \bibinfo {author} {\bibfnamefont {P.}~\bibnamefont
  {Mlinari\ifmmode~\acute{c}\else \'{c}\fi{}}},\ }\bibfield  {title} {\bibinfo
  {title} {Greedy emulators for nuclear two-body scattering},\ }\href
  {https://doi.org/10.1103/k77q-f82l} {\bibfield  {journal} {\bibinfo
  {journal} {Phys. Rev. C}\ }\textbf {\bibinfo {volume} {112}},\ \bibinfo
  {pages} {024002} (\bibinfo {year} {2025})}\BibitemShut {NoStop}%
\end{thebibliography}%

\end{document}